\documentclass{aa}  

\usepackage{graphicx}
\usepackage{txfonts}
\begin{document}

   \title{Giant planet formation at the pressure maxima of protoplanetary disks}


   \author{ O. M. Guilera 
     \inst{1,2}
     \and
     Zs. S\'andor
     \inst{3}
   }

   \institute{ Planetary Science Group, Astrophysical Institute of La Plata, National Scientific and Technical Research Council and National University of La Plata, Paseo del Bosque s/n, 1900 La Plata, Argentina.
     \and
     Planetary Science Group, Faculty of Astronomical and Geophysical Sciences, National University of La Plata, Paseo del Bosque s/n, 1900 La Plata, Argentina.\thanks{\email{oguilera@fcaglp.unlp.edu.ar}}
       \and
       Department of Astronomy, E\"otv\"os Lor\'and University, H-1117 Budapest, P\'azm\'any P\'eter s\'etany 1/A, Hungary. \thanks{\email{Zs.Sandor@astro.elte.hu}}
   }
 
   \date{Received 2016; }

 
  \abstract
   {In the classical core-accretion planet formation scenario, rapid inward migration and accretion timescales of kilometer size planetesimals may not favor the formation of massive cores of giant planets before the dissipation of protoplanetary disks. On the other hand, the existence of pressure maxima in the disk could act as migration traps and locations for solid material accumulation, favoring the formation of massive cores.}
   {We aim to study the radial drift of pebbles and planetesimals and planet migration at pressure maxima in a protoplanetary disk and their implications for the formation of massive cores as triggering a gaseous runaway accretion phase.}
   {The time evolution of a viscosity driven accretion disk is solved numerically introducing a a dead zone as a low-viscosity region in the protoplanetary disk. A population of pebbles and planetesimals evolving by radial drift and accretion by the planets is also considered. Finally, the embryos embedded in the disk grow by the simultaneous accretion of pebbles, planetesimals and the surrounding gas.}
   {Our simulations show that the pressure maxima generated at the edges of the low-viscosity region of the disk act as planet migration traps, and that the pebble and planetesimal surface densities are significantly increased due to the radial drift towards pressure maxima locations. However, our simulations also show that migration trap locations and solid material accumulation locations are not exactly at the same positions. Thus, a planet's semi-major axis oscillations around zero torque locations, predicted by MHD and HD simulations, are needed for the planet to accrete all the available material accumulated at the pressure maxima.}
   {Pressure maxima generated at the edges of a low-viscosity region of a protoplanetary disk seem to be preferential locations for the formation and trap of massive cores.}


   \maketitle
%

\section{Introduction}
\label{sec:intro}

In the classical core-accretion scenario \citep{bodenheimer_pollack1986,pollack_etal1996} the formation of planets begins on the smallest scale, and lasts from sub-micron sized dust to Jupiter-like giant planets. This scenario successfully explains the formation of both terrestrial and giant planets. One key element of the theory is the formation of the meter-sized bodies via dust coagulation, since these objects are the building blocks of planetesimals. Terrestrial planets, and solid cores of giant planets form by subsequent collision and accretion of planetesimals. The giant planet formation is completed with a rapid gas accretion from the ambient disk in a process known as gaseous runaway, when the mass of the envelope equals the mass of the core when it has $\sim 10~\text{M}_{\oplus}$. Despite its attractive completeness in predicting both the formation of terrestrial and giant planets, the classical core-accretion still has two groups of unresolved problems: the existence of the various \emph{barriers} hindering the growth of dust particles, and the \emph{timescale problem of the giant planet formation}. 

In this paper we pose the latter problem, which is due to the combination of the relatively short lifetime of a protoplanetary disk with the fast type I inward migration of massive solid cores of giant planets.  Observations indicate that the life-time of gaseous protoplanetary disks is about a few million years \citep{haisch_etal2001, Mamajek2009, Pfalzner.et.al.2014}. Consequently, an approximately $10~\text{M}_{\oplus}$ solid core should form early enough to be able to rapidly accrete gas reaching the giant planet phase. However, during its growing phase the planetary core is also subject to the fast type I migration \citep{ward1997}, whose rate is linearly proportional to the core's mass. According to theoretical models, the time needed to reach the critical mass for a planetary core may exceed its migration timescale. Recent new developments also show the possibility of outward type I migration \citep[see][]{paardekooper_etal2010, paardekooper_etal2011}. However, rapid inward or outward type I migration is still a danger to the full development of a Jupiter-like planet. In both cases the formation of a giant planet might be inhibited, because fast inward migration results in the quick loss of the embryo, while due to a rapid outward migration the embryo quickly reaches the outer part of the disk, where the surface density of planetesimals is very low, thus the timescale of the planetesimal accretion becomes very long, too.

In order to overcome the above problems, an extension to the classical core-accretion scenario has recently been suggested, namely, that there are particular places for planet growth. These places are the planet traps, where the torque responsible for the type I migration vanishes. If the location of a planet trap coincides with, or it is close to a local pressure maximum of gas, where the inward radial drift of dust and planetesimals is stalled, the growing protoplanet may increase its mass quickly by accreting the solid material accumulated there. In the following we exactly deal with this situation by investigating the mass accretion process of a growing core. We intend to demonstrate with this investigation that the formation time of a giant planet at a density/pressure maximum is significantly reduced being much shorter than the lifetime of the protoplanetary disk.

Our paper is organized as the following: first we describe the physical background behind the development of planet traps and planetesimal accumulation. In Section 3 we present our model of giant planet growth in a time evolving gas and planetesimal disk having a dead zone. In Section 4 the paper continues with the description of our result, and finally closes with a Conclusion and Summary.


\section{Physical background: the concept of the density/pressure maximum}
\label{sec:sec1}

The theoretical existence of a planet trap has been first reported by \citet{masset_etal2006} demonstrating that a steep surface density jump (being a maximum in the gas surface density) in a protoplanetary disk can halt the type I migration of planetary cores of several Earth masses. Moreover, assuming the equation of state for gas in the form $P=\rho c_s^2$, being $c_s$ the local sound speed, at the surface density maximum a pressure maximum also develops. Since at a pressure maximum the gas orbital velocity becomes circular Keplerian, the drag force felt by a particle disappears, therefore dust and the most sensitive planetesimals to aerodynamical drag can accumulate there. A pressure maximum itself can be a place where planetesimals born, since approaching the pressure maximum, the relative velocities between dust particles decrease, thus their collisions will not be destructive any longer. Moreover, at a pressure maximum, several barriers of dust growth can be overcome via coagulation and sweep up growth of small particles \citep{brauer_etal2008b, windmark_etal2012, drazkowska_etal2013}. 

The ideal candidates for the developments of density/pressure maxima are the inner and the outer boundaries of the \emph{dead zone}. It is widely accepted that gas accretion in a protoplanetary disk is driven by the turbulence caused by the magneto-rotational instability (MRI), which needs ionized plasma to be triggered  \citep{balbus_hawley1991}. However, only the very inner part of the disk is ionized (weakly) in its full column by thermal ionization of alkali metals, which cannot be sustained anymore when the temperature drops below 900 K. In the absence of thermal ionization, the other ionization sources could be the X-rays from the stellar magnetosphere, or cosmic rays. At high gas surface density the gas is self-shielded against these ionization sources, therefore gas accretion happens only in an upper layer \citep{gammie1996}. This part of the disk with reduced accretion is called as the dead zone. When the surface density drops as the function of the distance from the star, the external ionizing radiation can penetrate through the disk, which will be ionized again in its full column. Due to the reduced/enhanced accretion at the boundaries of the dead zone, density and consequently pressure maxima will develop. If certain physical conditions are fulfilled \citep{Lyraetal2015}, these density maxima can be manifested as large scale vortices, being prone to the Rossby wave instability \citep{Lovelaceetal1999}. With the current observation techniques these vortices, if they are far enough from the star, can be observed, and indeed, the horseshoe-shaped patterns discovered in dust continuum by the Sub-millimeter Array \citep{brown_etal2009} and recently by the Atacama Large Millimeter/Submillimeter Array \citep{vandermarel_etal2013, casassus_etal2013, Fukagawaetal2013, Isellaetal2013, Perezetal2014} might be attributed to the large dust collecting Rossby vortices \citep{regalyetal2012} indifferently whether they are assumed to be generated at the outer edge of the dead zone, or of a gap opened by a giant planet. 

Another candidate for the development of a pressure maximum might be the water snowline \citep{kretke_lin2007,brauer_etal2008b}. The water snowline separates the regions of the disk in which the water is in vapour and solid form. As the temperature drops below a certain limit (being approximately $170$ K for typical disk conditions) the water vapour condenses out possibly to the surface of silicate dust grains. Thus when crossing the snowline moving away form the star, the solid-to-gas ratio increases suddenly, affecting the strength of the MRI driven turbulence, as the number of free electrons, thus the conductivity of the plasma decreases suddenly, too \citep{Sanoetal2000, IlgnerNelson2006}. This change in the gas turbulence is also reflected in the gas accretion rate, being lower outside the snowline than inside of it. Therefore similarly to the inner edge of the dead zone, a density and a corresponding pressure maximum may appear. 

In our work we consider two pressure maxima, one at the water snowline, and the other at the outer edge of the dead zone. We do not consider the inner edge of the dead zone as a possible location for giant planet formation, since the high temperature $T\sim 1000$~K would certainly inhibit the effective cooling and collapse of the gas envelope, which is needed to form a giant planet.

\section{Our planet formation model}
\label{sec:sec2}

According to the classical core accretion scenario the formation of giant planets begins with the sedimentation and coagulation of dust to the disk's midplane, which is followed by the formation of planetesimals being larger objects (in the size regime $\lesssim 100$ km) less sensitive to drag from the ambient gas disk. According to the most recent picture, planetesimals are the outcome of a process called as gravoturbulent formation. During this process mm to cm sized dust grains are concentrated by transient high pressure regions in sufficiently high amount that triggers the streaming instability followed by gravitational collapse of the dust aggregates. The further collisions of planetesimals leads to the formation of planetary embryos (among them the larger ones can become the solid cores of giant planets), which continue growing by the accretion of planetesimals and the surrounding gas. As we mentioned in Section~\ref{sec:intro}, there is a timescale problem associated to the formation of giant planets meaning that the building time of a giant planet is very close to the lifetime of protoplanetary disks. In last years, however, several works including different physical phenomena, demonstrated the possibility of the formation of giant planets before the disk dissipation. For example, \citet{Hubickyj.et.al.2005} showed that a reduction of the grain opacity of the planet's envelope significantly reduce the formation time of giant planets. On the other hand, \citet{Hori-Ikoma2011} and \citet{Venturini.et.al.2015, Venturini.et.al.2016} found that the pollution by icy planetesimals and the consequent enrichment of the envelope of the planet significantly reduces the critical core mass and speeds up the formation of giant planets. Moreover, models for planetary population shyntesis based in the  classical core accretion scenario can reproduce several of the main observed characteristics of the exoplanet population \citep{Ida-Lin2004, Ida-Lin2013, Alibert.et.al.2013}.

On the other hand, in the past few years a new alternative model, which is also based in the accretion of solid material, has been proposed for the formation of giant planets. The basic assumption of this model is that the core of a giant planet can be formed rapidly as a seed of hundred kilometers accretes  cm-sized particles, known as pebbles. \citet{Ormel&Klahr2010}, \citet{Lambrechts&Johansen2012} and \citet{Guillot.et.al.2014} showed that pebbles, strongly coupled to the gas (with Stokes numbers lower than the unity), can be accreted very efficiently to form massive cores very quickly. The main difference between pebble accretion and planetesimal accretion is that pebbles can be accreted by the full Hill sphere of the growing core while planetesimals can only be accreted by a fraction of the Hill sphere. Thus, pebble accretion rates can be significantly larger than planetesimal accretion rates. \citet{Lambrechts-et-al-2014}, \citet{Levison.et.al.2015}, and \citet{Bitsch.et.al.2015} showed that solar system giant planets could be formed by the pebble accretion mechanism.

According to \citep{Chambers2016}, one important parameter of pebble accretion is the amount of the remaining pebbles after planetesimal formation took place. If at the onset of pebble accretion the leftover pebble population is still significant, planetary systems with multiple gas giants beyond the snowline and small planets closer to the star can be formed. Otherwise, pebble accretion could not be effective enough, and no giant planets can be formed. In the latter case, the largest bodies have comparable sizes to Earth. Moreover, the outcome of planet formation is also sensitive to the sizes of planetesimals that form as a result of gravoturbulent collapse. If the largest planetesimals do not overgrow the critical size of 300 km before the depletion of the cm-sized pebble population, giant planet formation will be inhibited, too. Additionally, the formation size of planetesimals is still a debated issue. According to \citet{Morbidelli.et.al2009} planetesimals form large, having characteristic sizes of $\sim 100$ km. On the other hand, the accretion of sub-km sized planetesimals cannot be ruled out \citep{Weidenschilling2011}. 

In our study we investigated giant planet formation both by considering the growth of the protogiant cores by accretion of planetesimals in the size interval between 0.1 - 100 km, and by cm-sized pebble accretion. As we will show later on, giant planet formation is possible in both cases including also the case when pebble accretion would be ineffective.

In a series of previous works \citep{Guilera2010, Guilera2011, Guilera2014}, we developed a model which calculates the formation of gaseous giant planets embedded in a time evolving protoplanetary disk. In this work, we incorporate some modifications to our previous model with the aim to study the formation of massive cores (which are the precursors of giant planets) at the pressure maxima of protoplanetary disks. 

In our model, the protoplanetary disk is characterized by two components: a gaseous component, evolving due to an $\alpha$-viscosity driven accretion, and a solid component represented by a planetesimal population being subject to accretion by the planets, and radial drift due to gas drag. The protogiant planets embedded in the disk grow by accretion of planetesimals and gas. 

\subsection {Initial radial profiles for gas and solid material}
\label{sec:sec2-1}

In our disk model the computational domain is defined between 0.1 au and 1000 au, using 5000 radial bins logarithmically equally spaced as using a classical 1D radial model. The gaseous component is characterized by the corresponding surface density $\Sigma_g(R)$, where $R$ is the radial coordinate, and the solid component is characterized by the planetesimal surface density $\Sigma_p(R)$. Moreover, our model allows us to study a discrete planetesimal size distribution too, thus in a more general view the planetesimal surface density can be characterized by $\Sigma_p(R,r_p)$, where $r_p$ represents the different sizes of the discrete distribution.     

In order to define the initial surface density profiles, we follow the suggestions of \citet{Andrews2010}, who studying the Ophiuchus star-forming region found that the gas surface density of the disks observed can be represented by
\begin{eqnarray}
  \Sigma_g &=& \Sigma_g^0 \left( \frac{R}{R_c} \right)^{-\gamma} e^{-(R/Rc)^{2-\gamma}}, \label{eq:eq1-sec2-1}
\end{eqnarray}
where $R_c$ is a characteristic radius, $\gamma$ represents the surface density gradient and $\Sigma_g^0$ is a parameter function of the disk mass, 
\begin{eqnarray}
  M_d= \int_{0}^{\infty} 2 \pi R \Sigma_g(R)~dR.  
\label{eq:eq2-sec2-1}
\end{eqnarray}
Integrating Eq.~(\ref{eq:eq2-sec2-1}), one can find that $\Sigma_g^0= (2-\gamma)M_d/(2 \pi R_c^2)$. 

A common assumption of planet formation models is that the metalicity along the disk is the same as that of the central star, and that dust sediments and coagulates very quickly to form a mid-plane planetesimal disk. Following this hypothesis, the initial planetesimal surface density is given by 
\begin{eqnarray}
  \Sigma_p &=& \eta_{\text{ice}}(R) \Sigma_p^0 \left( \frac{R}{R_c} \right)^{-\gamma} e^{-(R/Rc)^{2-\gamma}}, 
\label{eq:eq3-sec2-1}
\end{eqnarray}
where $\eta_{\text{ice}}(R)$ takes into account the sublimation of water-ice given by
\begin{eqnarray}
  \eta_{\text{ice}}= 
  \begin{cases}
    1 & \text{ if $R \ge R_{\text{ice}}$},  \\
    \\
    {\dfrac{1}{\beta}} & \text{ if $R < R_{\text{ice}}$},
  \end{cases} 
\label{eq:eq4-sec2-1}
\end{eqnarray}   
with $R_{\text{ice}}= 2.7$~au called as iceline (or snowline). For the Solar System, the factor $\beta$ could have taken values between $\sim 2$ and $\sim 4$ \citep{Hayashi1981, Lodders2003}. In this work, we adopted a value of $\beta= 3$. For a numerical convenience, we smoothed the discontinuity at $R= R_{\text{ice}}$ by 
\begin{eqnarray}
  \eta_{\text{ice}}= \frac{1}{\beta} + \frac{1}{2} \left( 1 - \frac{1}{\beta} \right) \left[ 1 + \tanh \left( \frac{R-R_{\text{ice}}}{\Delta_{\text{ice}}} \right) \right], 
\label{eq:eq5-sec2-1}
\end{eqnarray}  
where $\Delta_{\text{ice}}= \text{H}_g(R_{\text{ice}})$, being $\text{H}_g$ the scale height of the gas disk. 

On the other hand, $\Sigma_p^0$ is given by 
\begin{eqnarray}
  \Sigma_p^0= z_0 \Sigma_g^0,
\label{eq:eq6-sec2-1}
\end{eqnarray} 
with $z_0= 0.0153$ being the initial abundance of heavy elements, called also as the dust-to-gas ratio \citep{Lodders.et.al.2009}. Finally, we adopted some typical values for $\gamma$ and $R_c$ given by \citet{Andrews2010}: $\gamma= 1$ and $R_c= 25$~au. 

\subsection {Evolution of the gas disk with a wide range viscosity reduction}
\label{sec:sec2-2}

As we mentioned above, the gas surface density of the disk $\Sigma_g$ evolves as  an  $\alpha$ accretion disk \citep{Pringle1981}  
\begin{eqnarray}
  \frac{\partial \Sigma_g}{\partial t}= \frac{3}{R}\frac{\partial}{\partial R} \left[ R^{1/2} \frac{\partial}{\partial R} \left( \nu \Sigma_g R^{1/2}  \right) \right], 
\label{eq:eq1-sec2-2}
\end{eqnarray}
where $\nu= \alpha c_s \text{H}_g $ is the kinematic viscosity given by the dimensionless parameter $\alpha$ \citep{Shakura1973}. Usually, the parameter $\alpha \sim 10^{-3} \hdots 10^{-2}$ is a constant value along the disk. In order to reproduce the effect of the dead zone, we have chosen $\alpha$ to be a particular function of $R$ (see later). The sound speed is given by
\begin{eqnarray}
  c_s= \sqrt{ \frac{\gamma_g k_B \text{T}}{\mu_{\text{H}_2} m_{\text{H}_2}} },
  \label{eq:eq2-sec2-2}
\end{eqnarray}  
where $\gamma_g= 5/3$, $k_B$ is the  Boltzmann-constant, and $\mu_{\text{H}_2}$ and $m_{\text{H}_2}$ are the molecular weight and mass of molecular hydrogen, respectively. The radial temperature profile is given as the following power-law function
\begin{eqnarray}
  \text{T}= 280 \left( \frac{R}{1~\text{au}} \right)^{-1/2} ~ \text{K}.
  \label{eq:eq3-sec2-2}
\end{eqnarray}
Regarding the geometry of the disk, we will consider two cases, a flat and a flared disk. In the flat case the aspect ratio is constant $h= 0.05$, so $\text{H}_g= 0.05~R$ while for the flared disk we assumed that $\text{H}_g= c_s/\Omega_k \propto R^{5/4}$, being $\Omega_k$ the keplerian frecuency. 

In order to generate the inner and the outer pressure maximum at the water snowline and at the outer edge of the dead zone, respectively, we apply a reduction of the $\alpha$ parameter between them. Denoting by $\alpha_{\text{back}}$ the background $\alpha$-viscosity parameter, and by $\alpha_{\text{dz}}$ its reduced value the functional form of $\alpha(R)$ is given as
\begin{eqnarray}
\alpha(R) & = &  \Bigg[\left( \alpha_{\text{back}} - \alpha_{\text{dz}} \right) \Bigg\{ 1 - 0.5 \Bigg[ 1 + \nonumber \\
&& \tanh \left( \frac{R - R_{\text{in-dz}}}{c_{\text{in-dz}} \text{H}_g(R_{\text{in-dz}})} \right) \Bigg] \Bigg\} + \frac{\alpha_{\text{dz}}}{2} \Bigg]  + \nonumber \\
 && \Bigg[  \left( \alpha_{\text{back}} - \alpha_{\text{dz}} \right) \Bigg\{ 1 - 0.5 \Bigg[ 1 + \nonumber \\
 && \tanh \left( \frac{R_{\text{out-dz}}-R}{c_{\text{out-dz}} \text{H}_g(R_{\text{out-dz}})} \right) \Bigg] \Bigg\} + \frac{\alpha_{\text{dz}}}{2} \Bigg], 
  \label{eq:eq1-sec2-2-1}
\end{eqnarray}   
where $R_{\text{in-dz}}$ and $R_{\text{out-dz}}$ are the locations of the water snowline and outer edge of the dead zone, respectively, and $c_{\text{in-dz}}$ and $c_{\text{out-dz}}$ are constants that define the width of the transition in the viscosity profile. In the following we refer to this region as the dead zone, but we should keep in mind that the real dead zone may extend much closer to the star, until the thermal ionization dominates the accretion. On the other hand, the MRI driven turbulent viscosity qualitatively behaves similarly at the inner edge of the dead zone and at the water snowline, as the viscosity is suddenly reduced with increasing $R$.

\begin{figure}[t]
  \centering
  \includegraphics[width= 0.475\textwidth]{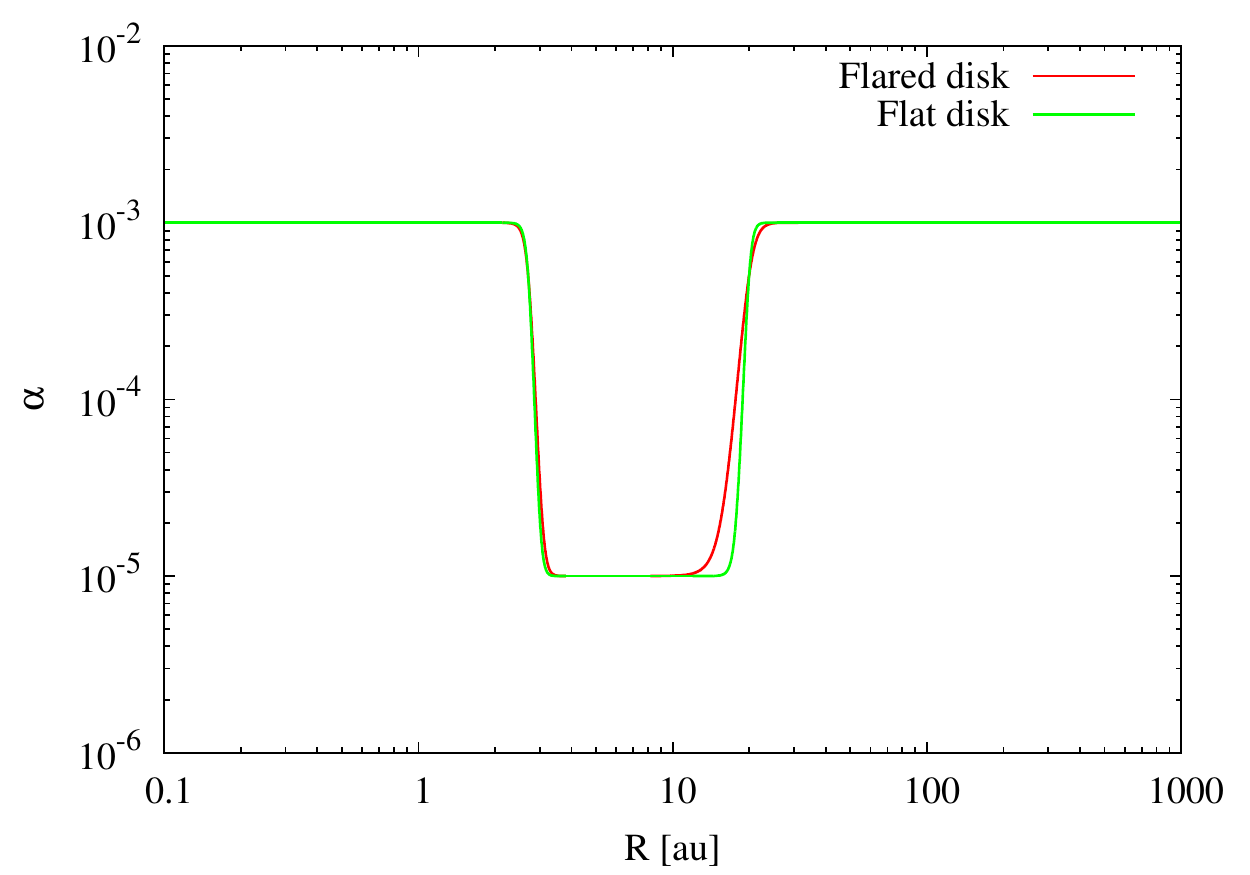}
  \caption{Alpha-viscosity parameter as function of the radial coordinate for a flat and a flared disk, with $\alpha_{\text{back}}= 10^{-3}$, $\alpha_{\text{dz}}= 10^{-5}$, $R_{\text{in-dz}}= 2.7$~au and $R_{\text{out-dz}}= 20$~au. We consider a disk of $0.05~\text{M}_{\odot}$, and for both disk $c_{\text{in-dz}}= c_{\text{out-dz}}= 1$. For a flat disk, $\text{H}_g(R_{\text{in-dz}})= 0.135$~au and $\text{H}_g(R_{\text{out-dz}})= 1$~au, while for a flared disk $\text{H}_g(R_{\text{in-dz}})= 0.165$~au and $\text{H}_g(R_{\text{out-dz}})= 2.013$~au. (Color version online).}
  \label{fig:fig1-sec2-2-1}
\end{figure}

Fig.~\ref{fig:fig1-sec2-2-1} represents the radial profiles of the alpha-viscosity parameter for a flat and a flared disk. We note that the differences in the widths of the transition regions are due to the fact that they are expressed in terms of the scale height of gas (see Eq.~\ref{eq:eq1-sec2-2-1}), which is larger for a flared disk than for a flat disk. These differences play an important role in the evolution of the gas and planetesimal surface densities, as will be discussed in next sections.

Finally, we mention that Eq.~(\ref{eq:eq1-sec2-2}) is solved using a full implicit Crank-Nicholson method considering zero torques as boundary conditions. For each time-step, we do not allow changes greater than 10\% for the gas surface density in each radial bin. 

\begin{table}[t]
\caption{Free parameters adopted in this work.}
\label{tab:tab1-sec3}    
\centering               
\begin{tabular}{c|c}   
\hline\hline             
$M_d$ & 0.03, 0.05, and 0.1 $\text{M}_{\odot}$  \\  
\hline                       
$R_{\text{in-dz}}$ & 2.7 au \\
\hline                       
$R_{\text{out-dz}}$ & 20 au \\
\hline
$\alpha_{\text{back}}(\alpha_{\text{dz}})$ & $10^{-2}(10^{-4})$, and $10^{-3}(10^{-5})$ \\
\hline
$r_p$ & 1 cm, 0.1 km , 1 km, 10 km, and 100 km \\ 
\hline\hline             
\end{tabular}
\end{table}

\subsection {Evolution of the planetesimal population}
\label{sec:sec2-3}

The numerical treatment of the evolution of the planetesimal population is described in detail in \citet{Guilera2014}. Here, we will only discuss the most relevant properties and some modifications of it with respect to our previous work. In our model planetesimals are subject to radial drift due to the drag force arising from the ambient gaseous disk. Moreover, planetesimals are also accreted by the growing protoplanets, and affected by mutual collisions, though this latter effect is not included in our model. 

The drag force between planetesimals and the gas depends on the relative velocities between gas and planetesimals and on the ratio between the planetesimal radii and the mean free path of the gas molecules. Similarly to our previous work, we consider three different regimes of the drag force, Epstein, Stokes and quadratic ones. The separate treatment of these regimes is important, because while big planetesimals are always in the quadratic regime, small planetesimals can change their regimes along the disk. The radial drift velocities of planetesimals are given by 
\begin{eqnarray} 
v_{\text{mig}}= \left\{ 
\begin{array}{cc}
 \frac{2R\eta}{t_{\text{stop}}^{\text{Eps}}}\left(\frac{s^2_{\text{Eps}}}{1+s^2_{\text{Eps}}}\right)
 & \text{Epstein regime}, \\
\\
 \frac{2R\eta}{t_{\text{stop}}^{\text{Sto}}}\left(\frac{s^2_{\text{Sto}}}{1+s^2_{\text{Sto}}}\right)
 & \text{Stokes regime}, \\
\\
 \frac{2R\eta}{t_{\text{stop}}^{\text{qua}}} & \textrm{quadratic regime},
\end{array} \right.
\label{eq:eq1-sec2-3}
\end{eqnarray}
where $t_{\text{stop}}$ and $s$ are the stopping time and the Stokes number (for the corresponding regime), respectively \citep[see][]{Guilera2014}, and $\eta= (v_g - v_k)/v_k$ is the fraction by which the gas deviates from the keplerian circular velocity given by  
\begin{eqnarray}
  \eta=  \frac{1}{2} \left( \frac{\text{H}_g}{R} \right)^2 \frac{\text{d} \ln P}{\text{d} \ln R}, 
  \label{eq:eq2-sec2-3}
\end{eqnarray}
where $P$ is the gas pressure in the midplane of the disk. Considering a local isothermal equation state for the gas, $P= c_s^2 \rho_g$, where $\rho_g= \Sigma_g / (\sqrt{2\pi}\text{H}_g)$ is the volumetric gas density at the midplane. Eq.~(\ref{eq:eq2-sec2-3}) can be expressed as   
\begin{eqnarray}
  \eta=  \frac{1}{2} \left( \frac{\text{H}_g}{R} \right)^2 \frac{\text{d} \ln (c_s^2 \rho_g)}{\text{d} \ln R}, 
  \label{eq:eq3-sec2-3}
\end{eqnarray}
thus, if the gas density is a decreasing function of the distance $R$ to the central star, the radial drift of the planetesimals is always inward (Eq.~\ref{eq:eq1-sec2-3}). However, if there is some local maximum in the gas density, planetesimals can also be drifted outwards, when $\text{d} \ln (c_s^2 \rho_g)/\text{d} \ln R > 0$. 

Finally, as the consequence of the mass conservation the evolution of the planetesimal surface density is described by the following advection equation
\begin{eqnarray}
\frac{\partial \Sigma_{\text{p}}(R,r_{p_j})}{\partial t} - \frac{1}{R} \frac{\partial}{\partial R} \left[ Rv_{\text{mig}}(R,r_{p_j})\Sigma_{\text{p}}(R,r_{p_j}) \right] = \mathcal{F}(R,r_{p_j}), 
\label{eq:eq4-sec2-3}
\end{eqnarray}  
where $\mathcal{F}$ represents the sink terms due to the accretion by the growing embryos or cores, and $r_{p_j}$ emphasizes the fact that Eq.~(\ref{eq:eq4-sec2-3}) is solved independently for each planetesimal size, when a planetesimal size distribution is considered. In this case, the total planetesimal surface density is given by   
\begin{eqnarray}
  \Sigma_{\text{p}}(R)= \sum_j \Sigma_{\text{p}}(R,r_{p_j}). 
  \label{eq:eq5-sec2-3}
\end{eqnarray}
Eq.~(\ref{eq:eq4-sec2-3}) is solved using a full implicit upwind-downwind mix method considering zero density as boundary conditions. For each time-step, we do not allow changes greater than 10\% for the planetesimal surface density in each radial bin. 

\begin{figure}[t]
  \centering
  \includegraphics[width= 0.475\textwidth]{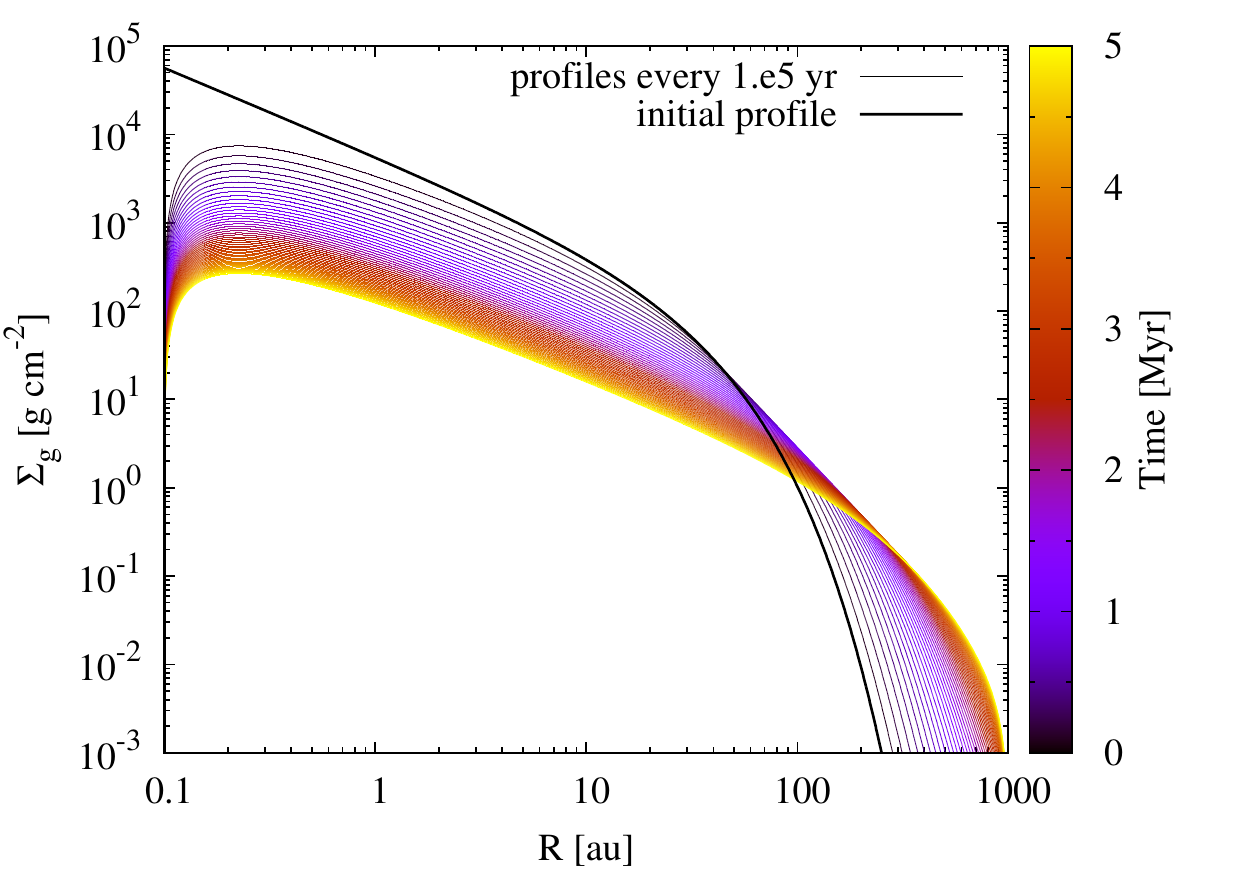}
  \caption{Time evolution of the gas surface density radial profiles. The simulation correspond to a disk of $M_d= 0.1~\text{M}_{\odot}$, using $\alpha= 10^{-3}$. The simulation is stopped after 5 Myr of viscous evolution. (Color version online).}
  \label{fig:fig1-sec3-1}
\end{figure}

\begin{figure*}[t]
    \centering
    \includegraphics[width= 0.475\textwidth]{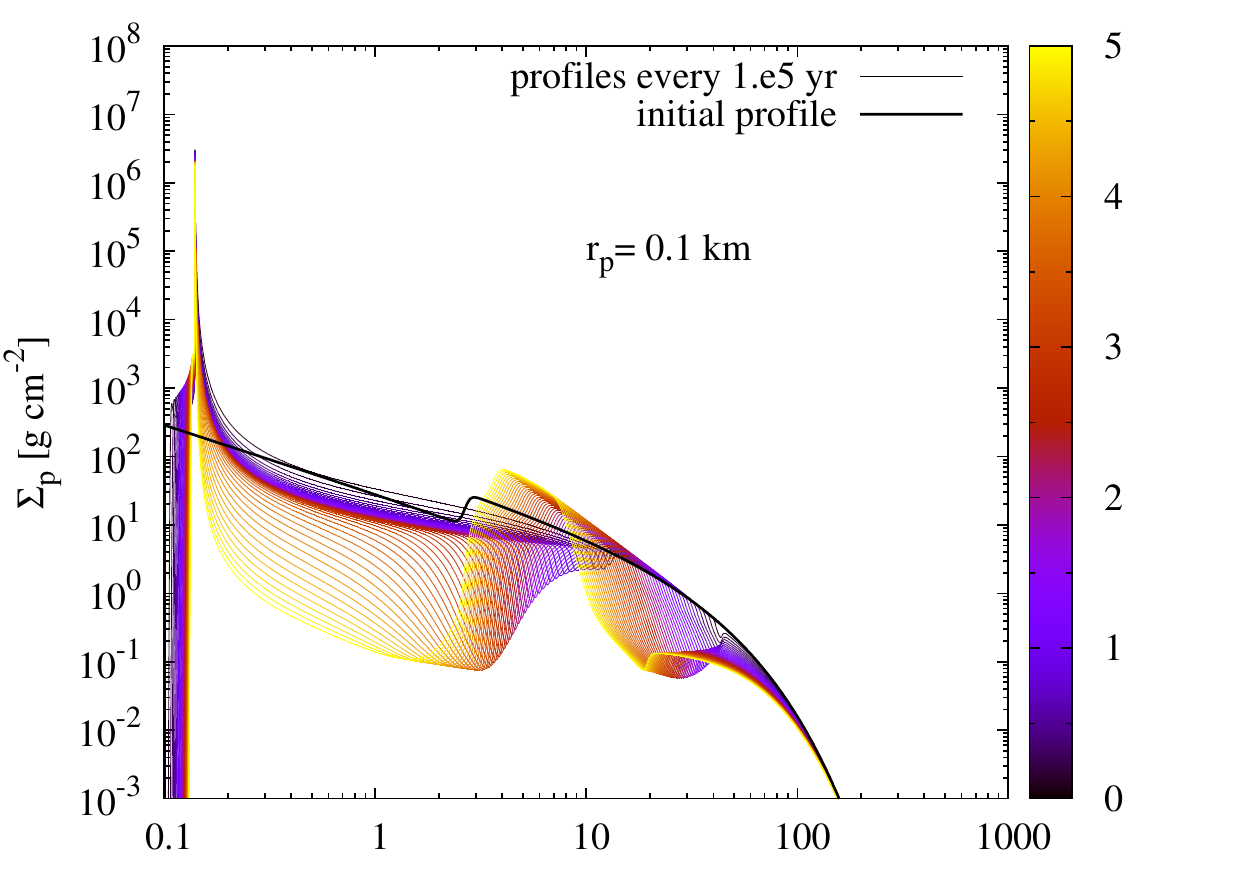} 
    \centering
    \includegraphics[width= 0.475\textwidth]{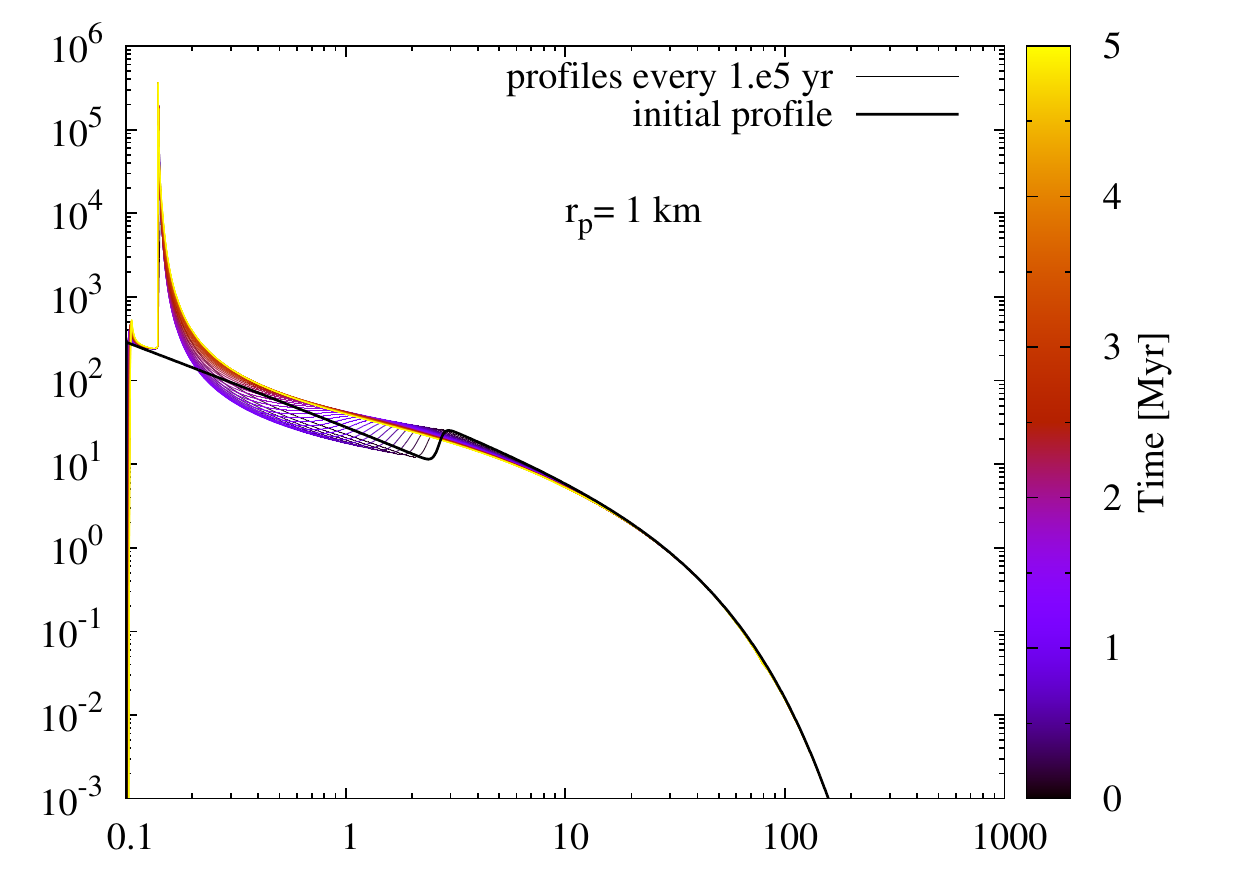} \\
    \centering
    \includegraphics[width= 0.475\textwidth]{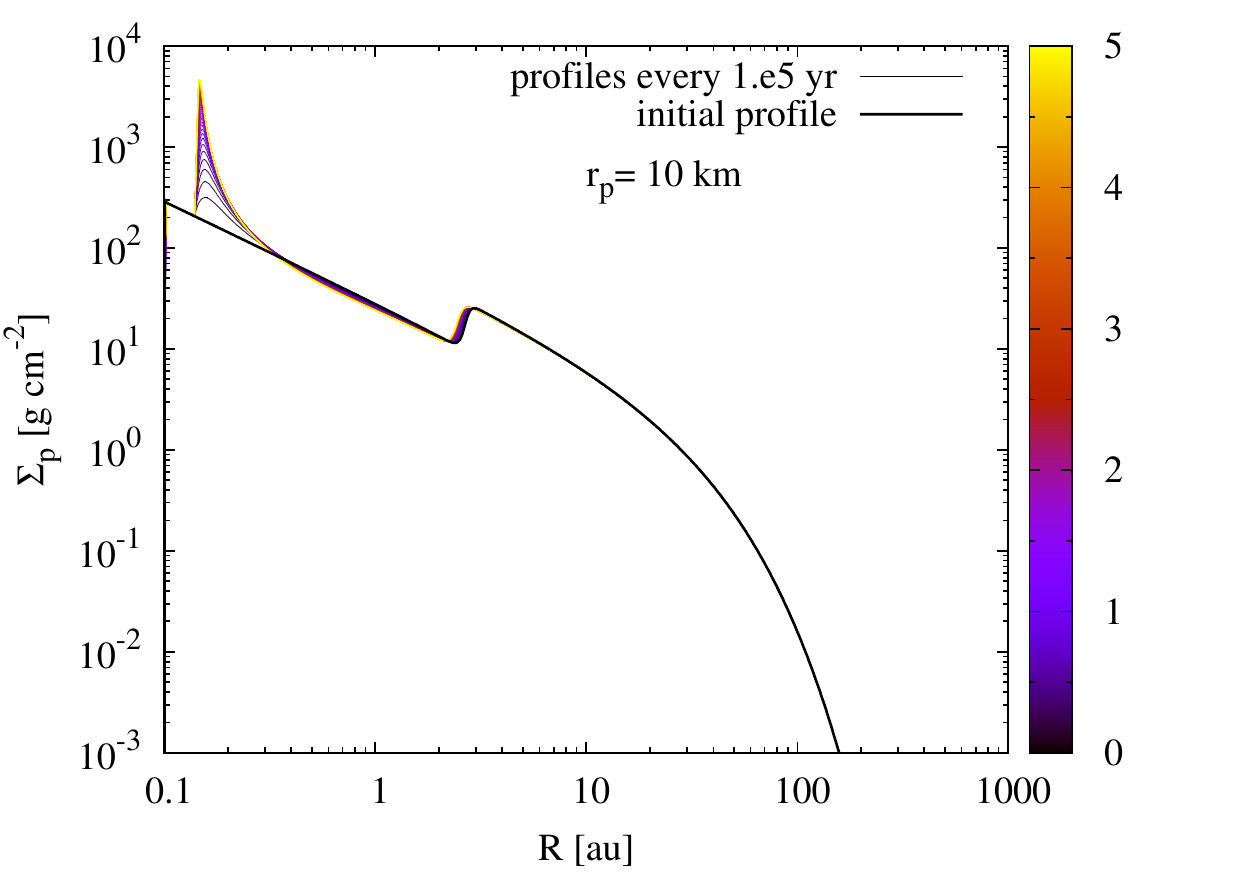} 
    \centering
    \includegraphics[width= 0.475\textwidth]{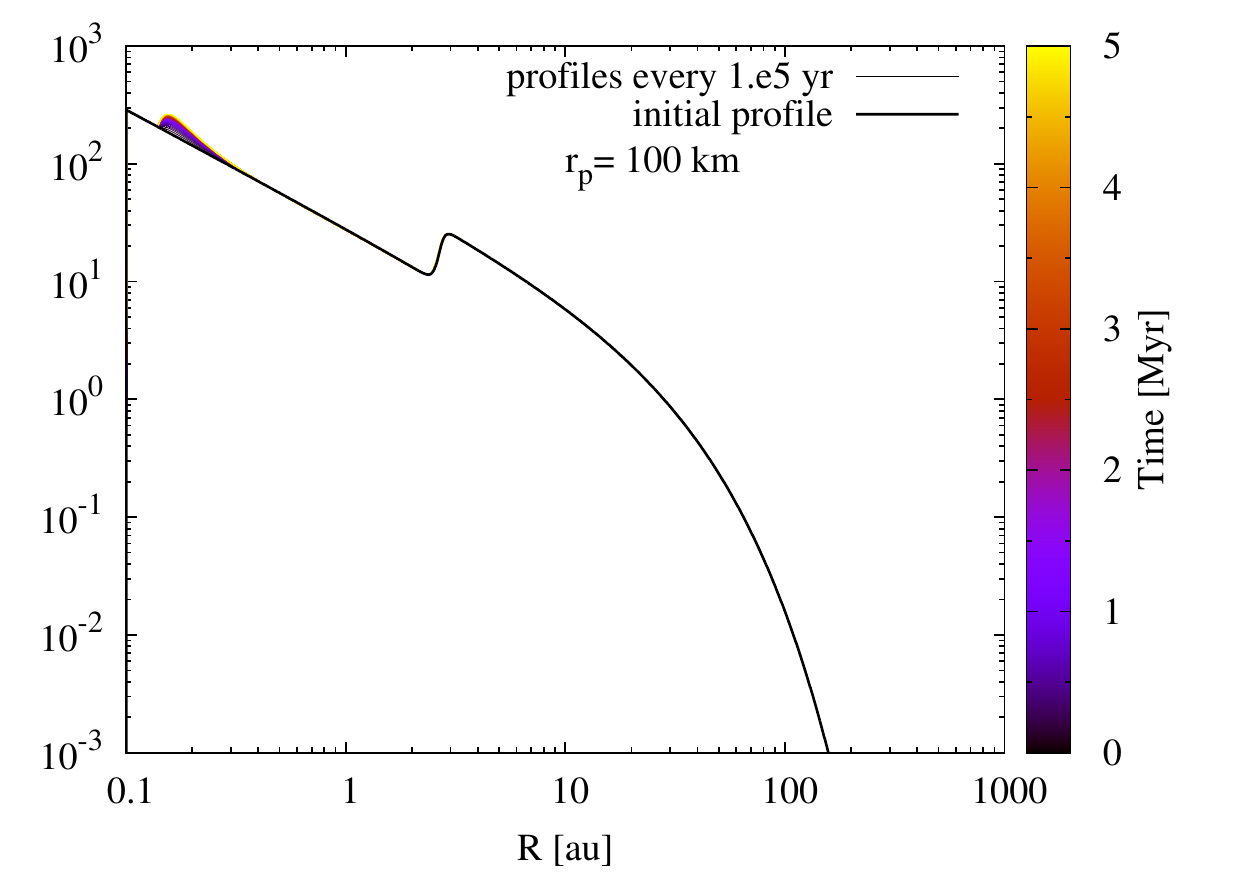}
  \caption{Time evolution of the planetesimal surface density radial profiles for planetesimal populations of different radii. Simulations correspond to a disk of $M_d= 0.1~\text{M}_{\odot}$, using $\alpha= 10^{-3}$. Simulations are stopped after 5 Myr of viscous evolution. (Color version online).}
  \label{fig:fig2-sec3-1}
\end{figure*}

\subsection {Growth of the proto planets}
\label{sec:sec2-4}

In our model, the planetary embryos embedded in the disk grow by the concurrent accretion of planetesimals and the surrounding gas \citep[see][for a detailed explanation]{Guilera2010, Guilera2014}.

The mass increase of a core in the oligarchic growth regime due to the accretion of planetesimals \citep{Inaba2001} is described by
\begin{eqnarray}
\frac{dM_C}{dt}= \frac{2\pi \Sigma_p(a_P)R_H^2}{T_{\text{orb}}}P_{coll},
\label{eq:eq0-sec2-4} 
\end{eqnarray}
where $M_{C}$ is the core's mass, $\Sigma_p(a_p)$ is the surface density of solids at the location of the planet, $R_H$ is the Hill radius, and $T_{\text{orb}}$ is the orbital period. $P_{coll}$ is a collision probability, which is a function of the core radius $R_C$, the Hill radius of the planet, and the relative velocity between the planetesimals and the planet $v_{\text{rel}}$, thus $P_{coll}=P_{coll}(R_C,R_H, v_{\text{rel}})$. Actually, as we also consider the drag force that planetesimals experience on entering the planetary envelope \citep[following][]{Inaba&Ikoma-2003}, the collision probability is function of the enhanced radius $\tilde{R}_C$ instead of $R_C$.  

The gas accretion rate and the thermodynamic state of the planet envelope are calculated by solving the standard equations of transport and structure \citep[see][for details]{Benvenuto&Brunini2005, Fortier.et.al.2007, Fortier2009, Guilera2010},
\begin{eqnarray}
\frac{\partial r}{\partial m_r} & = & \frac{1}{4\pi r^2  \rho}, \nonumber 
\\ 
\frac{\partial  P}{\partial  m_r} & = & - \frac{G m_r}{4\pi  r^4}, \nonumber 
\\ 
\label{eq:eq01-sec2-4}
\\
\frac{\partial L_r}{\partial m_r} & = & \epsilon_{pl} - T \frac{\partial S}{\partial t}, \nonumber 
 \\ 
\frac{\partial T}{\partial m_r} &  = & - \frac{G m_r T}{4\pi r^4 P} \nabla, \nonumber 
\end{eqnarray}
where $\rho$ is the envelope density, $G$ is the universal gravitational constant, $\epsilon_{pl}$ is the energy release rate by  planetesimal accretion, $S$ is the entropy per unit mass, and $\nabla= d\ln T/d\ln P$ is the dimensionless temperature gradient, which depends on the type of energy transport. 

We also incorporated in our model the prescription of type I migration for a locally isothermal disk to calculate the change in semi-major axis of the planetary embryo 
\begin{eqnarray}
  \frac{da_P}{dt}= -2a_P\frac{\Gamma}{\textrm{L}_P},
  \label{eq:eq1-sec2-4}
\end{eqnarray}
where $a_P$ represents the planetary embryo's semi-major axis and $\textrm{L}_P= \textrm{M}_P \sqrt{G \textrm{M}_{\star} a_P}$ its angular momentum. $\Gamma$ is the total torque, which is given by:
\begin{eqnarray}
  \Gamma= (1.364 + 0.541 \delta) \left( \frac{\textrm{M}_P ~ a_P ~ \Omega_P}{\textrm{M}_{\star} c_{s_P}} \right)^2 \Sigma_{g_P} ~ a_P^4 ~ \Omega_P^2,
  \label{eq:eq2-sec2-4}
\end{eqnarray}
where $\Omega_P$, $ c_{s_P}$ and $\Sigma_{g_P}$ are the values of the keplerian frequency, the sound speed, and the gas surface density at the position of the planet, respectively \citep[see][]{Tanaka.et.al.2002}. The factor $\delta$ is defined by $\delta= d\log\Sigma_g/d\log R$ evaluated at $R= a_P$. To follow the orbital migration and mass growth of a planetary embryo, Eq.~(7), (14), (\ref{eq:eq0-sec2-4}), (\ref{eq:eq01-sec2-4}), and (\ref{eq:eq1-sec2-4}) have to be numerically solved together self-consistently. 

Despite the developments of the last years aiming at improving analytic formulae for type I migration in more and more realistic disk models \citep[e.g.][]{paardekooper_etal2010, paardekooper_etal2011, Bitsch.et.al.2013, Bitsch.et.al.2014a, Bitsch.et.al.2014b, Benitez-llambay.et.al.2015}, also including the works \citet{Dittkrist.et.al.2015} and \citet{Bitsch.et.al.2015}, in our study we use the torque prescription given by Eq.~(\ref{eq:eq2-sec2-4}) to be consistent with our locally isotherm disk model. We also emphasize that in our formation scenario the embryo mainly increases its mass being trapped in the zero torque location, and the mass growth during its migration is not significant, thus the outcome of simulations maybe independent of the migration timescale.

\section{Results}
\label{sec:sec3}

The aim of this work is to study the formation of massive planetary cores due to the accumulation of solids in form of pebbles and planetesimals at pressure maxima developed in the disk. We recall that in this work we assume an inner pressure maximum that appears at the water snowline, and an outer pressure maximum that develops at the outer edge of the dead zone. We performed different sets of simulations varying the mass of the disk, the values of the $\alpha$ viscosity parameter inside and outside of the dead zone, and the size of the pebbles and planetesimals (considering a single-sized pebbles/planetesimal population). The combinations of such parameters have been also considered for a flared and a flat disks. Table~\ref{tab:tab1-sec3} summarizes the free parameters used in this work.

\subsection {Disk evolution without planets}
\label{sec:sec3-1}
 
As a first step, we analyze the evolution of the disk without planets, namely, the evolution of $\Sigma_g(R,t)$ and $\Sigma_p(R,t)$. It is important to note that in contrast to previous works \citep[e.g.][]{Matsumura.et.al.2009}, for the sake of simplicity, we assume that neither the location of the snowline nor the outer edge of the dead zone evolves in time. Our choice for fixed pressure maxima/migration traps is commented in a more detailed way in Section~\ref{sec:sec4}. We intend to study the accumulation of planetesimals at the the pressure maxima developed due to the the viscosity reduction.

In Fig.~\ref{fig:fig1-sec3-1} and Fig.~\ref{fig:fig2-sec3-1} the time evolution of the surface density of gas $\Sigma_g(R,t)$ and planetesimals $\Sigma_p(R,t)$ are shown for a disk without a dead zone (pebble accumulation is analyzed in Sec.~\ref{sec:sec3-3}). We consider four different simulations in each using a single size distribution for planetesimal radii. While the time evolution of $\Sigma_g(R,t)$ is the same for the four simulations (Fig.~\ref{fig:fig1-sec3-1}), there are significant differences in the time evolution of $\Sigma_p(R,t)$ due to the different drift rates for the different planesimal sizes (Fig.~\ref{fig:fig2-sec3-1}). While big planetesimals, of 10 km and 100 km of radius, do not suffer a significant inward drift except in the inner part of the disk, small planetesimals, particularly planetesimals of 100 m of radius, undergo a significant radial drift along all the disk. We note that due to the inner boundary condition there is an accumulation of planetesimal in the inner edge of the disk due to the generation of a gas pressure maximum. We run all our simulations for 5 Myr, a characteristic time for protoplanetary disk life times \citep{Mamajek2009, Pfalzner.et.al.2014}. It is generally accepted that EUV/FUV/X-ray photo-evaporation plays an important role in disk dissipation.  It has been showed that after a few Myr of viscous evolution, photo-evaporation  becomes significant and the protoplanetary disk is dispersed in a time-scale of $10^{5}$~yr, \citep{Alexander.et.al.2006, gortietal2009, owenetal2011}. As we are interested in the first stages of planet formation, particularly in the formation of massive cores until the planet achieves the critical mass (when the mass of the envelope equals the mass of the core), we only considered the viscous evolution of the disk.    

\begin{figure}[t]
  \centering
  \includegraphics[width= 0.475\textwidth]{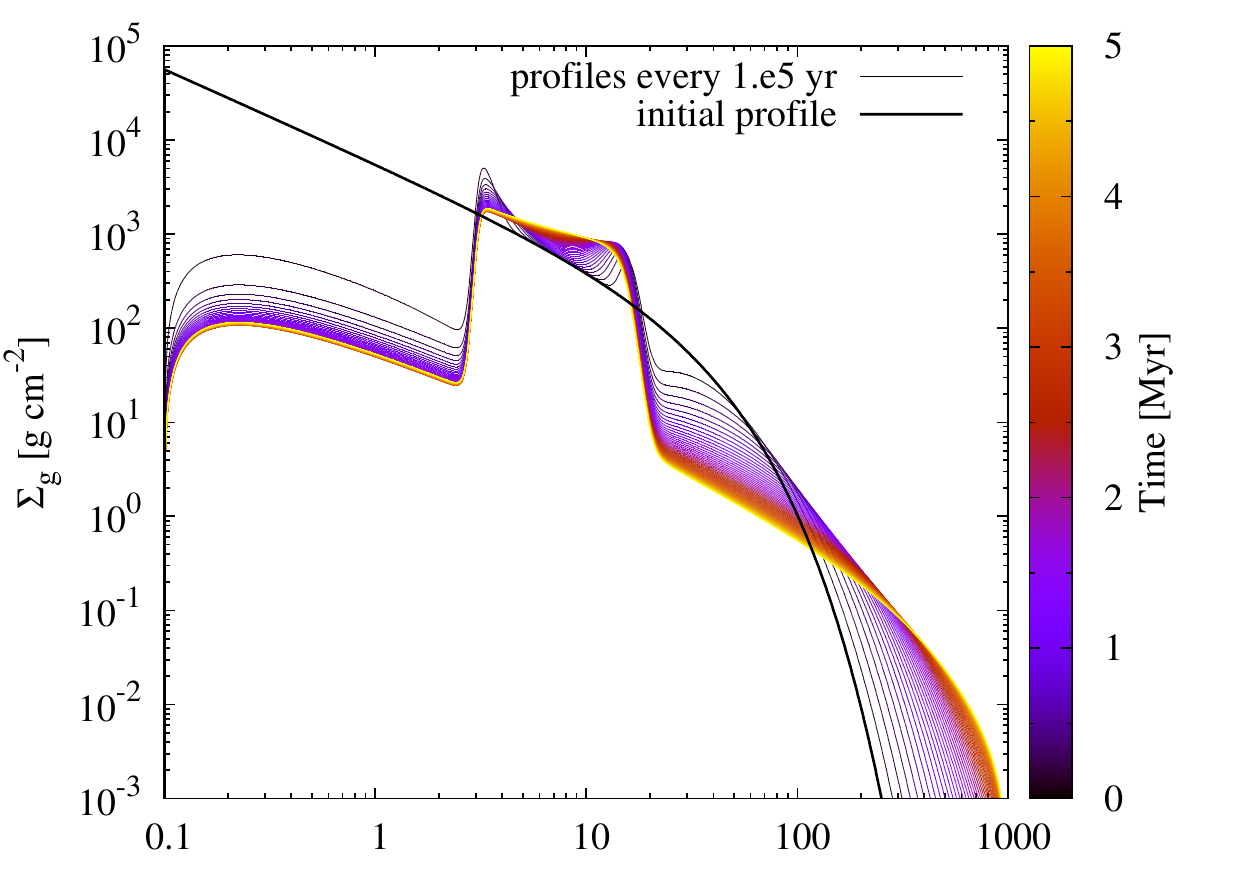} \\
  \includegraphics[width= 0.475\textwidth]{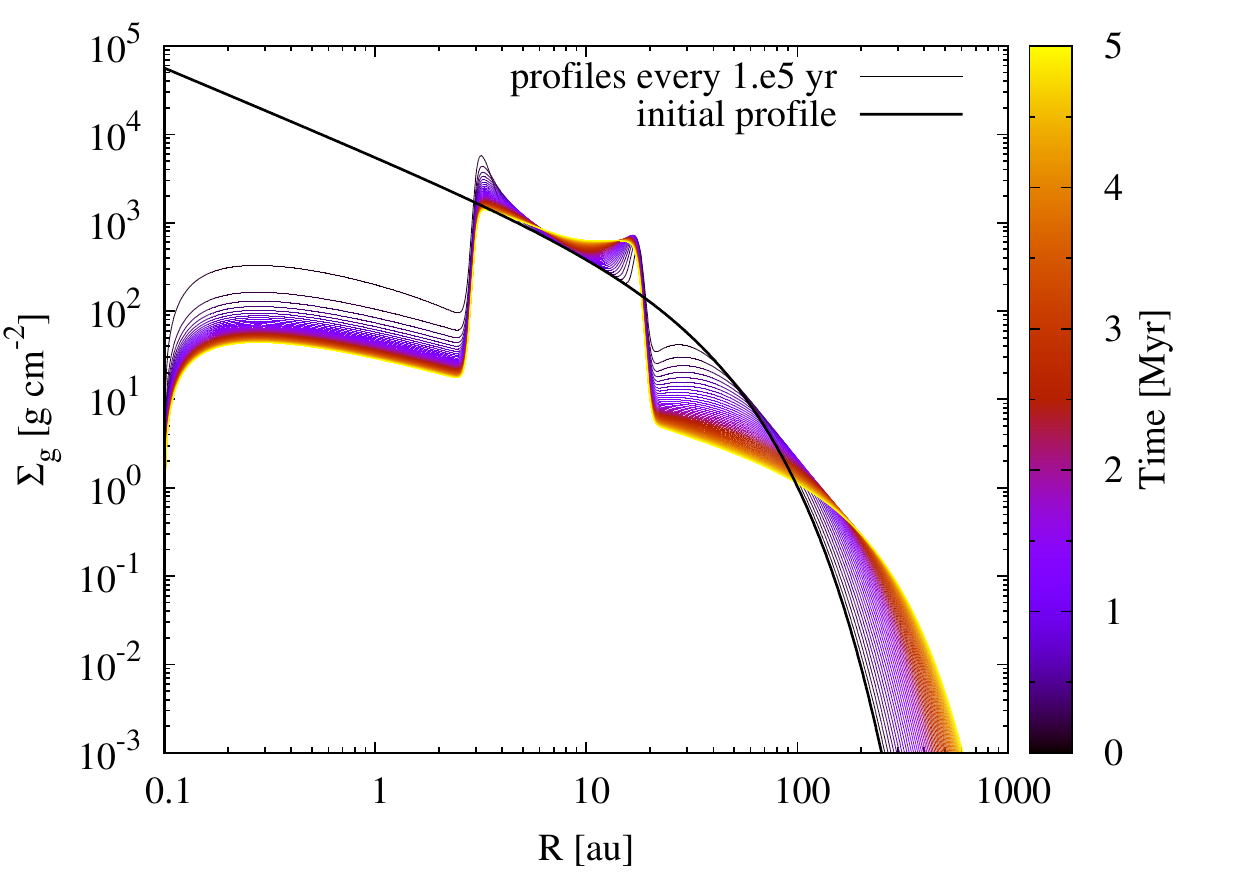}
  \caption{Time evolution of the gas surface density radial profiles assuming the existence of a dead zone for a flared disk (top panel) and a flat disk (bottom panel). Simulations corresponding to a disk of $M_d= 0.1~\text{M}_{\odot}$ using $R_{\text{in-dz}}= 2.7$~au, $R_{\text{out-dz}}= 20$~au, $\alpha_{\text{back}}= 10^{-3}$, and $\alpha_{\text{dz}}= 10^{-5}$. Simulations are stopped after 5 Myr of viscous disk evolution. (Color version online).}
  \label{fig:fig3-sec3-1}
\end{figure}

\begin{figure}[t]
  \centering
  \includegraphics[width= 0.475\textwidth]{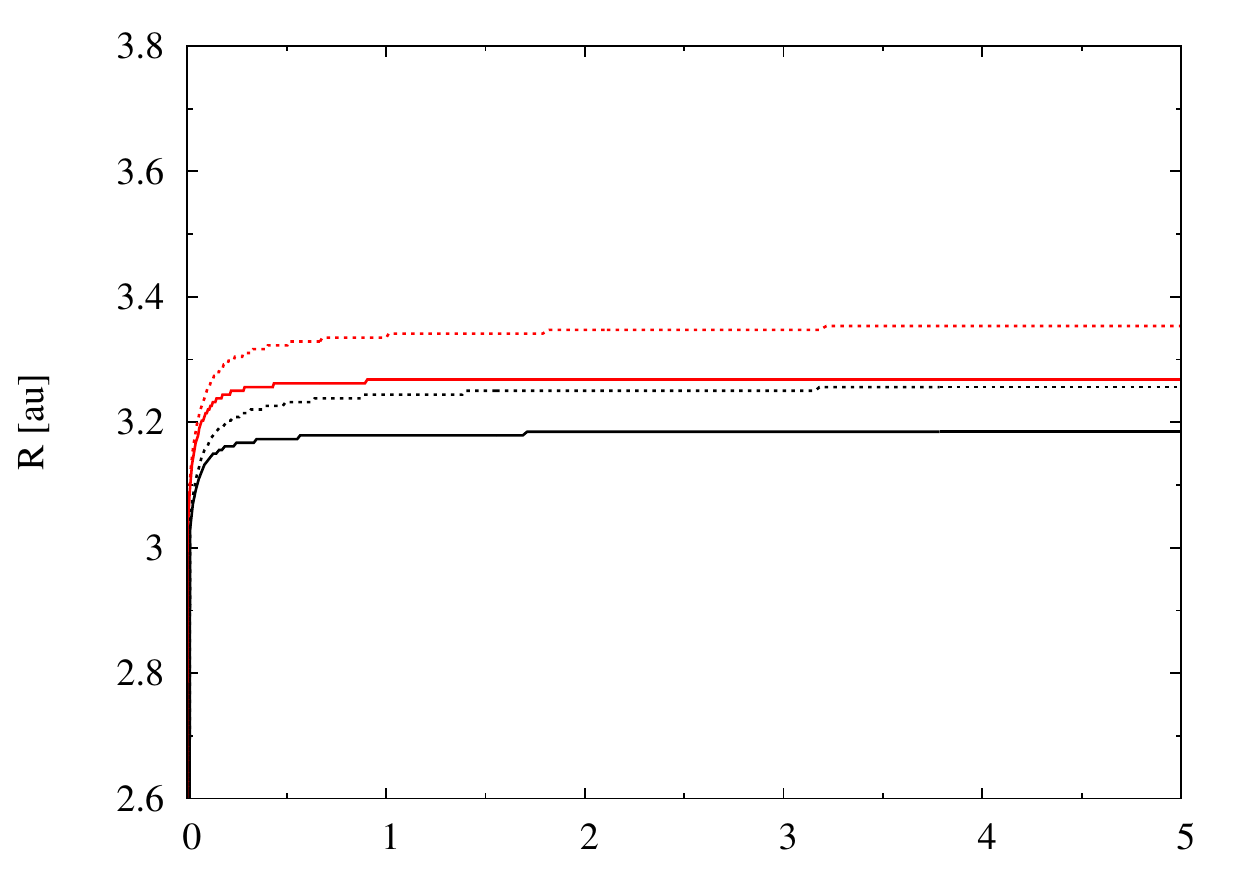} \\
  \includegraphics[width= 0.475\textwidth]{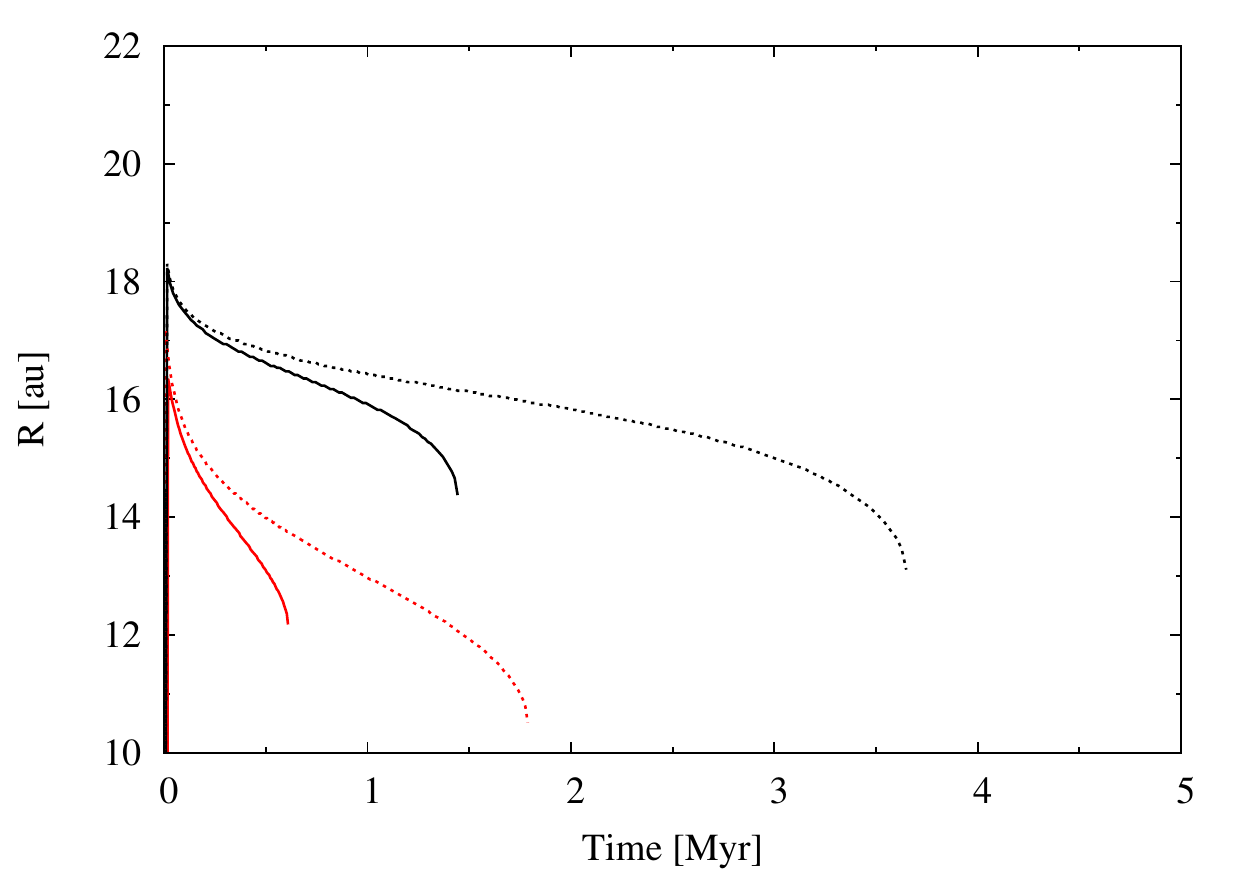}
  \caption{Time evolution of the zero torque locations (solid lines) and the locations of pressure maxima (dashed lines) at the inner egde (top panel) and outer edge (bottom panel) of the dead zone for a flared disk (red lines) and a flat disk (black lines). Simulations correspond to a disk of $M_d= 0.1~\text{M}_{\odot}$ using $R_{\text{in-dz}}= 2.7$~au, $R_{\text{out-dz}}= 20$~au, $\alpha_{\text{back}}= 10^{-3}$, and $\alpha_{\text{dz}}= 10^{-5}$. (Color version online).}
  \label{fig:fig4-sec3-1}
\end{figure}

As we mentioned before, we investigate in this work the possibility of the quick formation of massive planetary cores due to the accumulation of planetesimals around an inner and outer pressure maximum, respectively. To generate the inner and the outer pressure maximum, we implemented the functional form of the viscosity, given by Eq.~(\ref{eq:eq1-sec2-2-1}) when numerically solving Eq.~(\ref{eq:eq1-sec2-2}). Fig.~\ref{fig:fig3-sec3-1} shows the time evolution of the gas surface density for a flared disk (top panel) and a flat disk (bottom panel), considering that $R_{\text{in-dz}}= 2.7$~au, $R_{\text{out-dz}}= 20$~au, $\alpha_{\text{back}}= 10^{-3}$, and $\alpha_{\text{dz}}= 10^{-5}$. At a first glance, the time evolution of the gas density profiles are very similar. The evolution around the inner edge of the viscosity reduction is practically the same for the flared and the flat disk. However, at the outer edge of the dead zone the time evolution of the gas surface density turns out to be different. We can see that the gas density maximum (and the pressure maximum) disappear more quickly for the case of a flared disk. As we will show in next sections, this plays an important role for the formation of massive cores at the outer edges of the dead zones. One reason of this difference is that we express the width of the transition region for the viscosity at the dead zone's outer edge in terms of disk's scale height, which is larger for a flared disk (see Fig.~\ref{fig:fig1-sec2-2-1}). However, even adopting the same width, e.g. using $c_{\text{out-dz}}= 0.5$ for the flared disk, the pressure maximum also disappears faster than in the case of a flat disk. 

The two important locations in the disk which might play an important role in the quick formation of massive cores are the planet or migration traps and the pressure maxima. The growing cores are trapped at the zero torque locations, the most drag sensitive planetesimals are expected to be accumulated at pressure maxima. Fig.~\ref{fig:fig4-sec3-1} shows the time evolution of the locations of the migration traps and of the pressure maxima in the neighborhood of the inner (top panel) and outer (bottom panel) edge of the dead zone for the flared and the flat disk. We note that while the migration trap and pressure maximum at the inner edge of the dead zone (both for the flared and the flat disk) survives during the 5 Myrs of viscous evolution, they disappear after some time at the outer edge of the dead zone. Usually, the migration trap disappears earlier than the pressure maximum. As we mentioned before, we can also see that the migration trap and the pressure maximum at the outer edge of the dead zone are vanishing first for the  flared disk. It is also important to note that for the inner and outer edge of the dead zone, the locations of zero torque and maximum pressure do not exactly coincide. As we will show in next sections, this fact has important consequences for the presented scenario of planet formation.      

\begin{figure}[t]
  \centering
  \includegraphics[width= 0.475\textwidth]{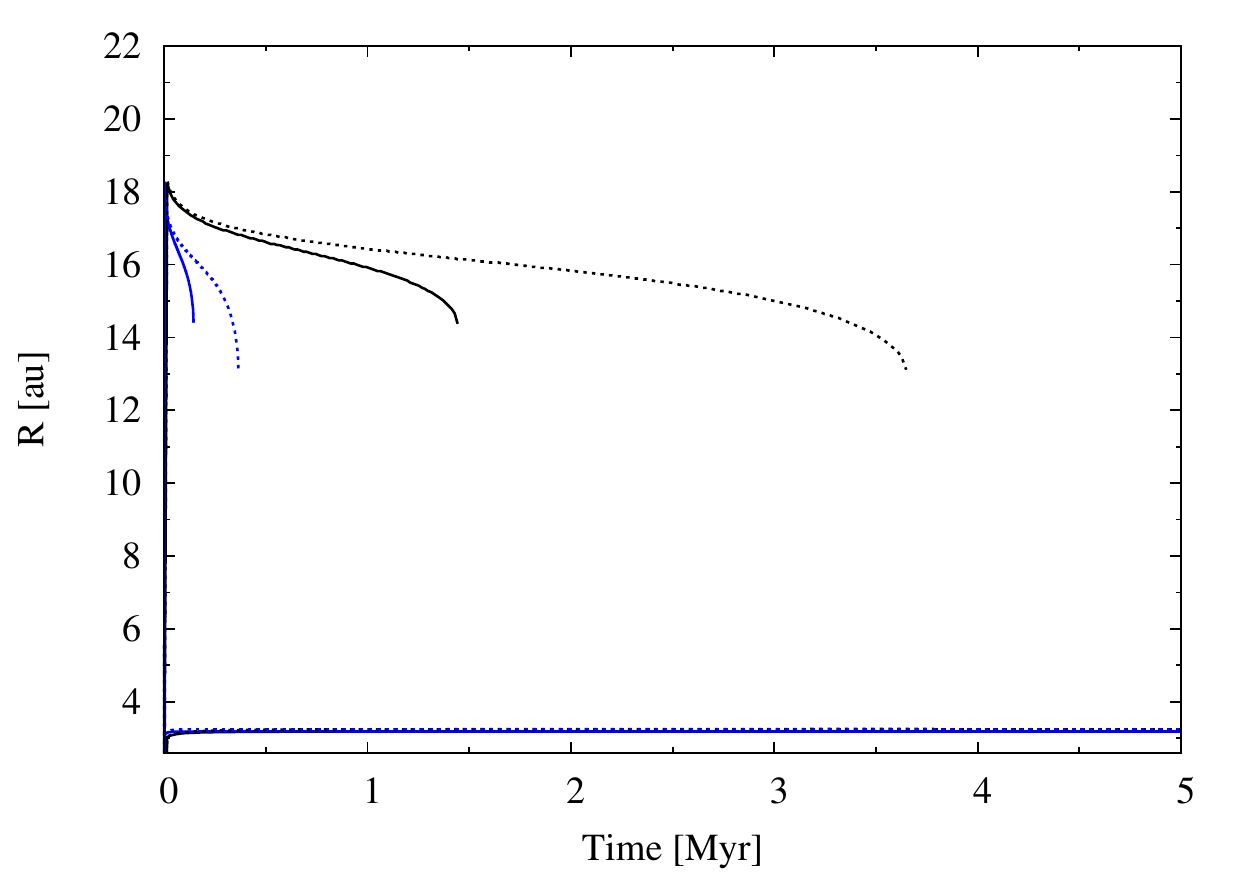} 
  \caption{Time evolution of the zero torque locations (solid lines) and pressure maximum locations (dashed lines) at the inner egde and outer edge of the dead zone for a flat disk using $\alpha_{\text{back}}= 10^{-3}$ and $\alpha_{\text{dz}}= 10^{-5}$ (black lines), and $\alpha_{\text{back}}= 10^{-2}$ and $\alpha_{\text{dz}}= 10^{-4}$ (blue lines). (Color version online).}
  \label{fig:fig5-sec3-1}
\end{figure}

A not less important parameter of our simulations is the value adopted for the $\alpha$-viscosity inside and outside of the dead zone. Fig.~\ref{fig:fig5-sec3-1} shows the time evolution of the locations of the migration trap and pressure maximum at the inner and the outer edge of the dead zone for a flat disk of $M_d= 0.01~\text{M}_{\odot}$ using different values for the $\alpha$-viscosity parameter inside and outside the dead zone. The evolution of the locations of the migration trap and the pressure maximum at the inner edge of the dead zone is practically the same for the different values of $\alpha_{\text{back}}$ and $\alpha_{\text{dz}}$. However, the migration trap and the pressure maximum at the dead zone's outer edge quickly disappear (in less than 0.5 Myr) for the case of $\alpha_{\text{back}}= 10^{-2}$ and $\alpha_{\text{dz}}= 10^{-4}$.   

\begin{figure*}[t]
    \centering
    \includegraphics[width= 0.475\textwidth]{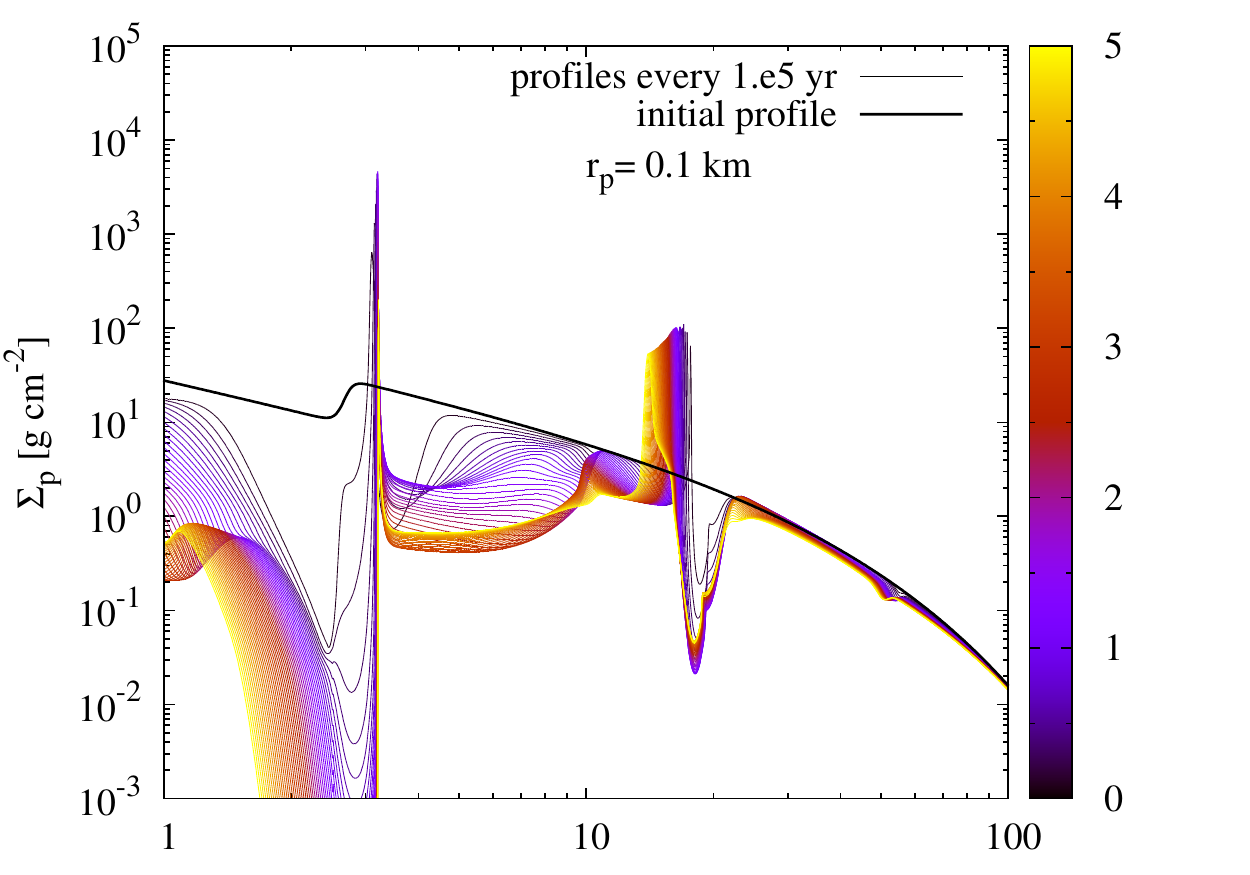} 
    \centering
    \includegraphics[width= 0.475\textwidth]{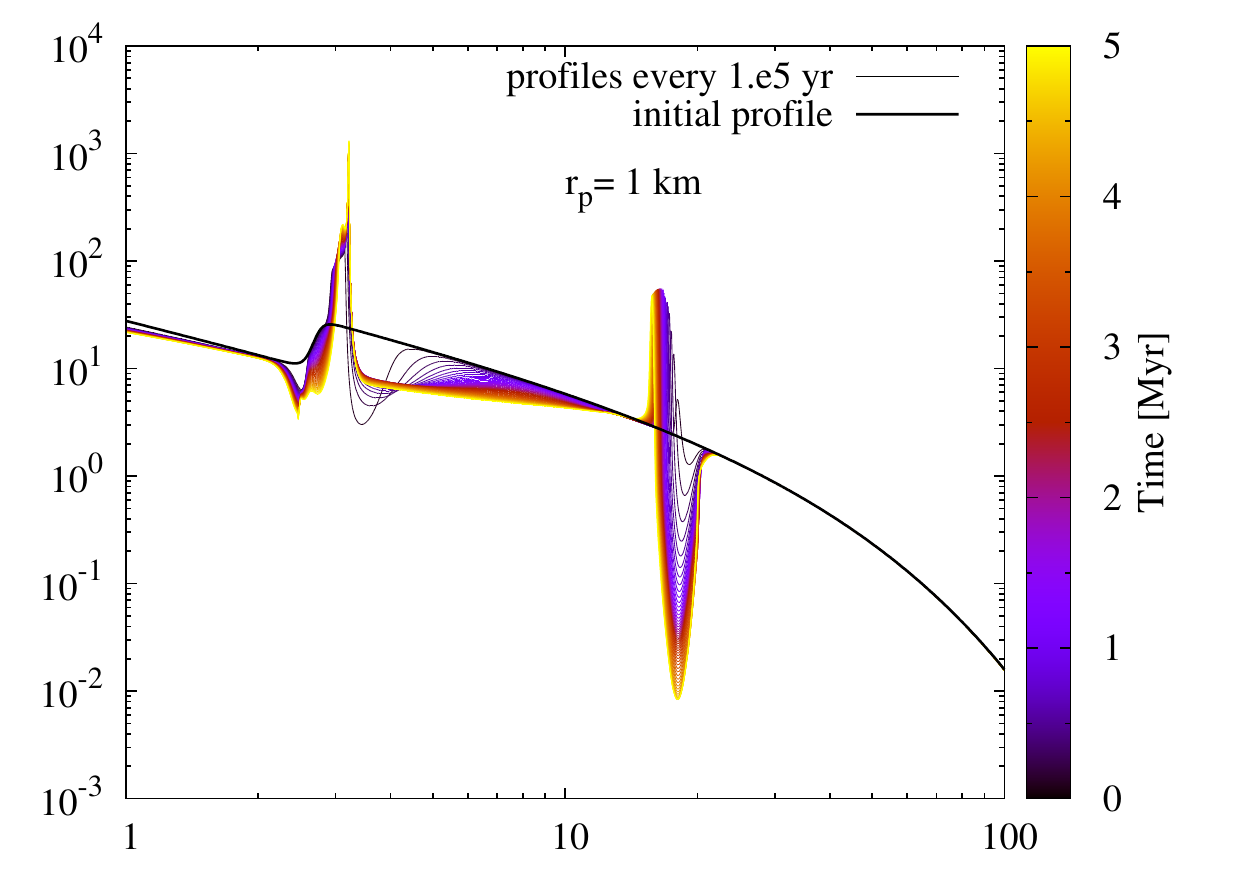} \\
    \centering
    \includegraphics[width= 0.475\textwidth]{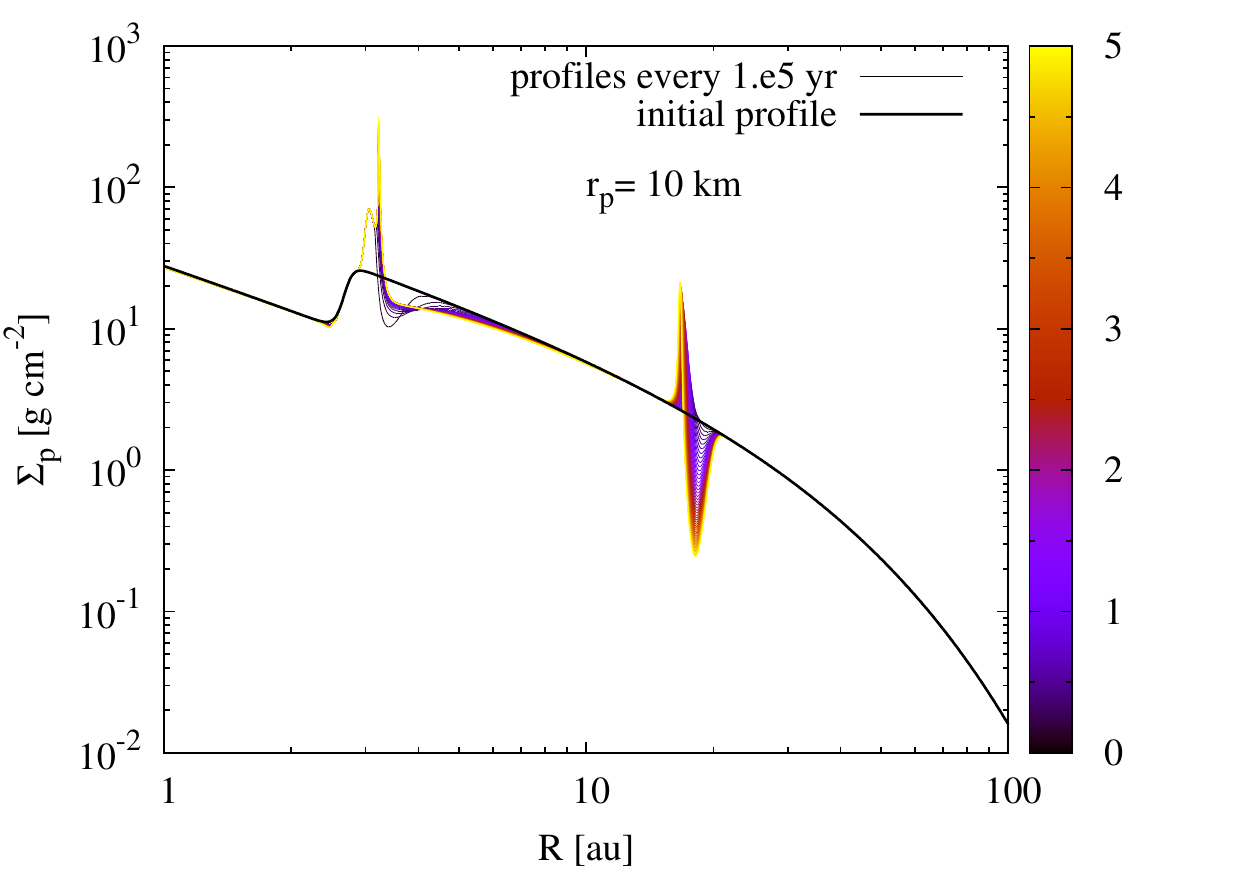} 
    \centering
    \includegraphics[width= 0.475\textwidth]{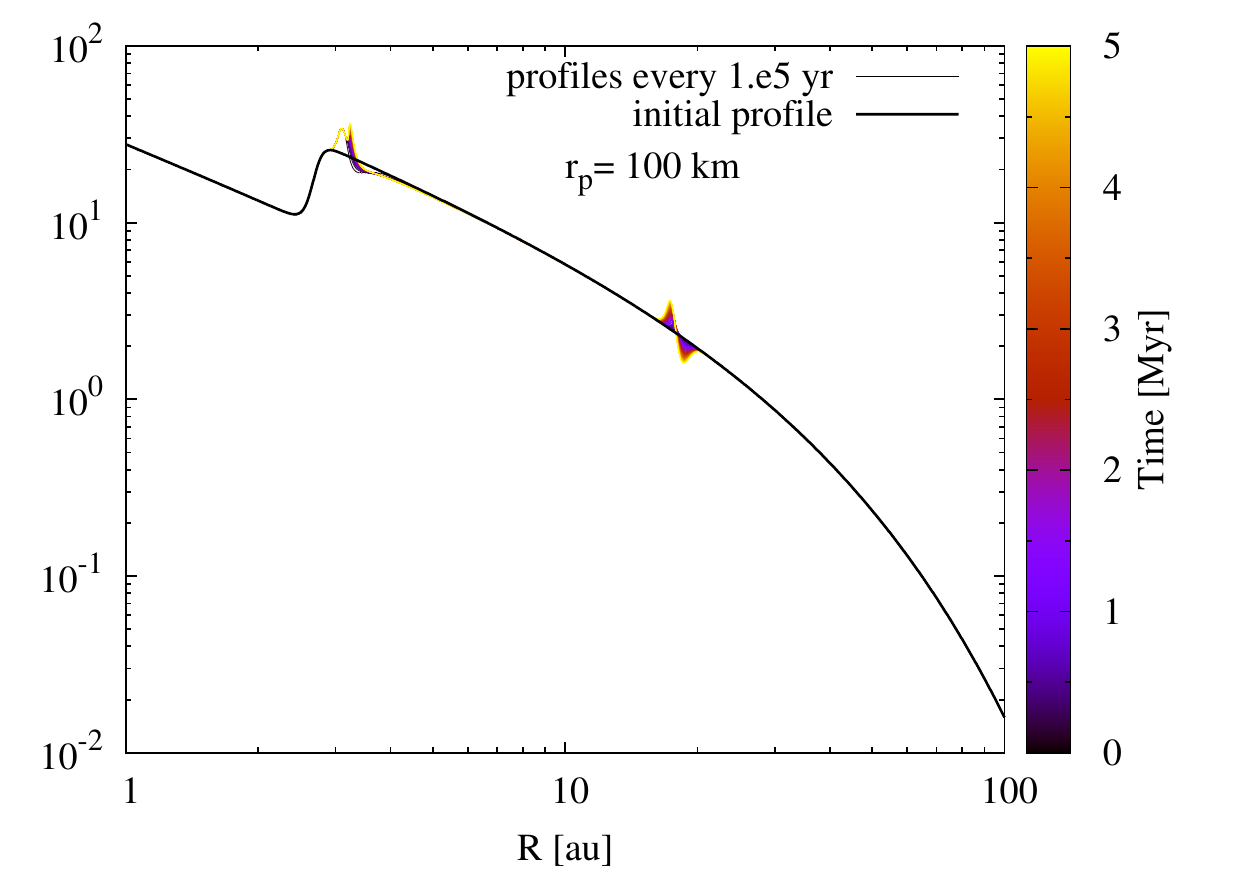}
  \caption{Time evolution of the planetesimal surface density radial profiles for planetesimal populations of different radii. Simulations correspond to a flat disk of $M_d= 0.1~\text{M}_{\odot}$, using $R_{\text{in-dz}}= 2.7$~au, $R_{\text{out-dz}}= 20$~au, $\alpha_{\text{back}}= 10^{-3}$, and $\alpha_{\text{dz}}= 10^{-5}$. Simulations are stopped after 5 Myr of viscous evolution. (Color version online).}
  \label{fig:fig6-sec3-1}
\end{figure*}

Finally, in Fig.~\ref{fig:fig6-sec3-1} we show the time evolution of the planetesimal surface density radial profiles for different planetesimal sizes when a dead zone is considered in the disk.  For smaller size planetesimals the accumulation of solids at the inner and outer edge of the dead zone is more significant due to the larger radial drift velocities. We also note that while the location of the planetesimal accumulation at the inner edge of the dead zone seems to be fixed, or moving slightly outwards, and to be the same for the different planetesimal sizes, the location of the planetesimal accumulation at the outer edge moves inward at different rates. In fact, this is a consequence of the inward motion of the pressure maximum at the outer edge of the dead zone (see Fig.~\ref{fig:fig4-sec3-1} bottom panel). However, it is interesting to analyze the relative migration between the location of the pressure maximum and the corresponding location of planetesimal accumulation. Planetesimals are trapped when $d \ln P/d \ln R= 0$ (Eq.~\ref{eq:eq2-sec2-3}). However, we can see in Fig.~\ref{fig:fig7-sec3-1} that this situation only happens for planetesimals of 0.1 km, which have the larger radial drift velocities, and not for all the time that the maximum pressure exists in the disk. For larger planetesimals the pressure maximum moves inwards faster than the average radial velocity of the bodies, thus they cannot be trapped there.

In next sections, we study whether a planet is able to accrete the above mentioned accumulations of planetesimals at the pressure maxima generated by the dead zone.  

\subsection {Formation of massive cores at pressure maxima}
\label{sec:sec3-2}

As we have shown in the previous section, the inner pressure maximum survives during all the evolution of the disk both for flared and flat disks, and  independently of the value of $\alpha_{\text{back}}$ and $\alpha_{\text{dz}}$ the pressure maximum associated to the outer edge of the dead zone vanishes at some time, which depends on the width of the viscosity transition and on the magnitude of the viscosity reduction. Moreover, while planetesimals are accumulated around the inner pressure maximum, large planetesimals could not accumulate at the outer edge of the dead zone, since they are drifted inwards slower than the pressure maximum moves. 

\subsubsection{In situ formation at pressure maxima}
\label{sec:sec3-2-1}

\begin{figure}[t]
  \centering
  \includegraphics[width= 0.475\textwidth]{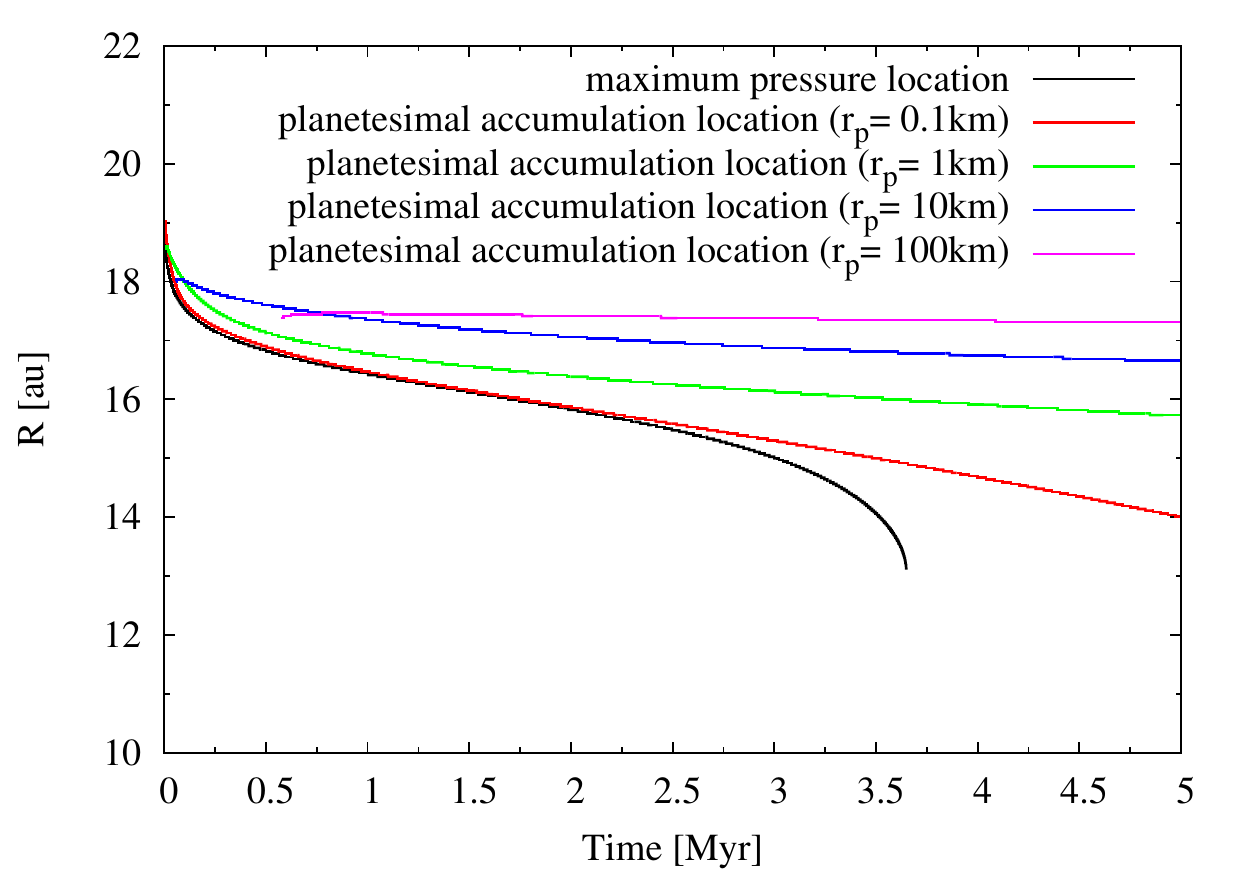} 
  \caption{Time evolution of the location of the pressure maximum planetesimal accumulations for the different planetesimal sizes at the outer edge of the dead zone for a flat disk of $M_d= 0.1~\text{M}_{\odot}$ using $\alpha_{\text{back}}= 10^{-3}$ and $\alpha_{\text{dz}}= 10^{-5}$. (Color version online).}
  \label{fig:fig7-sec3-1}
\end{figure} 

First we analyze the in situ formation of giant planets by fixing their positions to the locations of planetesimal accumulation. The inner pressure maximum efficiently accumulates planetesimals. However, as we have shown in the previous section, for the outer edge of the dead zone the pressure maximum migrates inward, thus not for all planetesimal sizes develops a significant accumulation of planetesimals. Moreover, depending on their sizes the generated planetesimal accumulation deviates differently from the location of the outer pressure maximum (see Fig.~\ref{fig:fig7-sec3-1}). Thus, for the outer pressure maximum we adopted different planet locations for the different planetesimal sizes. In the following we compare the in situ formation of a planet until achieving the critical mass in a model with a dead zone and without a dead zone. 

\begin{figure*}[ht]
    \centering
    \includegraphics[width= 0.475\textwidth]{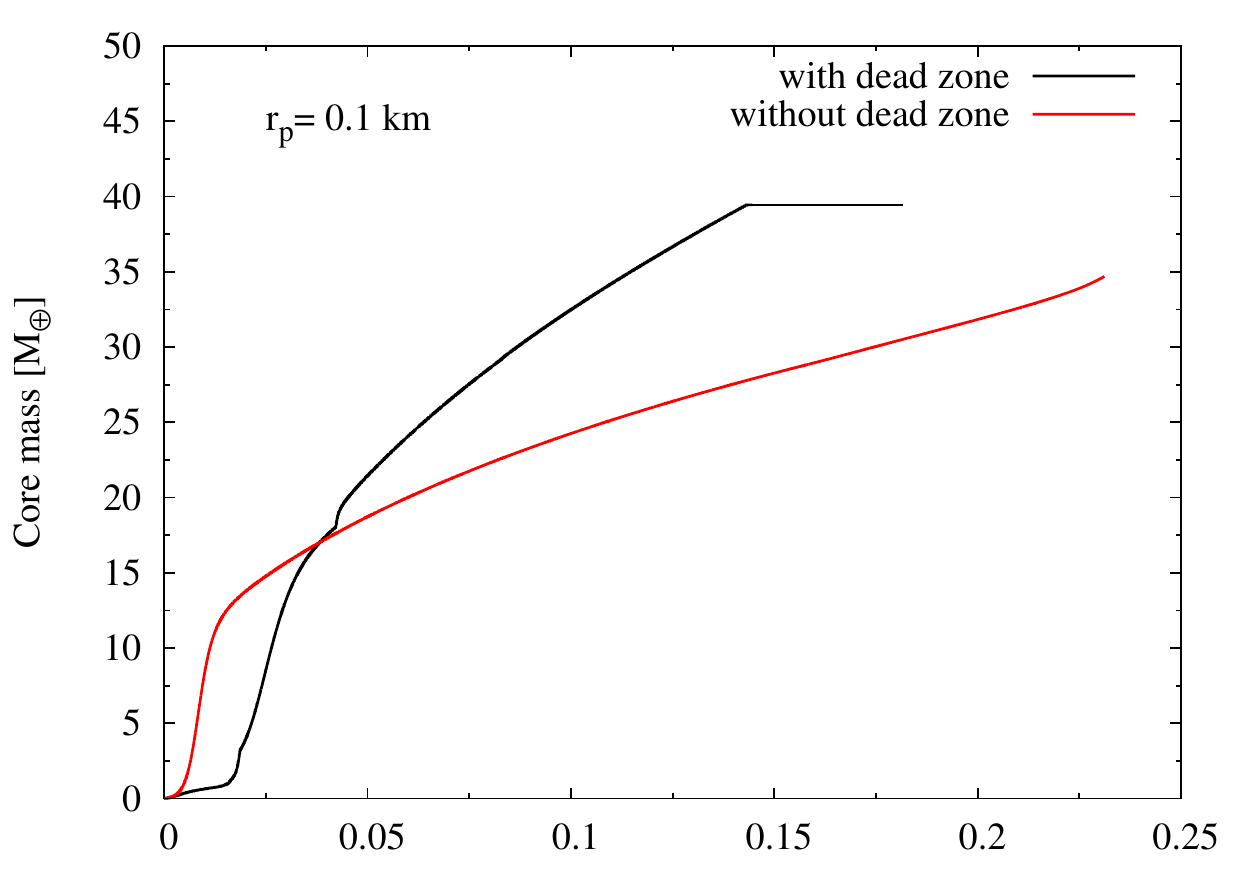} 
    \centering
    \includegraphics[width= 0.475\textwidth]{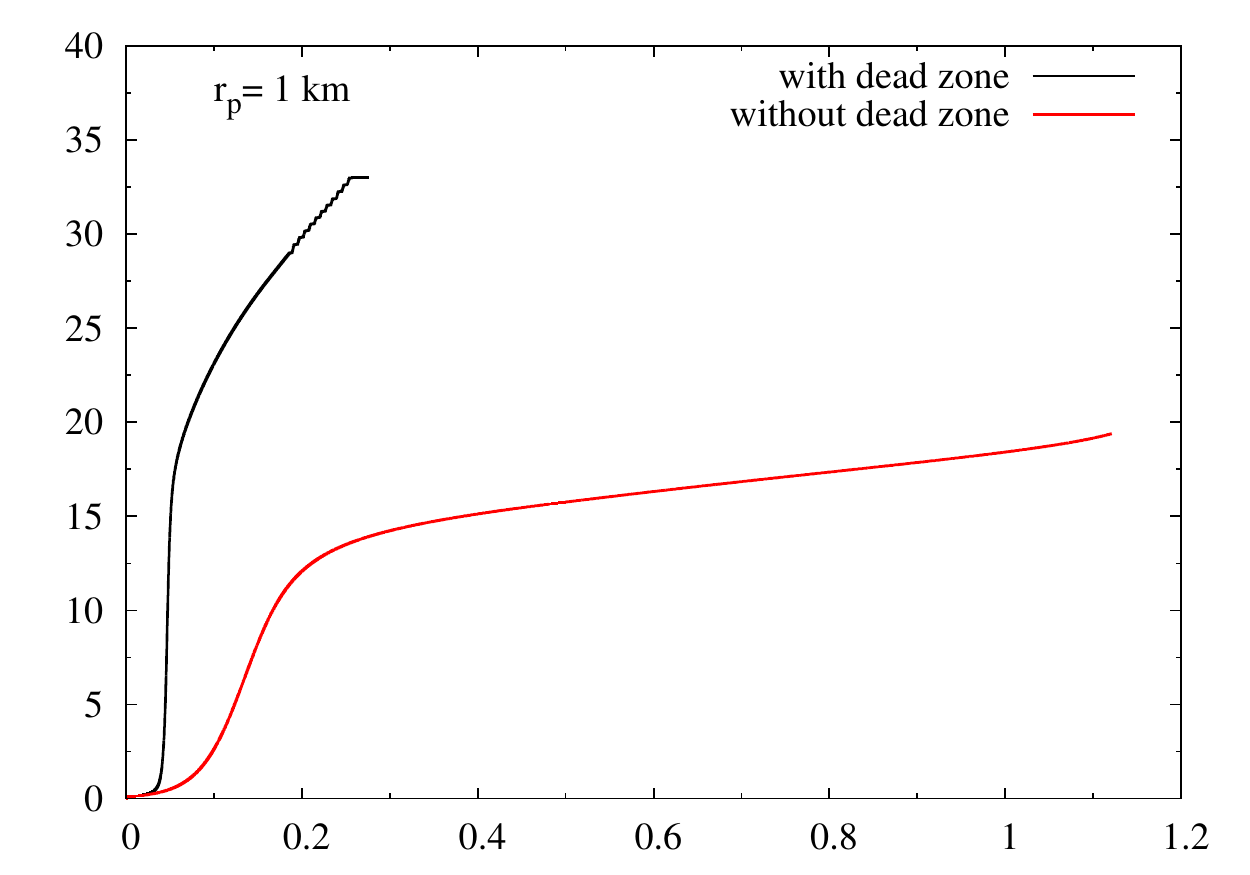} \\
    \centering
    \includegraphics[width= 0.475\textwidth]{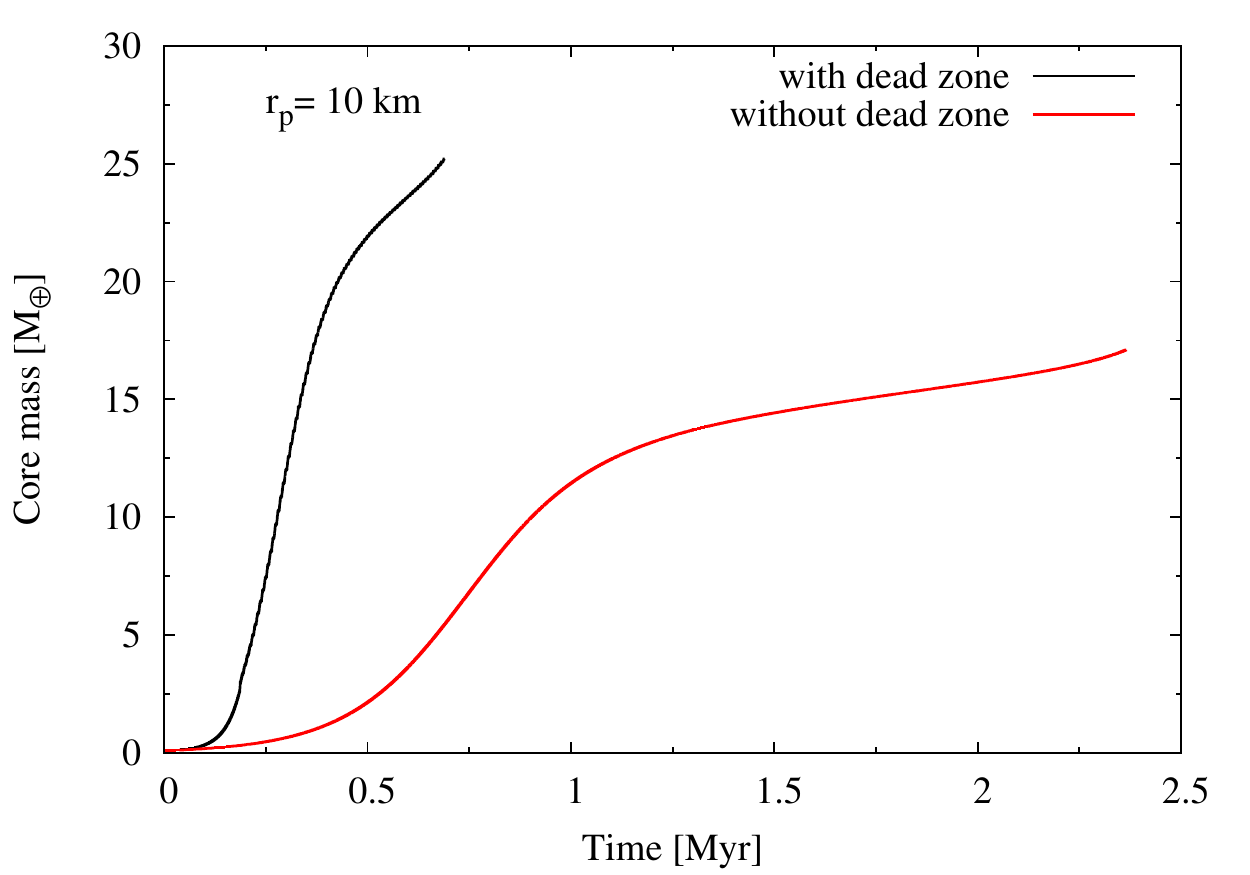} 
    \centering
    \includegraphics[width= 0.475\textwidth]{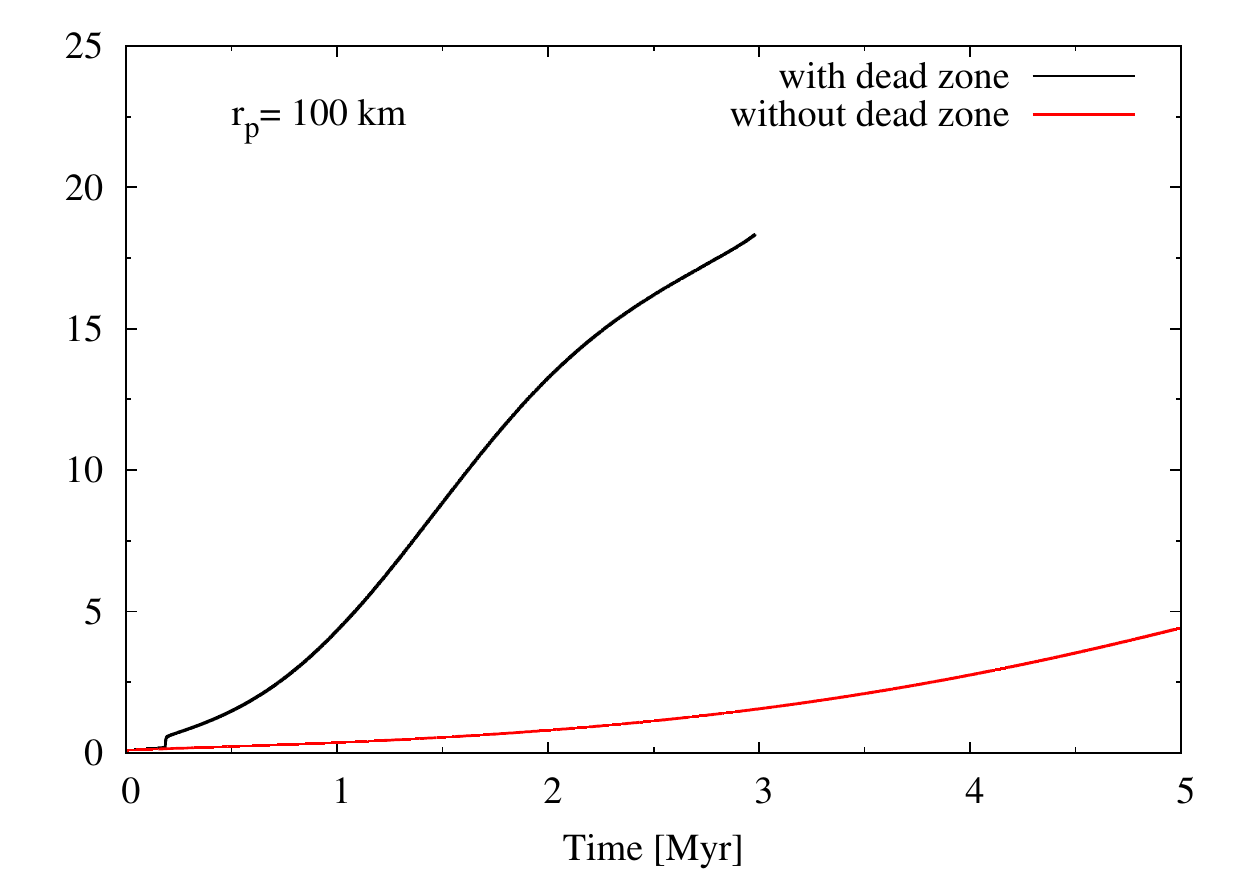}
  \caption{Time evolution of the core's mass in the in situ formation of a planet located at $\sim 3.2$~au for different planetesimal sizes. Simulations correspond to a flat disk of $M_d= 0.1~\text{M}_{\odot}$ considering a disk with (black lines) and without a dead zone (red lines). When a dead zone is considered we use $\alpha_{\text{back}}= 10^{-3}$, $\alpha_{\text{dz}}= 10^{-5}$, $R_{\text{in-dz}}= 2.7$~au and $R_{\text{out-dz}}= 20$~au, while if the dead zone is not taken account we use $\alpha= 10^{-3}$ along the disk. Simulations end when the planet achieves the critical mass or after 5 Myr of viscous disk evolution. (Color version online).}
  \label{fig:fig1-sec3-2-1}
\end{figure*}

\begin{figure*}[ht]
    \centering
    \includegraphics[width= 0.475\textwidth]{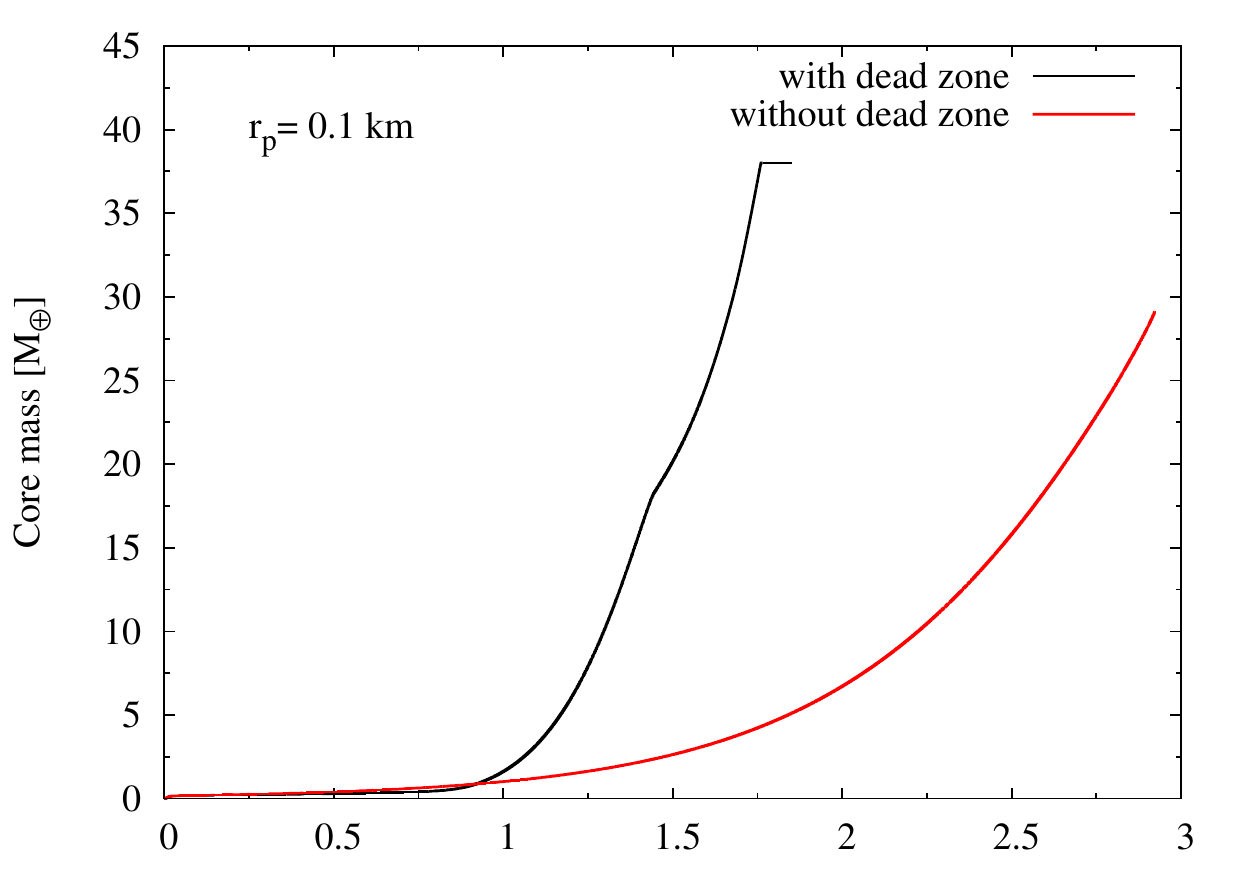} 
    \centering
    \includegraphics[width= 0.475\textwidth]{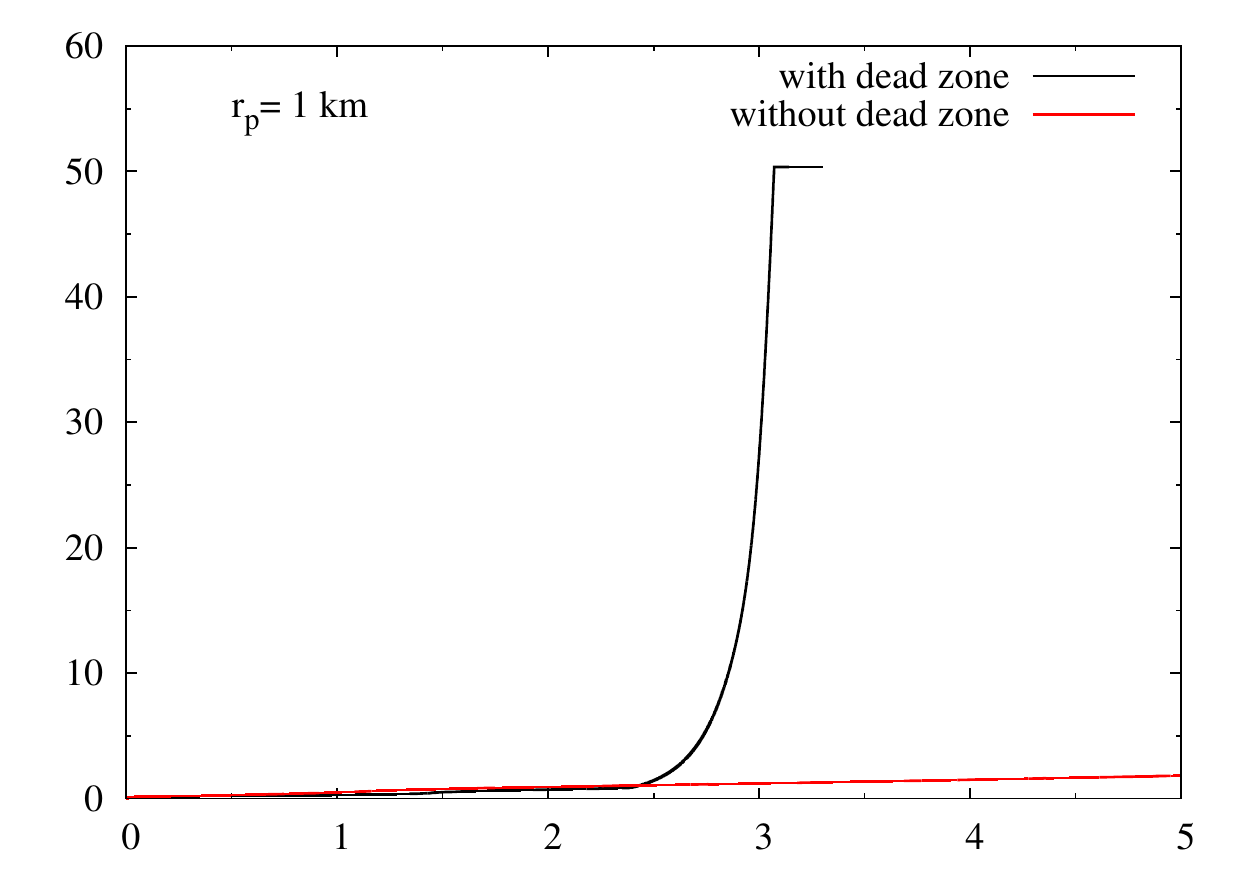} \\
    \centering
    \includegraphics[width= 0.475\textwidth]{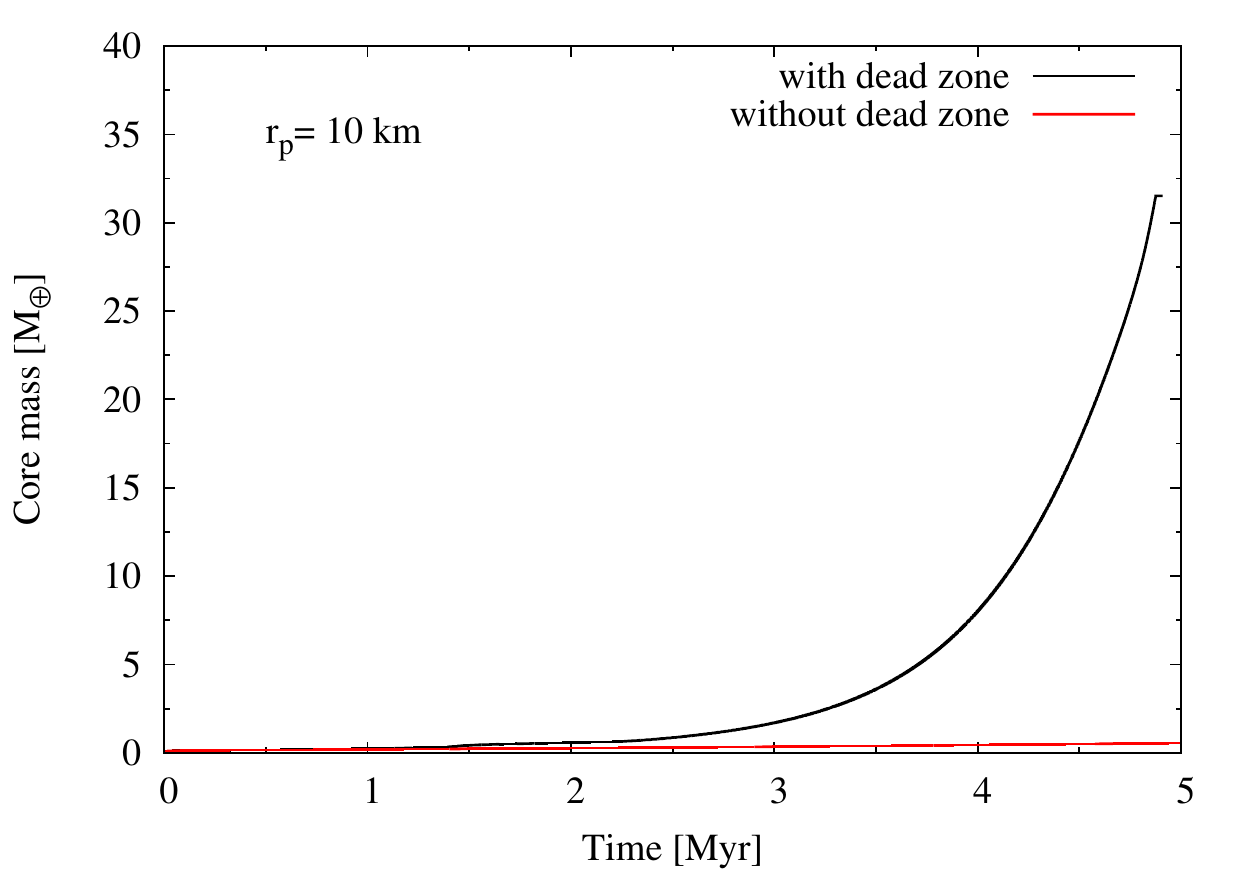} 
    \centering
    \includegraphics[width= 0.475\textwidth]{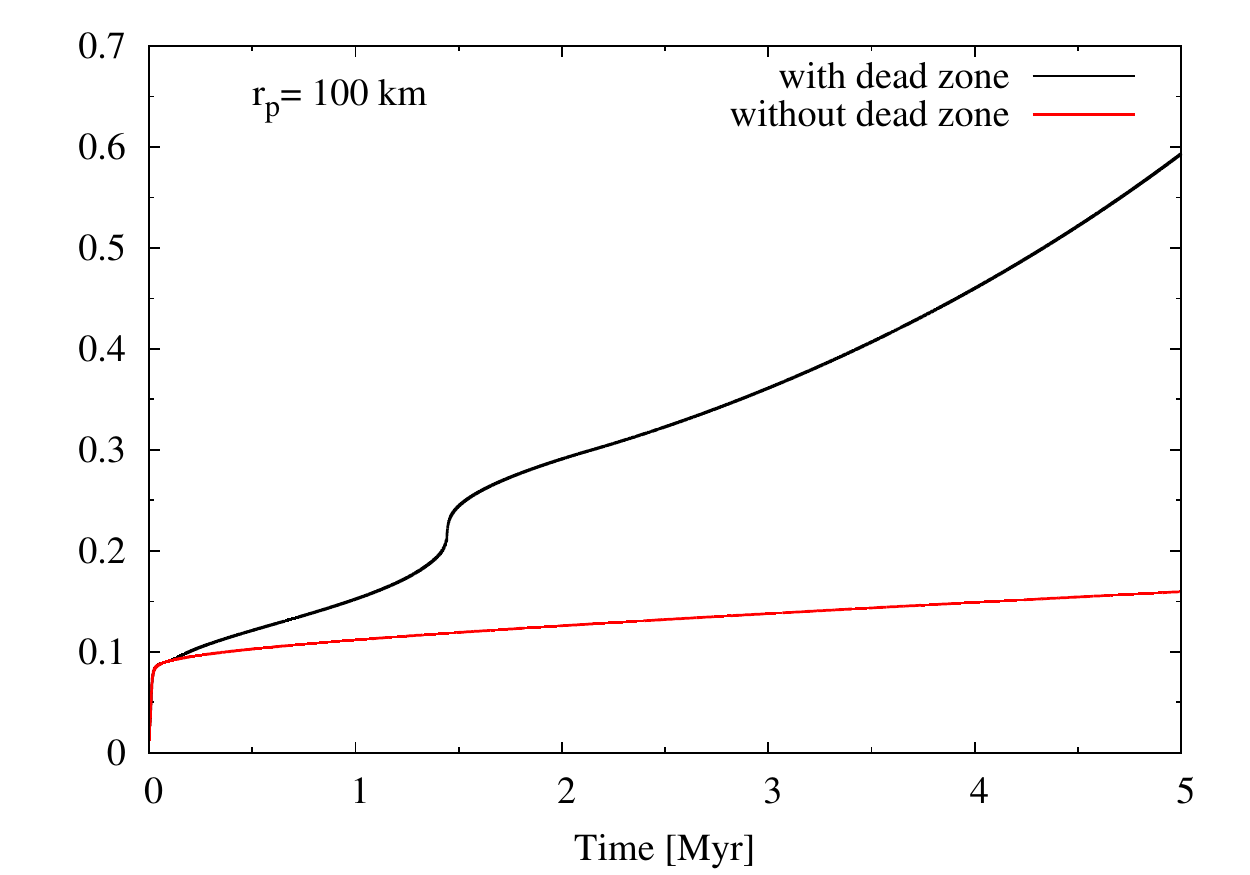}
  \caption{Time evolution of the core's mass in the in situ formation of a planet located near the outer edge of the dead zone. The planet is located at 16~au, 16.5~au, 17~au, and 17.5~au for planetesimals of 0.1 km, 1 km, 10 km, and 100 km of radius, respectively. Simulations correspond to a flat disk of $M_d= 0.1~\text{M}_{\odot}$ considering a disk with (black lines) and without a dead zone (red lines). When a dead zone is considered we use $\alpha_{\text{back}}= 10^{-3}$, $\alpha_{\text{dz}}= 10^{-5}$, $R_{\text{in-dz}}= 2.7$~au and $R_{\text{out-dz}}= 20$~au, while if the dead zone is not taken account we use $\alpha= 10^{-3}$ along the disk. Simulations end when the planet achieves the critical mass or after 5 Myr of viscous disk evolution. (Color version online).}
  \label{fig:fig2-sec3-2-1}
\end{figure*}

As a demonstrative example, we calculate the formation of two planets embedded in a flat disk with mass $M_d= 0.1~\text{M}_{\odot}$. For a disk without a dead zone, we use $\alpha= 10^{-3}$, and for the case with a dead zone we use $\alpha_{\text{back}}= 10^{-3}$, $\alpha_{\text{dz}}= 10^{-5}$, $R_{\text{in-dz}}= 2.7$~au and $R_{\text{out-dz}}= 20$~au. Initially, both embryos have a core mass of $M_c= 0.01~\text{M}_{\oplus}$ and an envelope mass of $\sim 10^{-13}~\text{M}_{\oplus}$. For all sizes of the planetesimal population, the inner planet has been located at $\sim 3.2$~au (roughly corresponding to the planetesimal accumulation location), while the outer planet is located at 16~au, 16.5~au, 17~au, and 17.5~au, considering a planetesimal population size of 0.1 km, 1 km, 10 km, and 100 km, respectively. Fig.~\ref{fig:fig1-sec3-2-1} and Fig.~\ref{fig:fig2-sec3-2-1} show the comparison of the core mass growth as function of time, for the inner and outer planet, respectively, between the cases with and without a dead zone for the different planetesimal radii.  We run our simulations until the planets achieve the critical, or also called cross-over mass, which happens when the mass of the core equals to the mass of the envelope, or for 5~Myr of viscous evolution. For the inner planet, except for the case of planetesimals of 100~km of radius, the planet achieved the critical mass before 5~Myr when a disk without a dead zone was considered. When a dead zone is considered in the disk, the inner planet achieves the critical mass for all planetesimal sizes. In all of the disk models with dead zone, the formation times (the time needed for the planet to achieve the critical mass) are significantly shorter than the time needed in a disk model without a dead zone. It is clear that the accumulation of planetesimals at the pressure maxima, and the consequent increment in the planetesimal surface density, significantly favors the formation of massive cores.      

Similar results have been found for the outer planet achieving the critical mass in less than 5~Myr for planetesimals of 0.1 km, 1 km, and 10 km of radius when a dead zone was considered in the disk. In all of the above cases the accumulation of planetesimals due to the pressure maxima generated by the dead zone clearly favors the formation of massive cores. 

We should note however that the above presented in situ scenario does not take into account the planet-disk interaction, so the results obtained are idealizations as considering a maximal growing rate for the planetary core, and therefore the shortest formation times.

\subsubsection{Formation with planet migration}
\label{sec:sec3-2-2}

The gravitational interaction between the gaseous protoplanetary disk and an embedded planet causes a migration of the planet due to exchange of angular momentum. Low mass planets, up to a tens of Earth masses, are subject to type I migration. In order to study the formation of massive cores in presence of a dead zone, as we mentioned in Sec.~\ref{sec:sec2-4}, we incorporated in our model the prescription for the type I migration given by \citep{Tanaka.et.al.2002}. 

\begin{figure*}[ht]
    \centering
    \includegraphics[width= 0.475\textwidth]{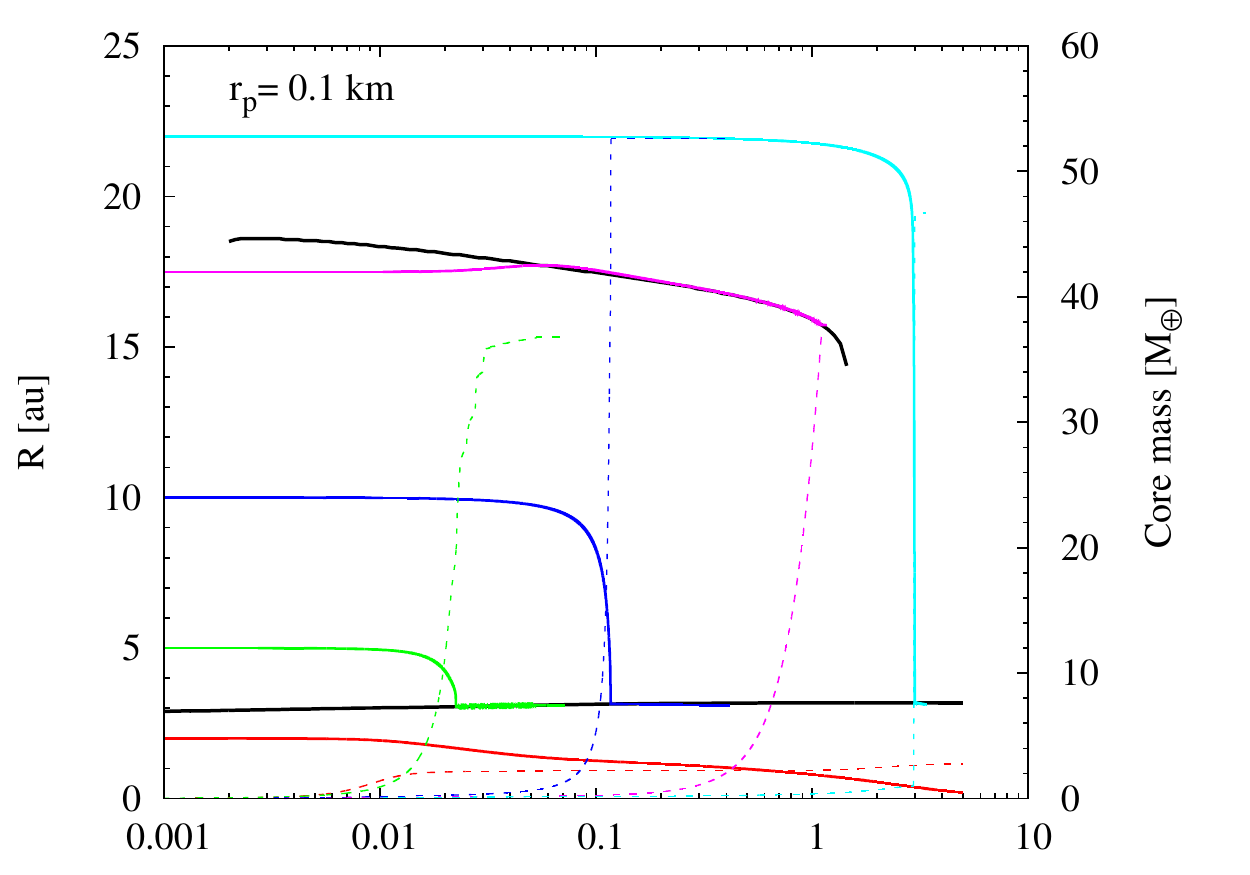} 
    \centering
    \includegraphics[width= 0.475\textwidth]{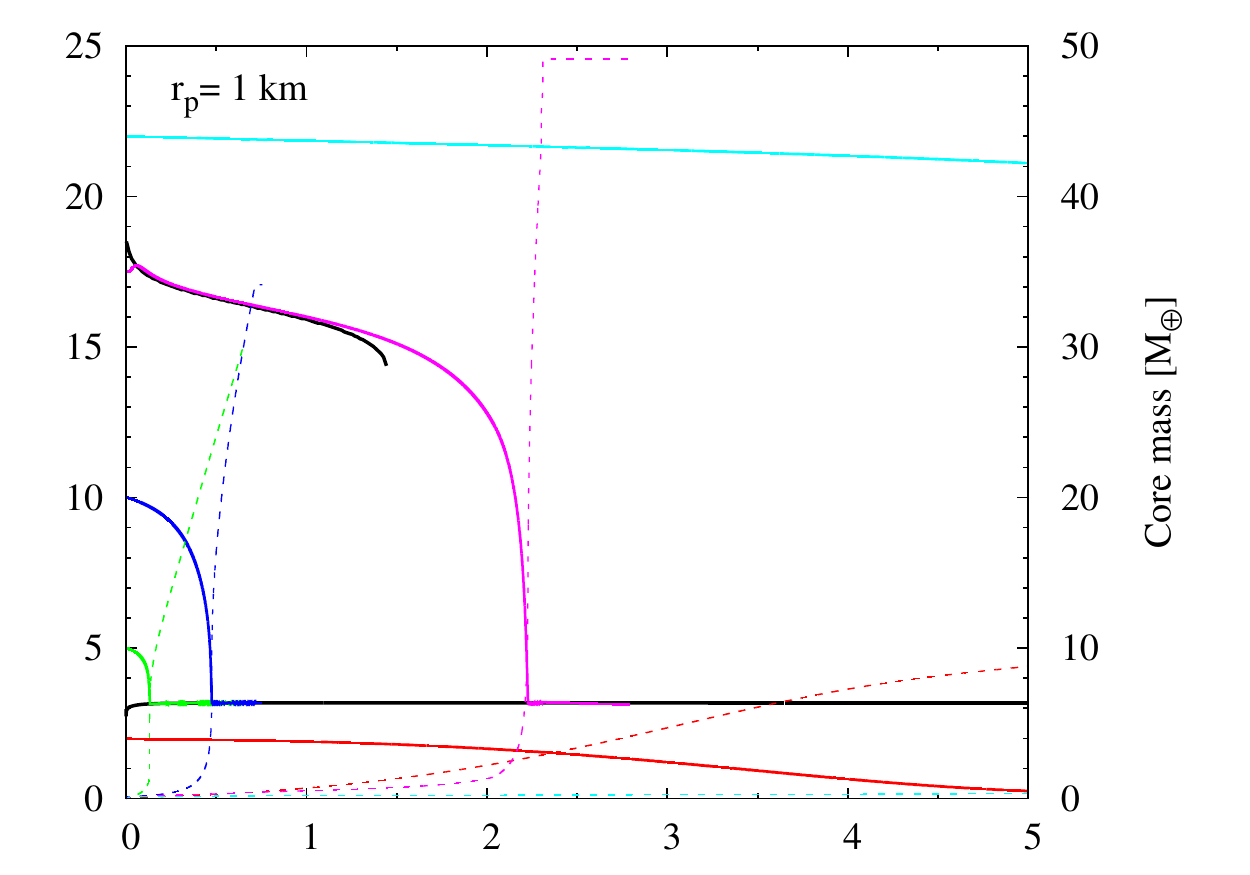} \\
    \centering
    \includegraphics[width= 0.475\textwidth]{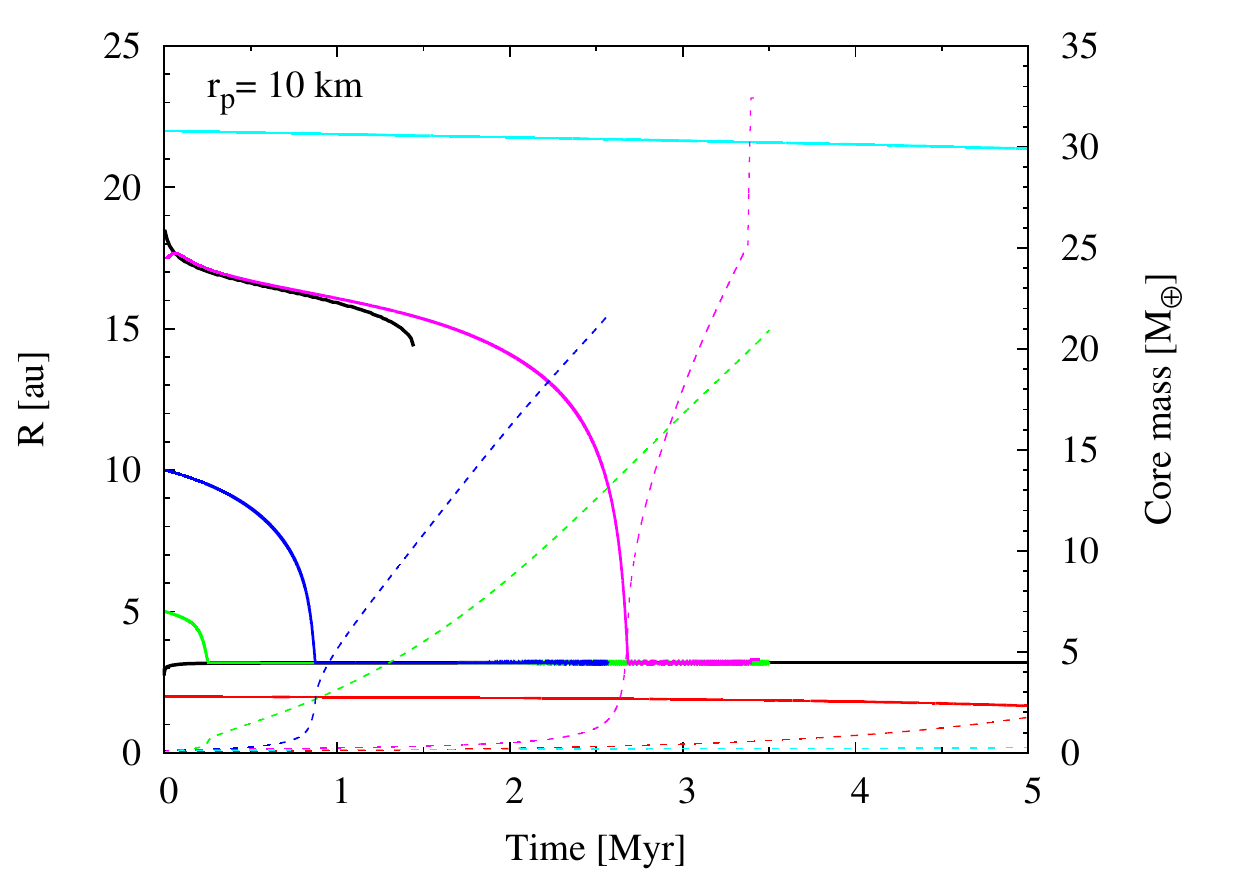} 
    \centering
    \includegraphics[width= 0.475\textwidth]{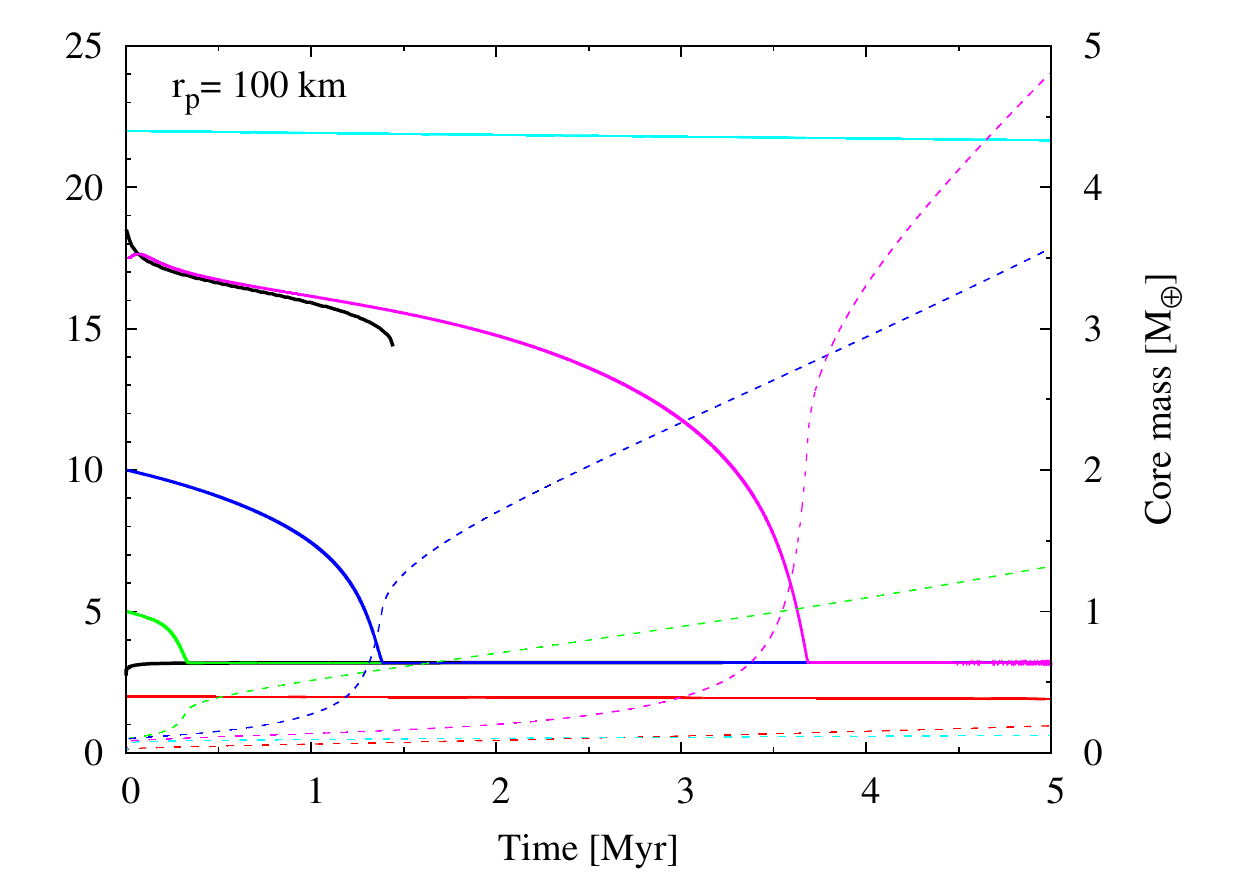}
    \caption{Time evolution of the planet's semi-major axis (left y-axis) and planet's core mass (right y-axis) for different initial planet locations: 2~au, 5~au, 10~au, 17.5~au, and 22~au. Black lines represent the position of the zero torque. Simulations correspond to a flat disk with mass of $M_d= 0.1~\text{M}_{\odot}$ using $\alpha_{\text{back}}= 10^{-3}$, $\alpha_{\text{dz}}= 10^{-5}$, $R_{\text{in-dz}}= 2.7$~au and $R_{\text{out-dz}}= 20$~au. Simulations end when the planet achieves the critical mass or after 5 Myr of viscous disk evolution. (Color version online).}
  \label{fig:fig1-sec3-2-2}
\end{figure*}

We analyze the formation of a giant planet whose core is located at different position in the disk by the accretion of planetesimals of different sizes (Fig,~\ref{fig:fig1-sec3-2-2}). We initially start the planetary core at 2~au, 5~au, 10~au, 17.5~au, and 22~au. The first and the last position of the planet are outside the dead zone, while the other three locations are inside of it, where we choose two positions near the inner and outer edges of the dead zone. We run our simulations for our fiducial model, i.e., $M_d= 0.1~\text{M}_{\odot}$, $\alpha_{\text{back}}= 10^{-3}$, $\alpha_{\text{dz}}= 10^{-5}$, $R_{\text{in-dz}}= 2.7$~au and $R_{\text{out-dz}}= 20$~au, considering a flat disk.     

If the starting position of the planetary core is at 2~au, for planetesimals of 0.1~km and 1~km of radius, the planet becomes massive enough to migrate inwards approaching the inner edge of the disk. For planetesimals of 10~km and 100~km of radius, the mass of the planet remains small enough to not suffer a significant migration. 

On the other hand, when the initial location of the core is at 5~au and 10~au, the formation history of the planet is similar. For planetesimals of 0.1~km of radius, the planet quickly becomes massive enough to obey a fast inward migration, and the planet quickly achieves the migration trap, namely the zero toque location, generated by the inner edge of the dead zone. Having trapped there, the planet is able to accrete  the already accumulated planetesimals at the pressure maximum. This phenomenon allows the planet to achieve the critical mass very quickly, in a time-scale of $10^{5}$~yr. For planetesimals of 1~km and 10~km, the planet grows and migrates inward until achieves the inner zero torque location in less than 1~Myr. Then the planet continues growing being trapped at the zero torque location until it achieves the critical mass. For planetesimals of 1~km of radius, the planet achieves the critical mass in less than 1~Myr, while for planetesimals of 10~km of radius the planet achieves the critical mass in a few Myr. Finally, for planetesimals of 100~km of radius the planet only achieves a few Earth masses.  

When the initial location of the planet is near to the outer edge of the dead zone, in our case at 17.5~au, thus inside of it, the planet initially migrates outward until it is getting trapped at the outer migration trap. As we show in Fig.~\ref{fig:fig4-sec3-1} and Fig.~\ref{fig:fig5-sec3-1}, the migration trap moves inward due to the diffusive evolution of the outer density/pressure maximum. Thus during its slow inward motion with the migration trap the planet grows its mass by accreting planetesimals. However, as shown in Fig.~\ref{fig:fig4-sec3-1} and Fig.~\ref{fig:fig5-sec3-1} the location of the pressure maximum deviates significantly from the migration trap's position at the outer edge of the dead zone, in a $\sim 1$~Myr timescale. Moreover, as shown in Fig. ~\ref{fig:fig6-sec3-1}, the locations of planetesimal accumulation and the pressure maximum coincide only for the smaller planetesimals, which have the largest radial drift velocities. For these reasons, the planet can reach the critical mass before the vanishing of the migration trap only for planetesimals with size of 0.1~km. For the rest of the planetesimal sizes, after the vanishing of the zero torque location the planet migrates inward until reaches the inner zero torque location, achieving the critical mass there. 

Finally, if the initial location of the planet is 22~au, beyond the outer edge of the dead zone, the planet becomes massive enough to undergo a significant inward migration only for small sized planetesimals. After vanishing the outer zero torque, the planet starts to migrate achieving the critical mass at the inner migration trap. We note that the planet significantly increases its mass when passing through the outer planetesimal accumulation location. However, for larger planetesimals, the planet does not become massive enough to perform a substantial migration.

\subsubsection{Oscillation of the planet's semi-major axis around zero torque location}
\label{sec:sec3-2-3}
  
As we have shown in the previous section, the migration of the planet towards the inner zero torque location, and its trapping there, favors the formation of a massive core. However, as we have also demonstrated in previous sections, the positions of the inner pressure maximum and of the migration trap do not coincide. Thus, the accumulation of planetesimals is at different location than the migration trap.      

On the other hand numerical magnetohydrodynamic simulations have revealed that at the inner edge of the dead zone the planet is not simply trapped but oscillates around the zero torque place \citep{FaureNelson2016}. The oscillation of the semi-major axis of a massive core at the outer edge of the dead zone has also been found in the hydrodynamic simulations of \cite{regaly_etal2013}. In both cases, beside the turbulence, the oscillation of the planet's semi-major axis might be the result of the planet-vortex interaction, since the sudden jumps in gas surface density are prone to the Rossby wave instability enabling the formation of a large vortex at the density jump's position \citep{ Lovelaceetal1999}.

Thus, following the above mentioned works, we mimic the oscillation of the growing protoplanet around the migration trap. For this, we implemented a noisy oscillation of the semi-major axis of the protoplanet in the form
\begin{eqnarray}
  a_P= a_{0} + \psi \Delta a \sin\left( \frac{t-t^*}{\xi T_{\text{orb}}} \right),
  \label{eq:eq1-sec3-2-3}
\end{eqnarray}
where $a_{0}$ is the zero torque location, $\Delta a$ is the oscillation amplitude, $t^*$ is the time at which the planet achieves the zero torque location, and $\psi$ and $\xi$ are random numbers that take values between $1 \pm f$ and  $1 \pm g$, respectively. The constants $f$ and $g$ ($0<f,g<1$), are chosen in a way that enables the planet to accrete the accumulated planetesimals. To fulfill this criterion, the oscillation amplitude should be set large enough for the planet to go through repeatedly the region where planetesimals are accumulated.
 
\begin{figure*}[t]
    \centering
    \includegraphics[width= 0.475\textwidth]{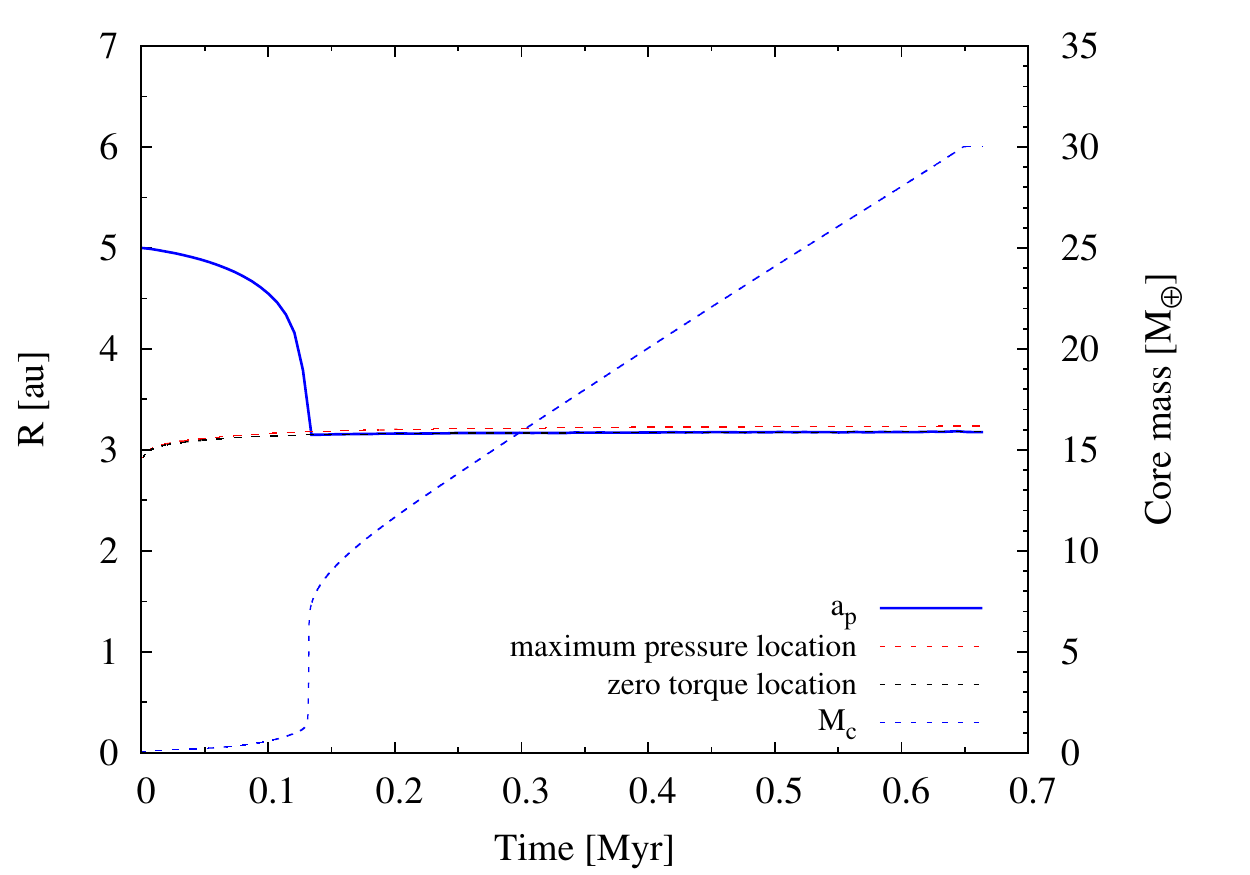} 
    \centering
    \includegraphics[width= 0.475\textwidth]{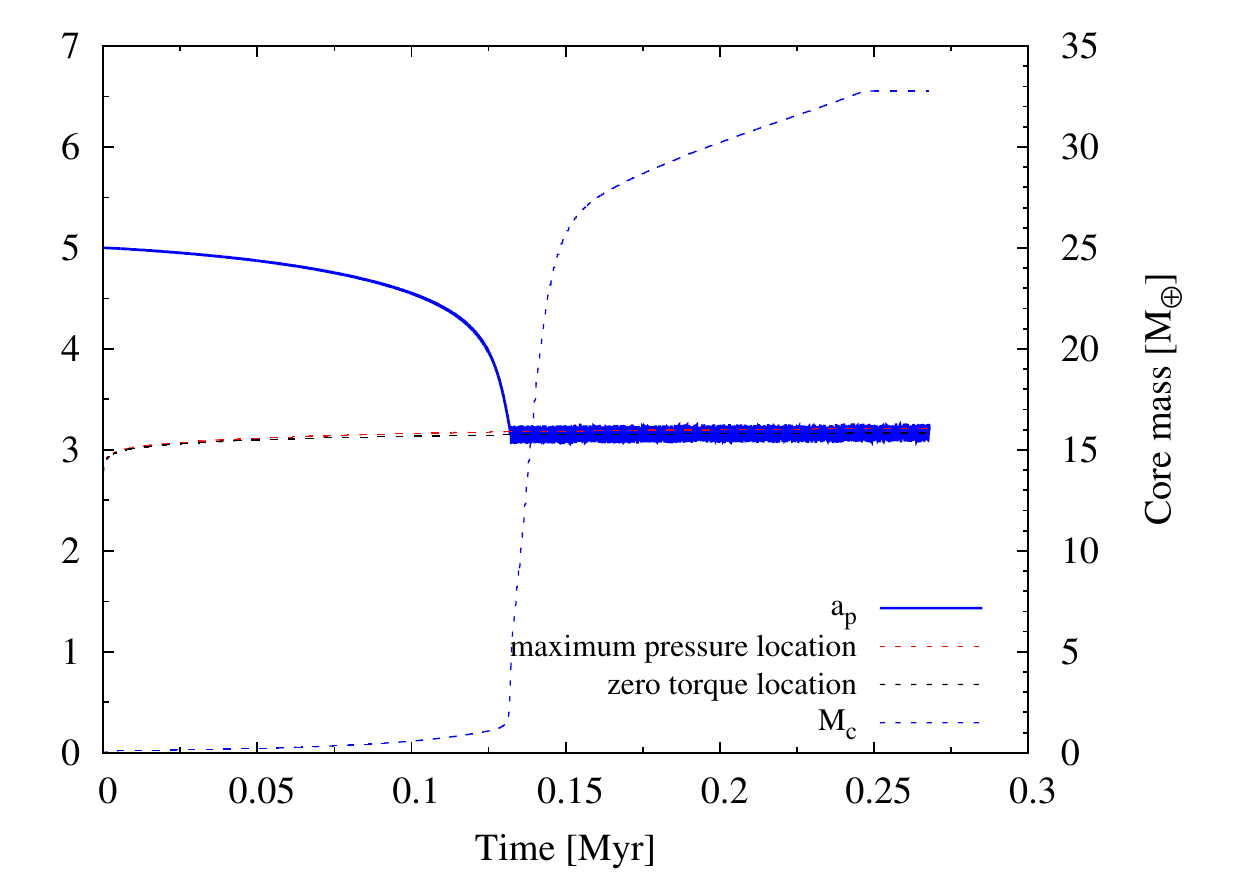} \\
    \centering
    \includegraphics[width= 0.475\textwidth]{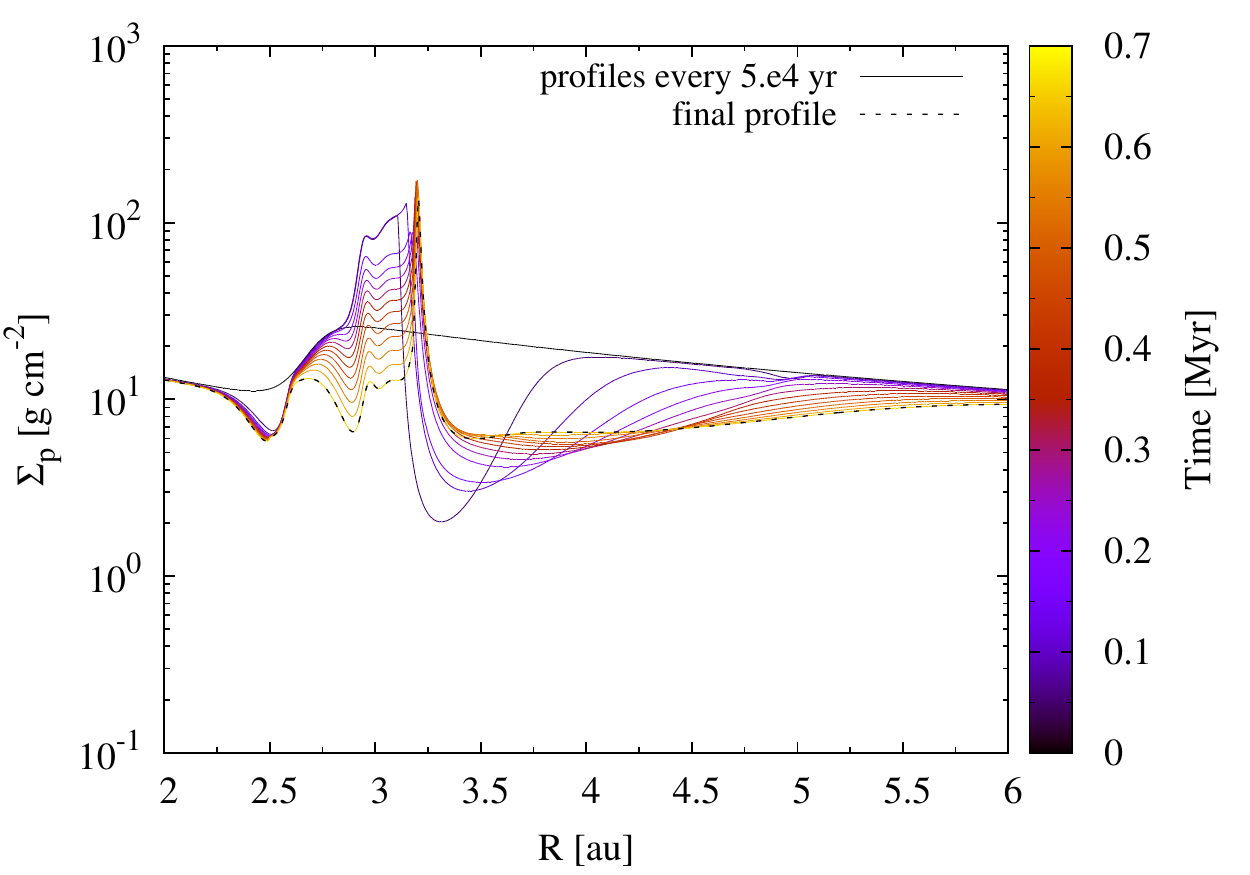} 
    \centering
    \includegraphics[width= 0.475\textwidth]{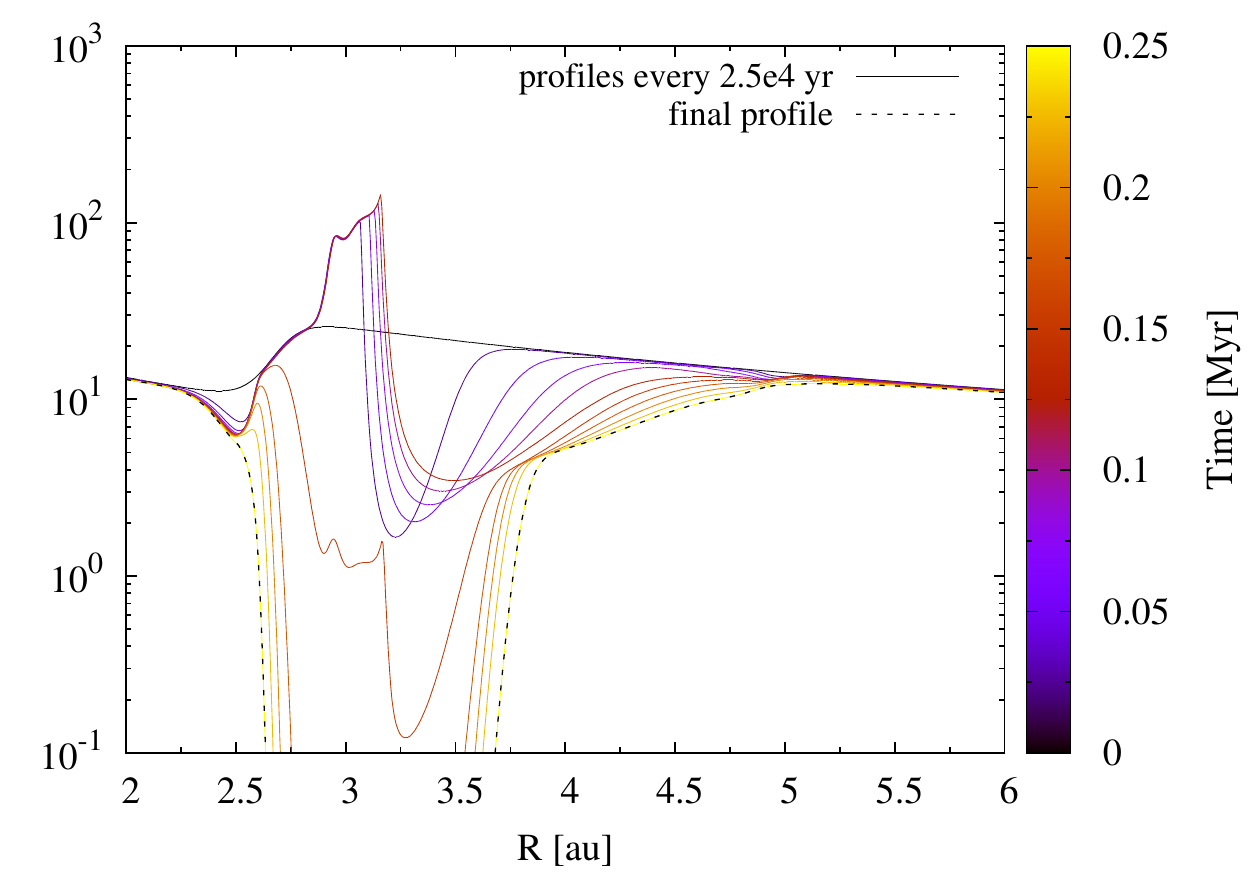}
    \caption{Top: time evolution of the planet's semi-major axis (left y-axis) and planet's core mass (right y-axis) being initially located at 5~au. Red dashed line represents the location of the pressure maximum, and black dashed line represents the zero torque's position. Bottom: time evolution (color palette) of the planetesimal surface density radial profiles. Left panel represents the case where the planet is trapped at the zero torque location, while right panel represents the case where the semi-major axis of the planet oscillates around the zero torque location after being trapped. Simulations correspond to a flat disk with mass $M_d= 0.1~\text{M}_{\odot}$ using $\alpha_{\text{back}}= 10^{-3}$, $\alpha_{\text{dz}}= 10^{-5}$, $R_{\text{in-dz}}= 2.7$~au, $R_{\text{out-dz}}= 20$~au, and planetesimals of 1~km of radius. Simulations end when the planet achieves the critical mass. (Color version online).}
  \label{fig:fig1-sec3-2-3}
\end{figure*}

Fig.~\ref{fig:fig1-sec3-2-3} (top-left panel) shows the time evolution of the semi-major axis of the planet's core and mass, started initially from 5~au, considering a population of 1~km sized planetesimals. The core grows and migrates quickly until reaching the migration trap. Having trapped there, the protoplanet continues growing until achieving the critical mass. However, as we can see, the the pressure maximum (red dashed line) is not at the same location as the migration trap. Analyzing the time evolution of the planetesimal surface density profiles for this simulations (bottom-left panel) one can see that while the planet is able to accrete a large quantity of the accumulated planetesimals, it is not able to accrete all the available mass and, a significant amount of planetesimals remains in the disk. When applying the random oscillation of the growing protoplanet around the zero torque location, the formation time of the critical mass core is reduced in more than 50\% with respect to the previous case. One can see that the random oscillation allows the growing planet to accrete all the mass accumulated at pressure maximum (bottom-right panel).       

\begin{figure*}[t]
    \centering
    \includegraphics[width= 0.475\textwidth]{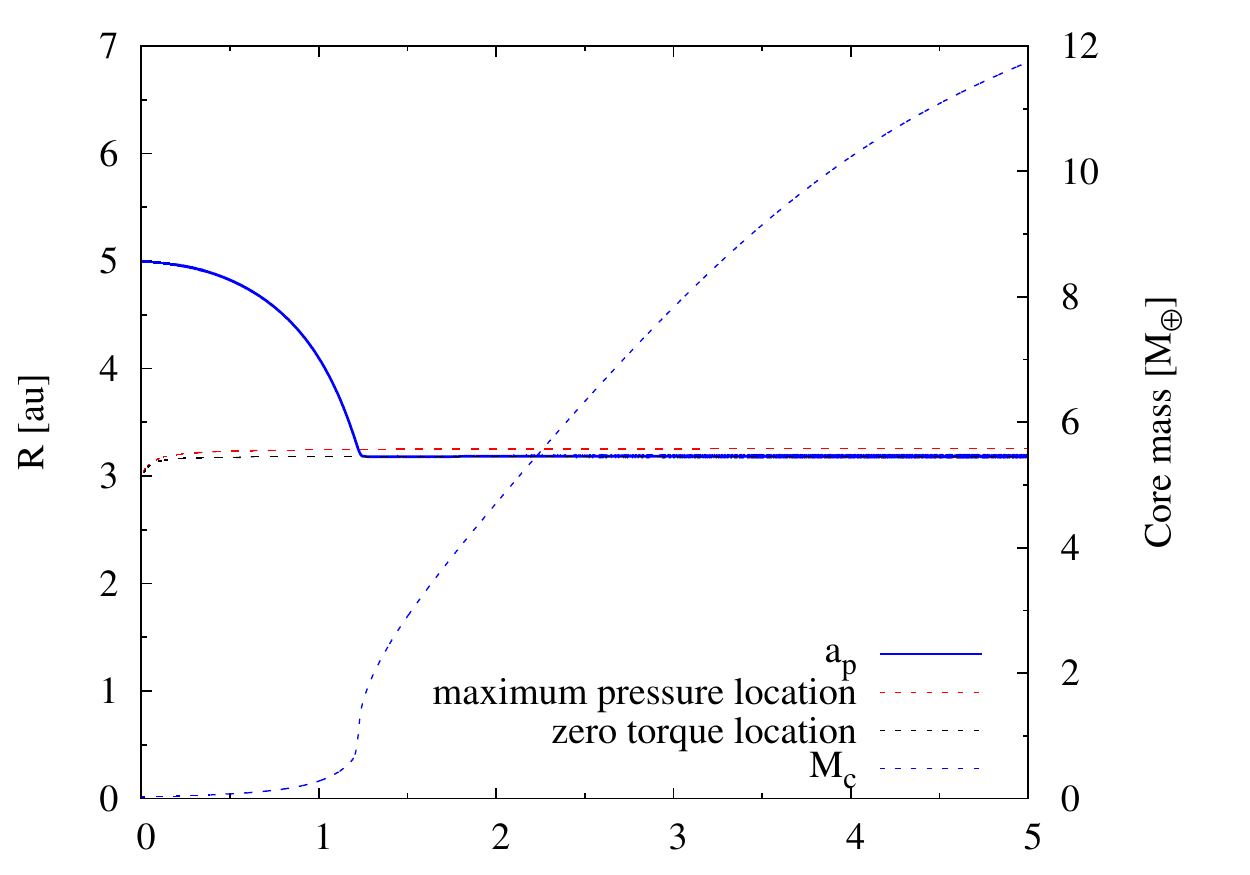} 
    \centering
    \includegraphics[width= 0.475\textwidth]{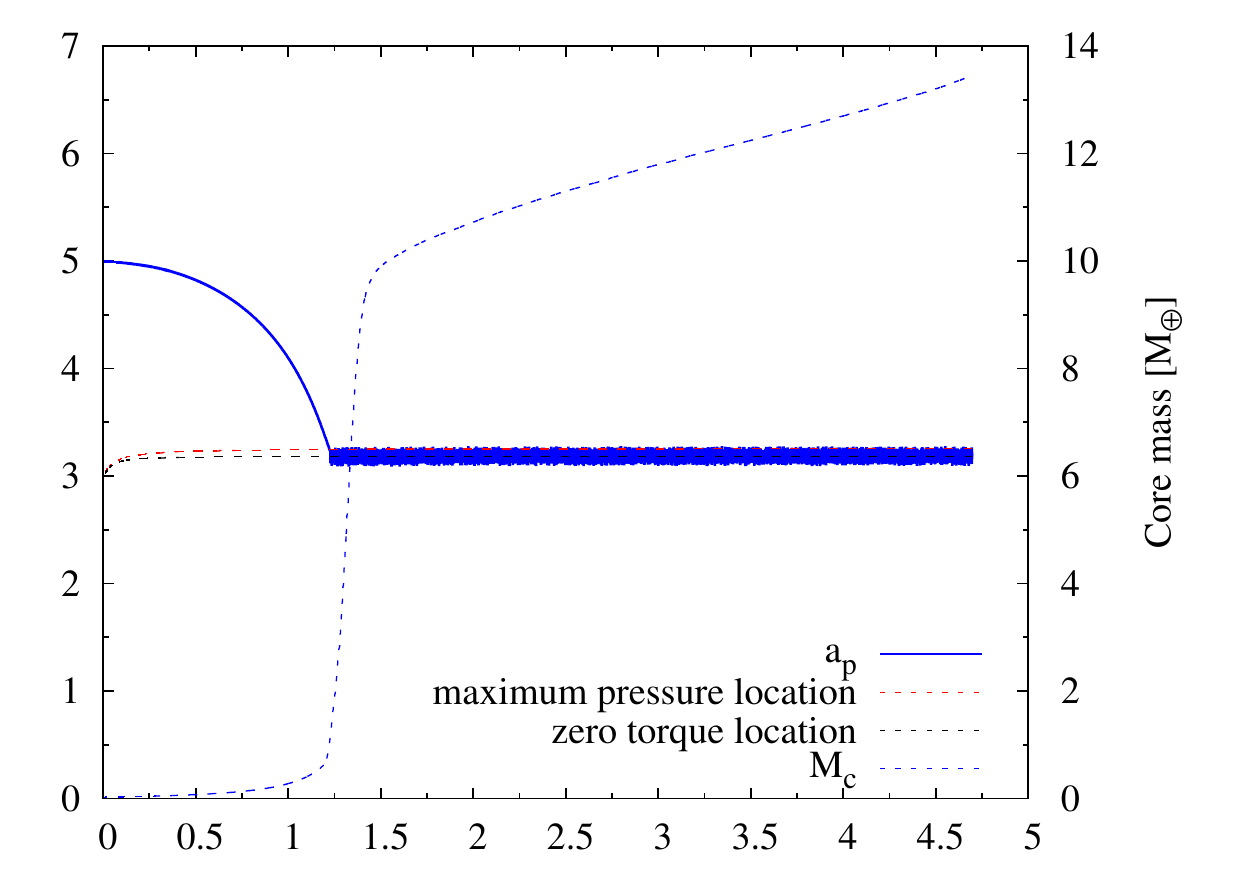} \\
    \centering
    \includegraphics[width= 0.475\textwidth]{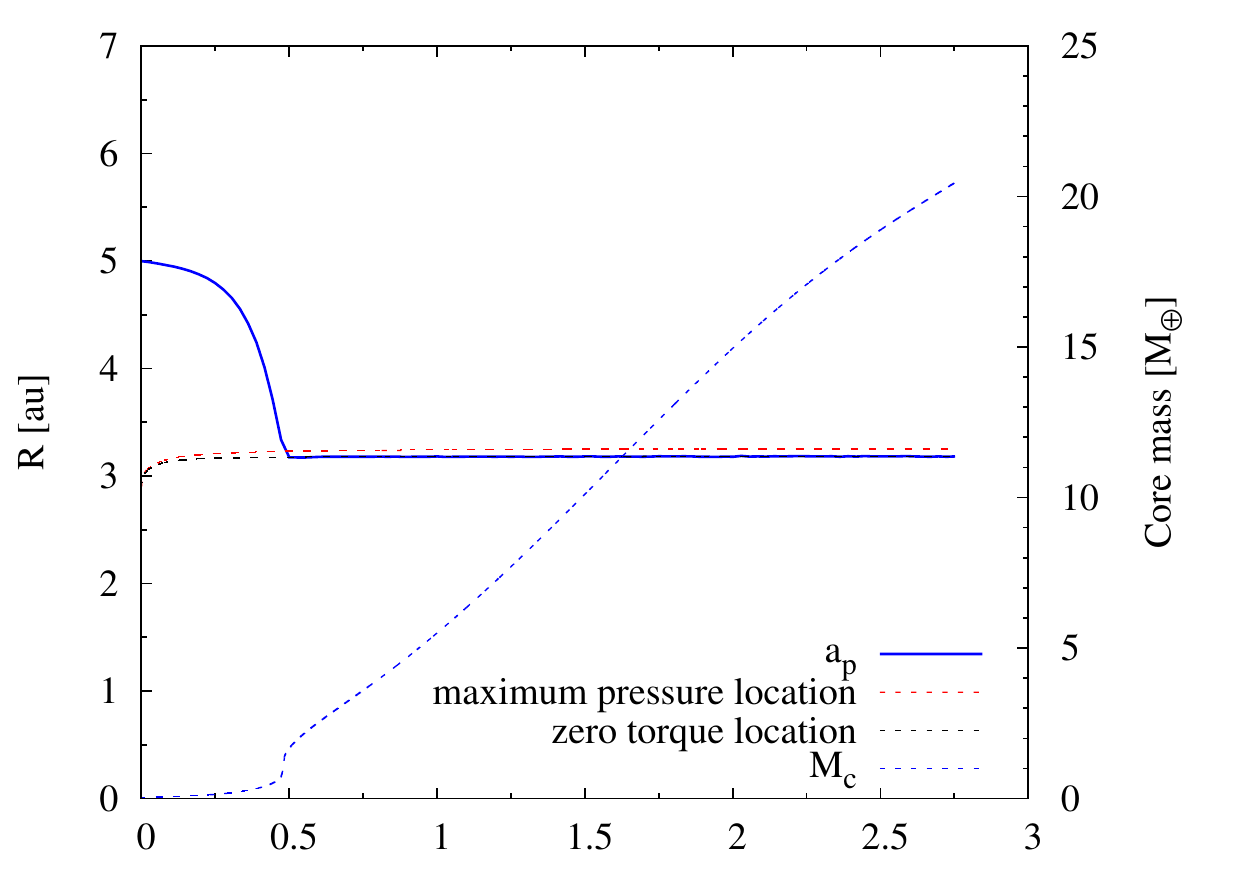} 
    \centering
    \includegraphics[width= 0.475\textwidth]{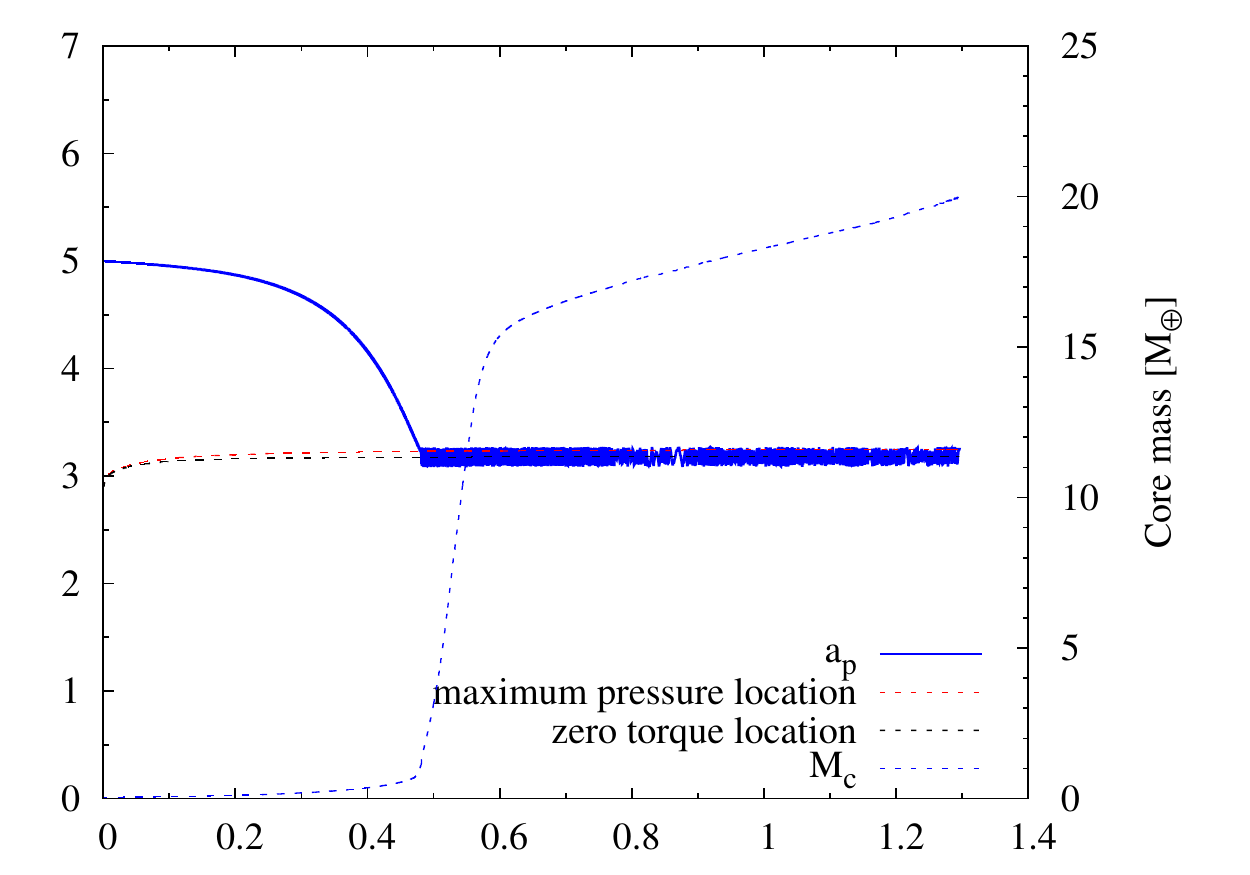} \\
    \centering
    \includegraphics[width= 0.475\textwidth]{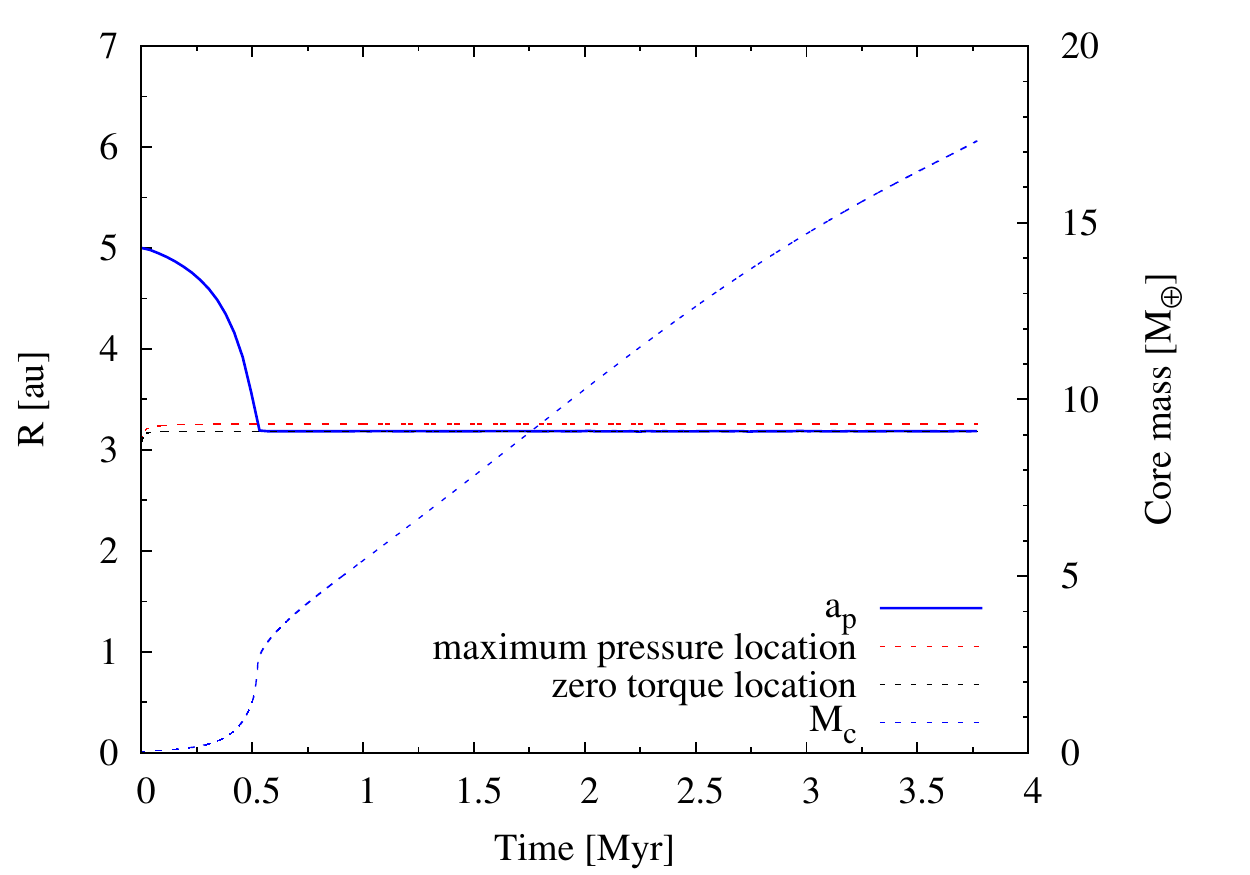} 
    \centering
    \includegraphics[width= 0.475\textwidth]{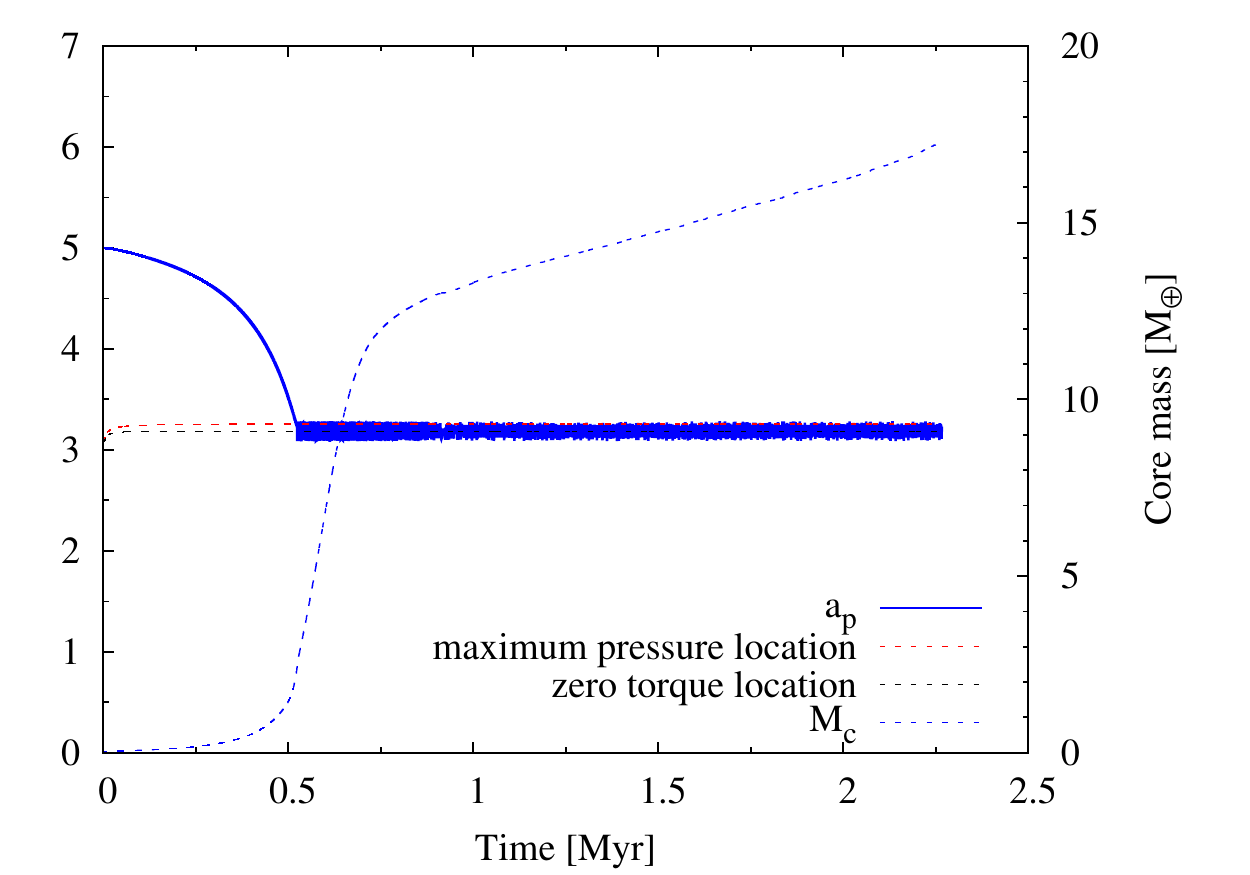}
    \caption{Same as top panel of Fig.~\ref{fig:fig1-sec3-2-3} for a disk with $M_d= 0.03~\text{M}_{\odot}$ (top panel) and a disk with $M_d= 0.05~\text{M}_{\odot}$ (middle panel). Bottom panel represents the case of a disk with mass $M_d= 0.05~\text{M}_{\odot}$ but considering that $\alpha_{\text{back}}= 10^{-2}$, $\alpha_{\text{dz}}= 10^{-4}$ (Color version online).}
  \label{fig:fig2-sec3-2-3}
\end{figure*}

This phenomenon is more significant for less massive disks. Fig.~\ref{fig:fig2-sec3-2-3} shows the results of the same simulations as before, but considering a disk with mass $M_d= 0.03~\text{M}_{\odot}$ (top panel) and with $M_d= 0.05~\text{M}_{\odot}$ (middle panel). Left and right panels show the behavior of the semi-major axis and the mass of the growing protoplanet as functions of time with and without the random oscillation around the zero torque location, respectively. For a disk of $M_d= 0.03~\text{M}_{\odot}$, the planet does not achieve the critical mass in less than 5~Myr unless the random oscillation around the zero torque location is considered. For a disk of $M_d= 0.05~\text{M}_{\odot}$, the formation time is practically halved in the case of random oscillation.

Finally, bottom panel of Fig.~\ref{fig:fig2-sec3-2-3} shows again same results but now for a disk of $M_d= 0.05~\text{M}_{\odot}$ considering $\alpha_{\text{back}}= 10^{-2}$ and $\alpha_{\text{dz}}= 10^{-4}$. In this case, the formation times are $\sim 1$~Myr greater than the case of $\alpha_{\text{back}}= 10^{-3}$ and $\alpha_{\text{dz}}= 10^{-5}$ (middle panel). This is due to the fact that the accumulation of planetesimals around the pressure maximum location is more effective for smaller values of the alpha parameter.   

\begin{figure*}[ht]
    \centering
    \includegraphics[width= 0.475\textwidth]{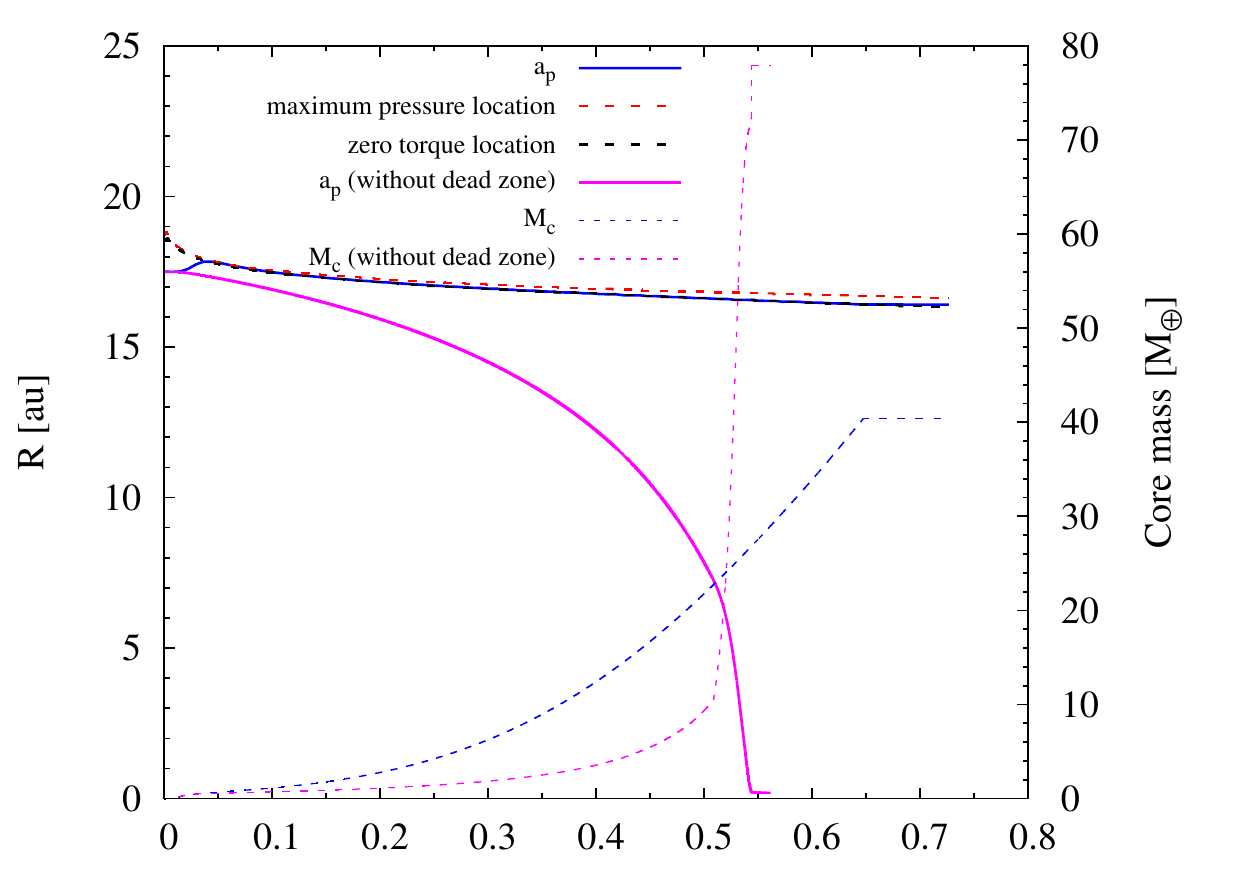} 
    \centering
    \includegraphics[width= 0.475\textwidth]{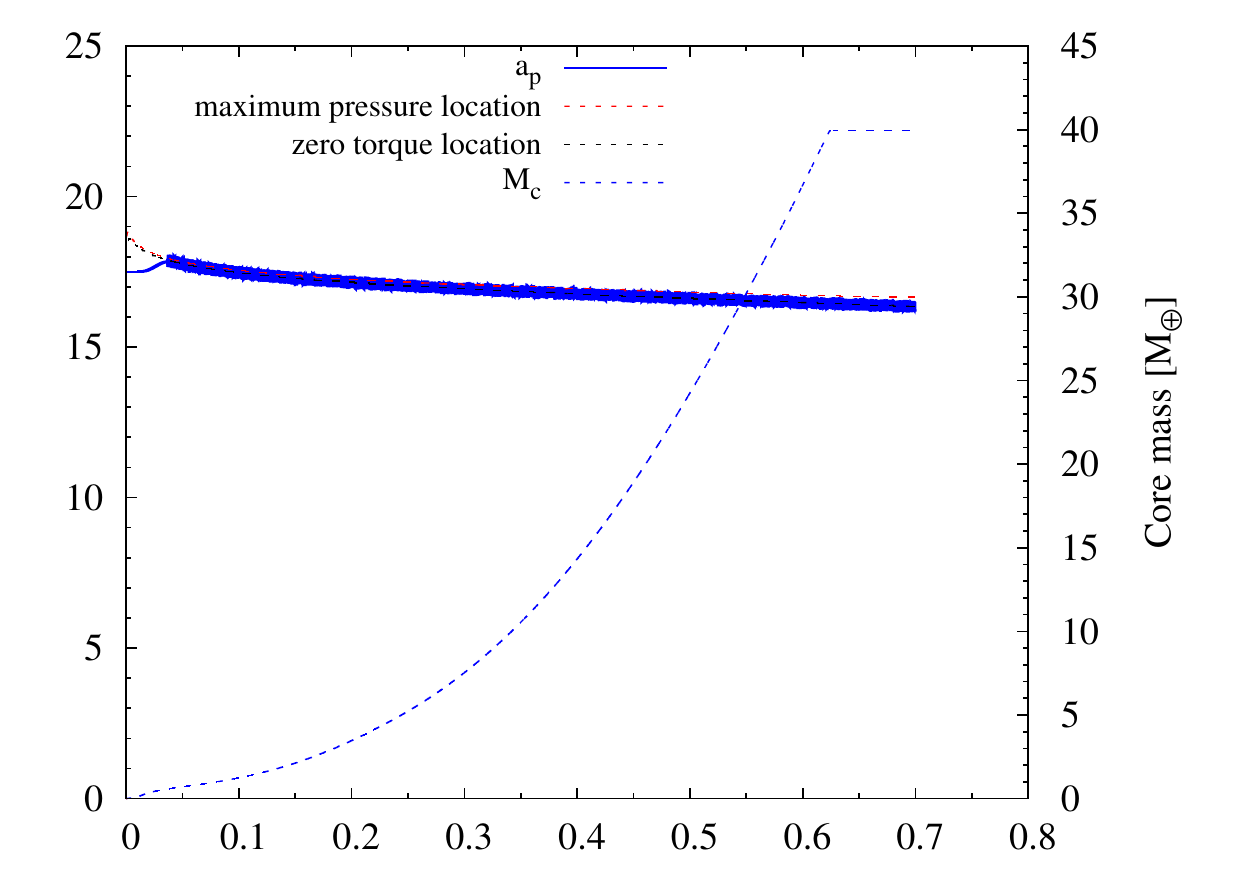} \\
    \centering
    \includegraphics[width= 0.475\textwidth]{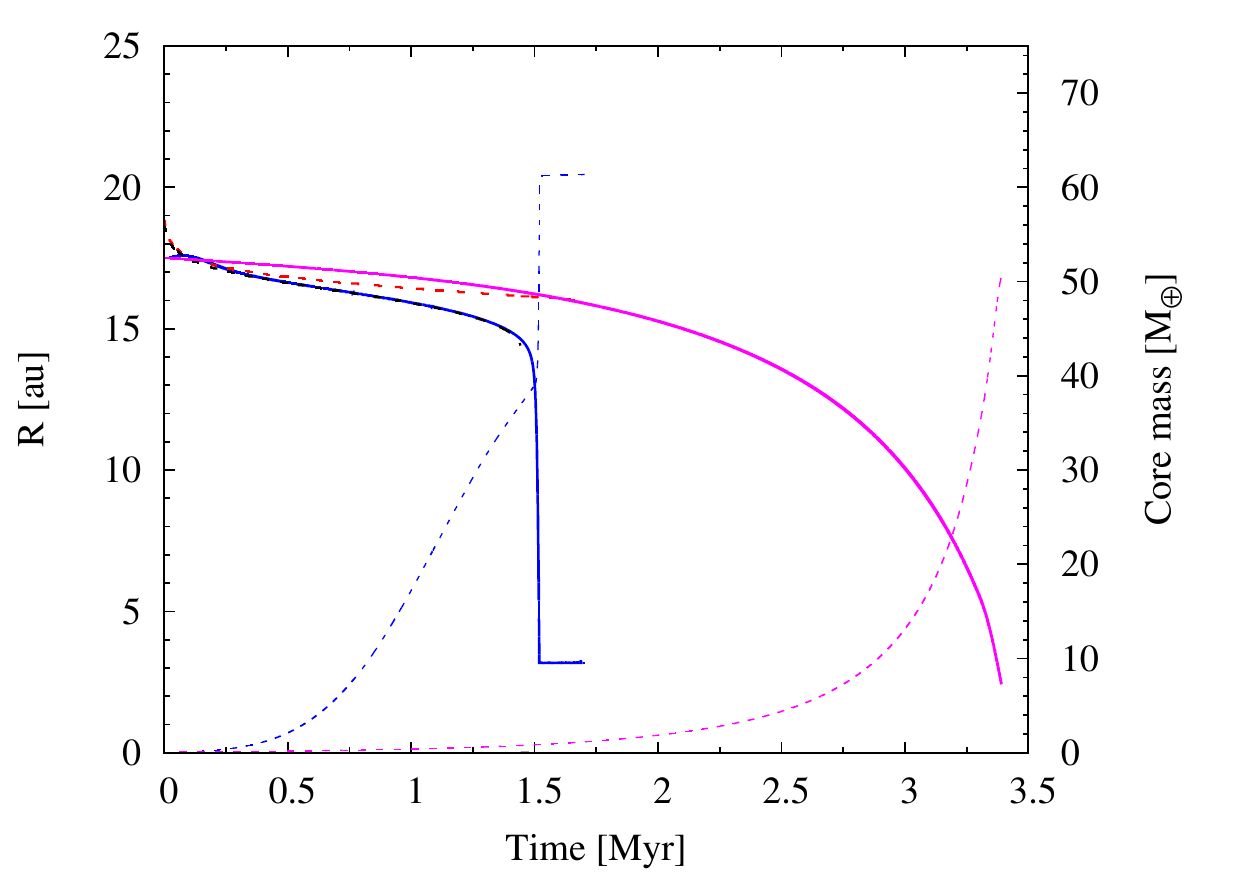} 
    \centering
    \includegraphics[width= 0.475\textwidth]{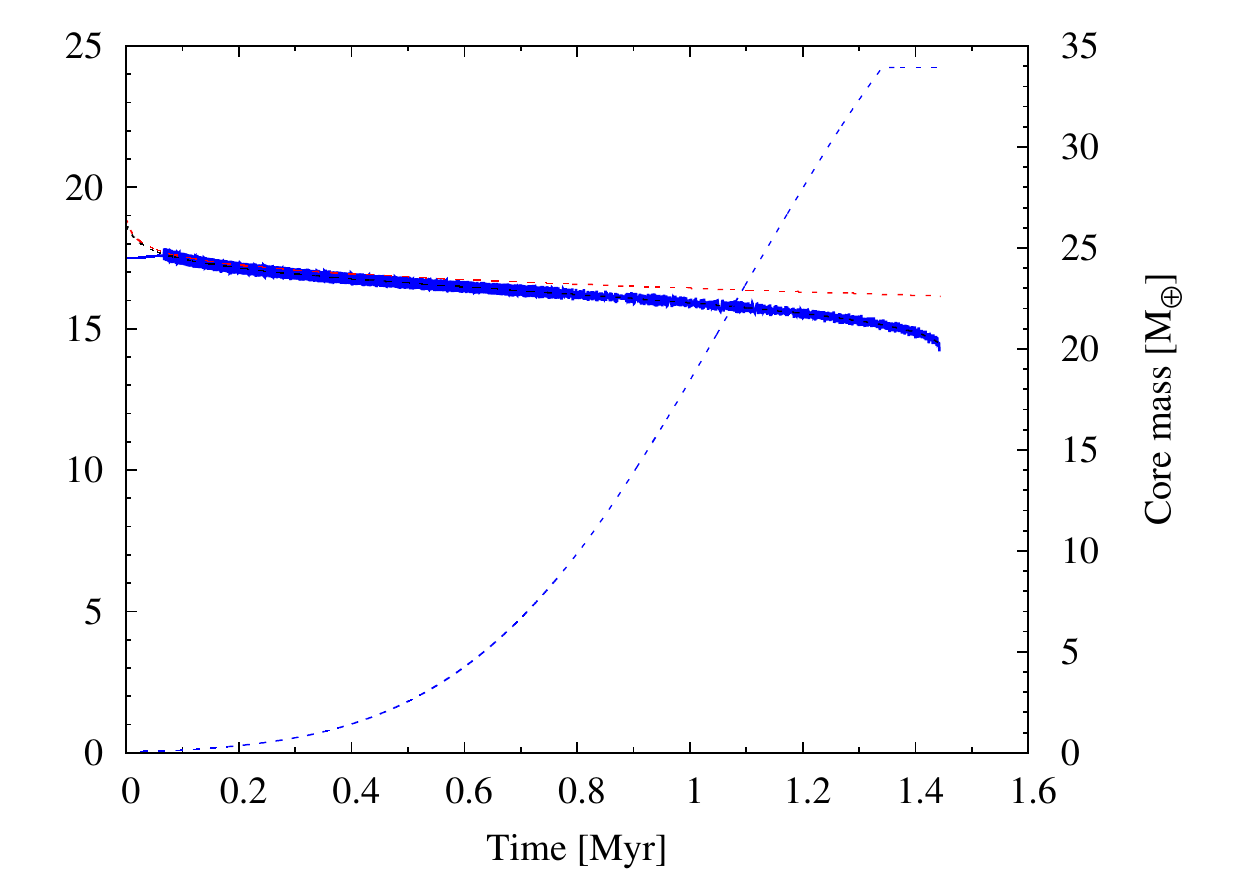} 
    \caption{Time evolution of the planet's semi-major axis (left y-axis ) and planet's core mass (right y-axis) being initially located at 17.5~au for a disk of $0.1~\text{M}_{\odot}$ (top panel) and a disk of $0.05~\text{M}_{\odot}$ (bottom panel). Red dashed line represents the location of the pressure maximum, and black dashed line represents the zero torque location. Left panel represents the case where the planet is trapped at the zero torque location (blue lines) and the case where a dead zone is not present in the disk (pink lines), while right panel represents the case where the semi-major axis of the planet randomly oscillates around the zero torque location after getting trapped. Simulations correspond to a flat disk of with $\alpha_{\text{back}}= 10^{-3}$, $\alpha_{\text{dz}}= 10^{-5}$, $R_{\text{in-dz}}= 2.7$~au and $R_{\text{out-dz}}= 20$~au for the disk with the dead zone and $\alpha= 10^{-3}$ for the disk without the dead zone, and using planetesimals of 100~m of radius. Simulations end when the planet achieves the critical mass. (Color version online).}
  \label{fig:fig3-sec3-2-3}
\end{figure*}

Regarding the outer pressure maximum, the planetesimal accumulation location coincides with the pressure maximum location only for planetesimals having 100 m as radius. For the other planetesimal sizes, the radial drift velocities are lower than the shifting velocity of the outer pressure maximum, so the initial planetesimal accumulations remain far away from the pressure maximum and zero torque location (Fig.~\ref{fig:fig4-sec3-1} bottom panel). Thus, except for planetesimals of 100~m of radius we need a big oscillation amplitude in order to accrete the planetesimal accumulation. For this reason, we only analyze the case of planetesimals of 100~m of radius. Fig.~\ref{fig:fig3-sec3-2-3} shows both the time evolution of the planet's semi-major axis (left y-axis) and the mass of the planetary core (right y-axis). When we do not consider a dead zone in the disk, the planet achieves the critical mass at the inner part of the disk due to type I migration for both disk masses of the disk ($0.1~\text{M}_{\odot}$ and $0.05~\text{M}_{\odot}$). When a dead zone is considered in the disk, but the planet does not oscillate around the zero torque location, for the massive disk the planet is able to achieve the critical mass before the disappearance of the zero torque location. However, for the disk of $0.05~\text{M}_{\odot}$, the planet is not able to achieve the critical mass at the outer region until the zero torque disappears, and it quickly migrates inward reaching the inner migration trap where achieves the critical mass. When the noisy oscillation of the planet's semi-major axis is considered, the planet achieves the critical mass before the zero torque disappears. Thus, the planet become a giant planet at wide orbit, even for a moderate-mass disk. However, for the disk of $0.1~\text{M}_{\odot}$ the formation history of the planet is practically the same indifferently whether the noisy oscillation of its semi-major axis has been considered or not around the zero torque location.

\subsection{Pebble accretion at pressure maxima}
\label{sec:sec3-3}

In this section we analyze the formation of massive cores by pebble accretion at the pressure maxima of the disk. To do so, we have considered a population of pebbles of 1 cm in size incorporating pebble accretion in our model of planet formation. As in \citet{Guilera.2016}, we adopt the pebble accretion rates given by \citet{Lambrechts-et-al-2014},
\begin{eqnarray}
  \frac{dM_C}{dt}\bigg|_{\text{pebbles}} =
  \begin{cases}
     2 \beta R_H^2\Sigma_p(a_P) \Omega_P, \quad \text{if } ~ 0.1 \le \text{St} < 1, \\
    \\
     2 \beta \left( \frac{\text{St}}{0.1} \right)^{2/3} R_H^2\Sigma_p(a_P) \Omega_P, \quad \text{if }~\text{St} < 0.1, 
  \end{cases}  
  \label{eq:eq1-sec3-3}
\end{eqnarray}
where $\text{St}= t_{\text{stop}} \Omega_k$ is the Stokes number, being $t_{\text{stop}}$ the stopping time which depends on the drag regime \citep{Rafikov2004, Chambers2008}, and where we introduce the factor $\beta= \text{min}(1, R_H/H_p)$ in order to take into account a reduction in the pebble accretion rates if the scale height of the pebbles, $H_p$, becomes greater than the Hill radius of the planet. The scale height of the solids at a given distance $R$ from the central star is given by \citep{Youdin2007}
\begin{eqnarray}
  H_p= H_g\sqrt{\frac{\alpha}{\alpha + \text{St}}}.
  \label{eq:eq2-sec3-3}
\end{eqnarray}

\begin{figure}[t]
  \centering
  \includegraphics[width= 0.475\textwidth]{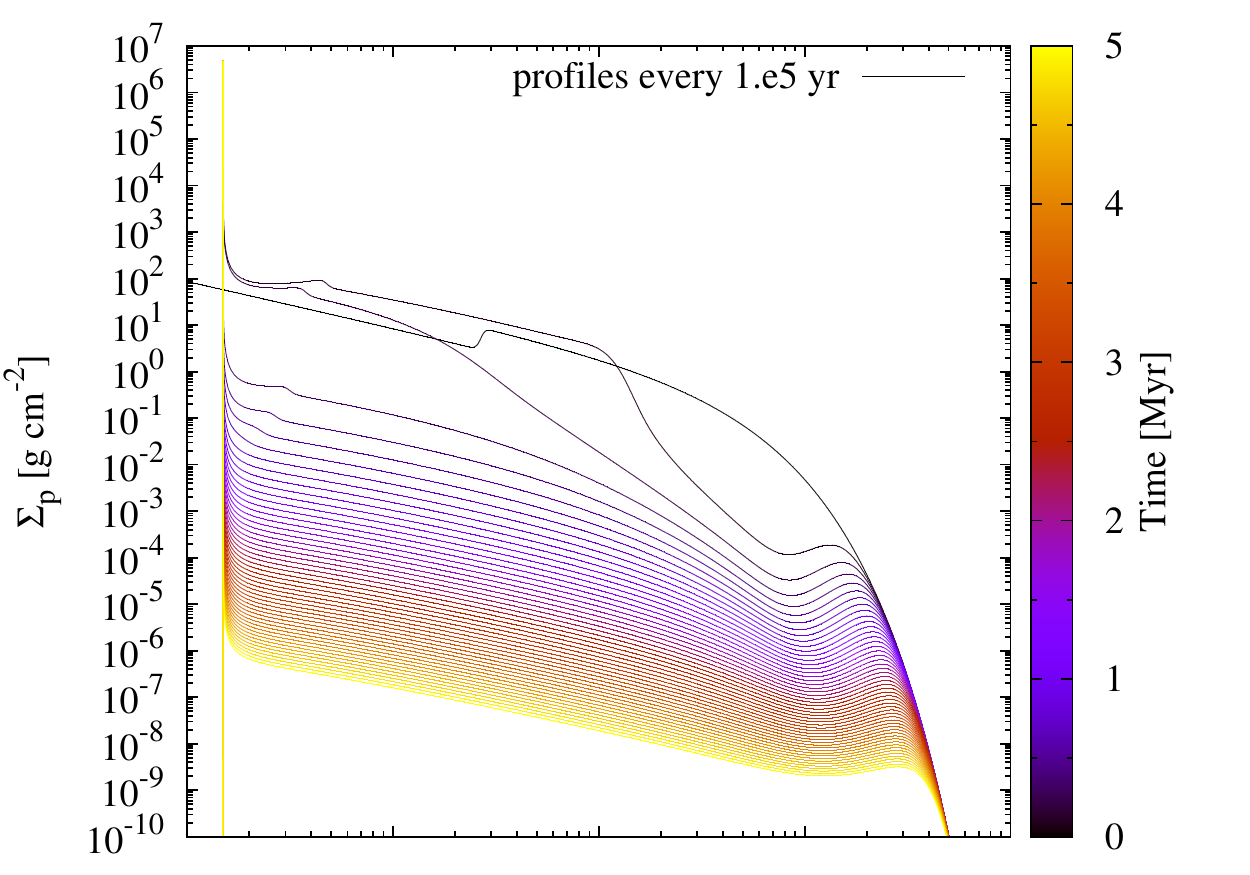} \\
  \includegraphics[width= 0.475\textwidth]{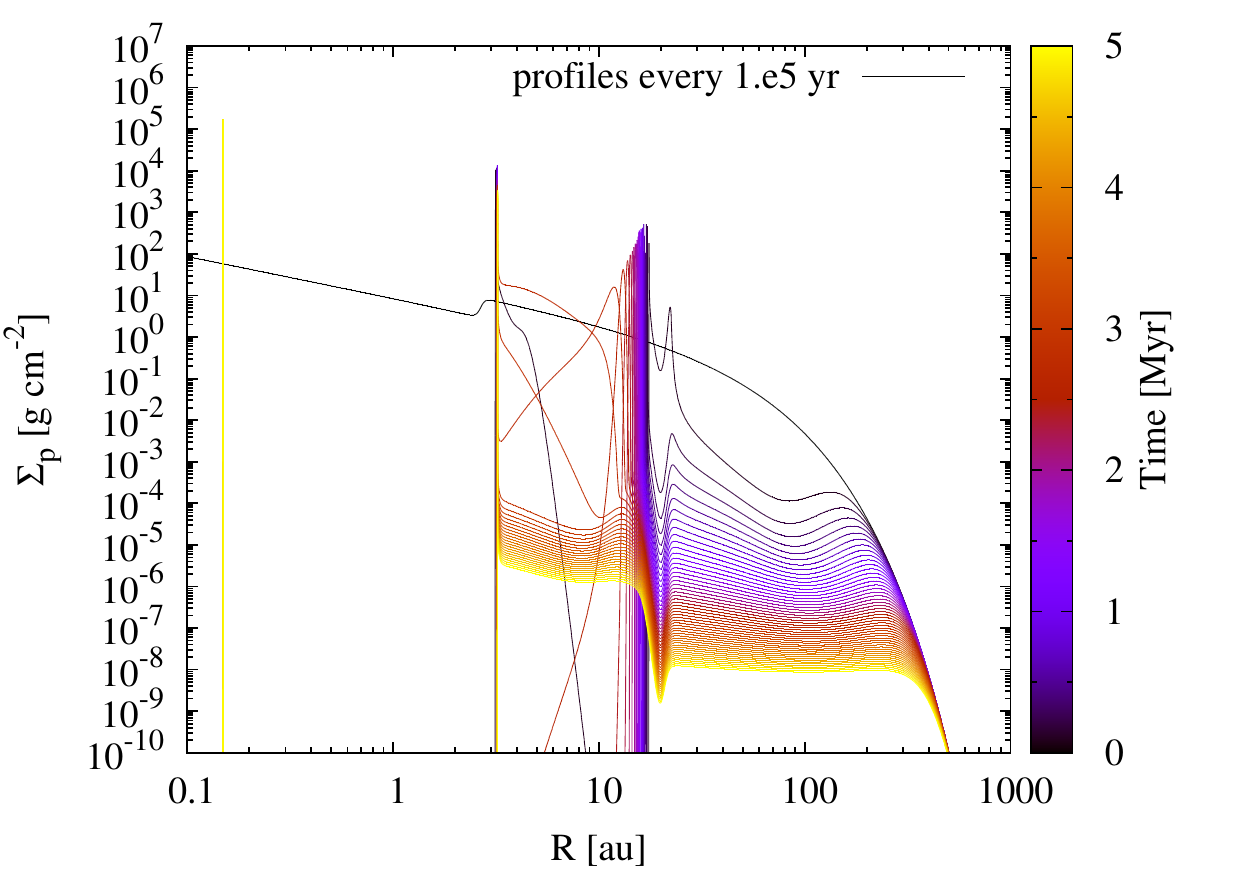}
  \caption{Time evolution of the pebble surface density radial profiles. The top panel corresponds to a disk without a dead zone while bottom panel corresponds to the case of a disk with a dead zone. Simulations correspond to a low-mass flat disk of $M_d= 0.03~\text{M}_{\odot}$ and using $R_{\text{in-dz}}= 2.7$~au, $R_{\text{out-dz}}= 20$~au, $\alpha_{\text{back}}= 10^{-3}$, and $\alpha_{\text{dz}}= 10^{-5}$, when the dead zone is considered. (Color version online).}
  \label{fig:fig1-sec3-3}
\end{figure}

\begin{figure}[t]
  \centering
  \includegraphics[width= 0.475\textwidth]{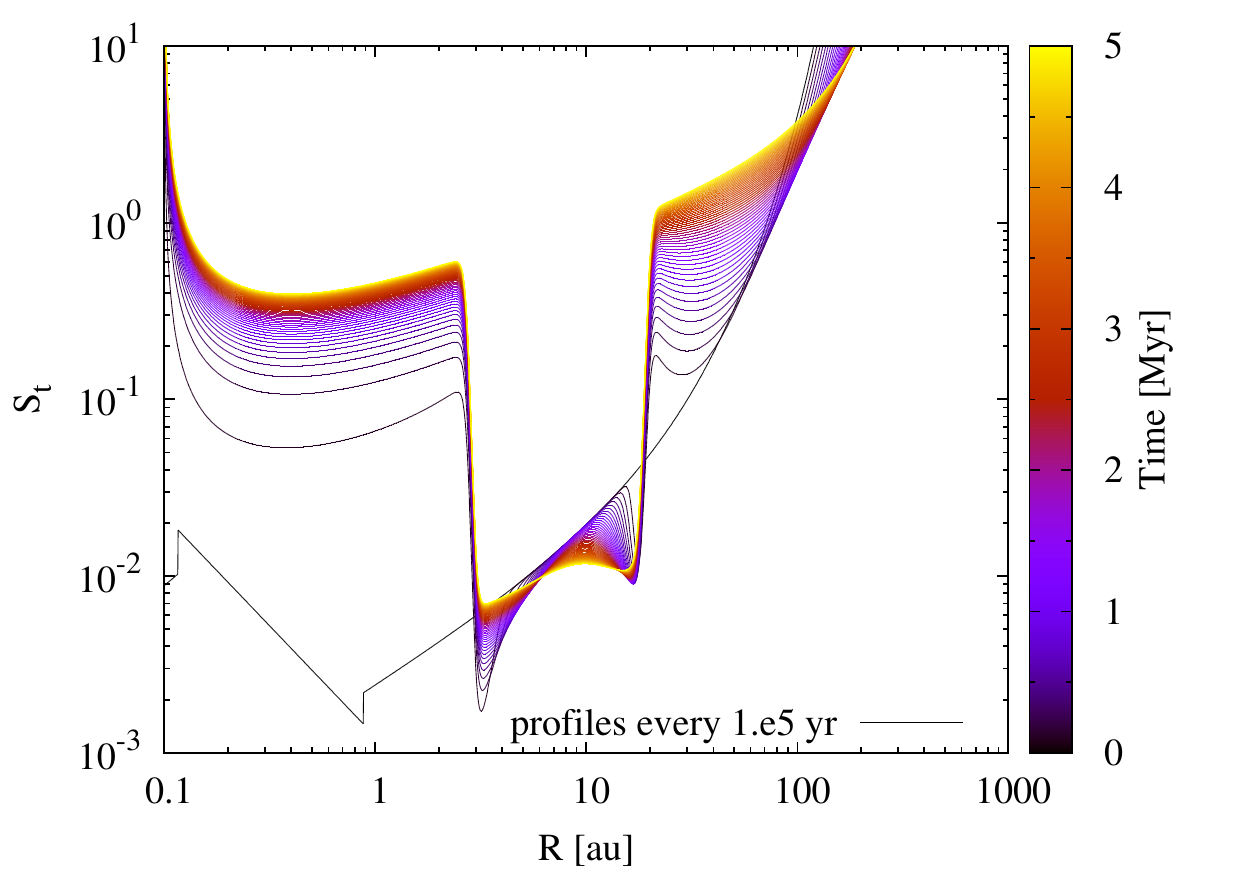} 
  \caption{Time evolution of the Stokes number as function of the distance to the central star for particles of 1 cm. Simulation correspond to a low-mass flat disk of $M_d= 0.03~\text{M}_{\odot}$ with a dead zone using $R_{\text{in-dz}}= 2.7$~au, $R_{\text{out-dz}}= 20$~au, $\alpha_{\text{back}}= 10^{-3}$, and $\alpha_{\text{dz}}= 10^{-5}$. (Color version online).}
  \label{fig:fig2-sec3-3}
\end{figure}

First, we calculate the evolution of a disk without any planet immersed in it. Fig.~\ref{fig:fig1-sec3-3} shows the time evolution of the surface density of the pebble population for a low-mass flat disk of $0.03~\text{M}_{\odot}$. The top panel represents the case where the dead zone is not considered. Initially, the migration of the pebbles from the outer part of the disk increases the pebble surface density in the planet formation region ($R \lesssim 10$~au). However, as time advances all the solid material is deposited in the pressure maximum generated by the inner boundary condition (see Fig.~\ref{fig:fig1-sec3-1}). The bottom panel shows the case where the dead zone is included in the model. We can see that the solid material is quickly accumulated at the pressure maxima of the disk. The pebbles between the star and the inner edge of the dead zone are concentrated at the pressure maximum generated by the inner boundary condition. The pebbles in the dead zone quickly migrate to the inner pressure maximum generated by the snowline (at $\sim 3$~au), increasing the surface density of solids there approximately by three orders of magnitude. The outer pressure maximum being developed at the outer edge of the dead zone concentrates all the inward drifting solid material of the outer disk. When this pressure maximum disappears, solids are drifted again toward the star until reaching the inner pressure maximum. In Fig.~\ref{fig:fig2-sec3-3}, we show that particles of 1 cm behave as pebbles (with $\text{St} \le 1$) for distances $R \lesssim 20~au$, including the region of the disk where pressure maxima are developed.  

\begin{figure}[t]
  \centering
  \includegraphics[width= 0.475\textwidth]{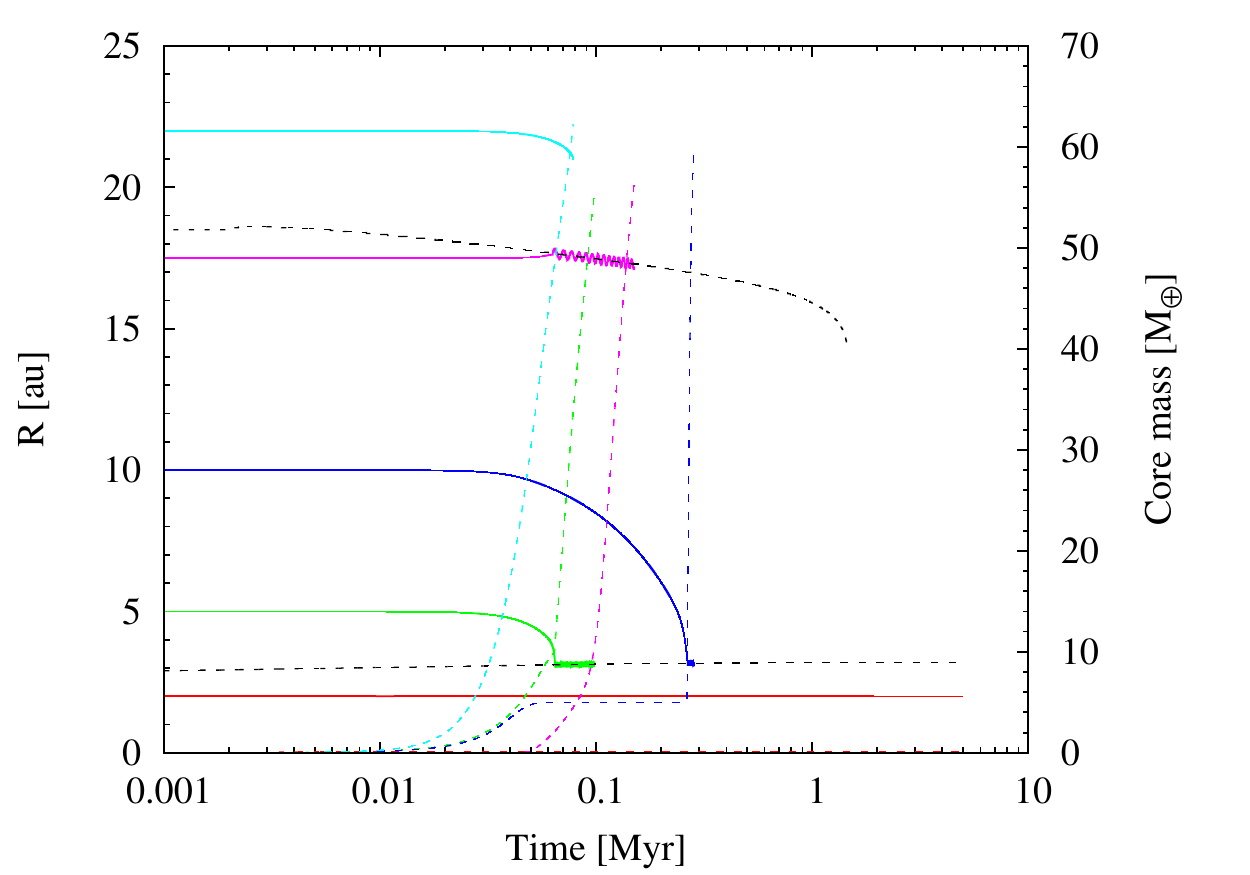}
  \caption{Time evolution of the planet's semi-major axis (solid lines) and planet's core mass (dashed lines) for different initial locations. Black dashed line represents the zero torque location. Simulations correspond to a low-mass flat disk of $M_d= 0.03~\text{M}_{\odot}$ with a dead zone using $R_{\text{in-dz}}= 2.7$~au, $R_{\text{out-dz}}= 20$~au, $\alpha_{\text{back}}= 10^{-3}$, and $\alpha_{\text{dz}}= 10^{-5}$. Simulations stop after 5~Myr or when the planet reaches the cross-over mass. (Color version online).}
  \label{fig:fig3-sec3-3}
\end{figure}

Finally, in Fig.~\ref{fig:fig3-sec3-3} we show the time evolution of the semi-major axis and the mass of the giant planet's core for different starting positions. When the starting location of the planet is at 2~au, the core does not grow nor migrate. This is due to the fact that pebbles at about 2~au are drifted very quickly to the inner radius of the disk, therefore all the outer solid material moving inward is collected at the the pressure maximum at the inner edge of the dead zone depleting the feeding zone of the core very quickly. The situation for the other cases is very different. When the initial position of the core is at 5~au and 10~au, the core grows rapidly reaching a few Earth masses due to the high pebble accretion rates. Thus the growing core also migrates quickly achieving the inner zero torque location in $\sim 10^5$~yr. When reaching the zero torque location, a rapid growth of the core begins due to accretion of the pebbles accumulated at the pressure maximum being close to the zero torque's place. Similar result is obtained when the core is initially placed at 17.5 au. In this case, the planet slightly migrates outward until reaching the outer zero torque location, then a rapid growth of the core begins due to accretion of the pebbles accumulated at the outer pressure maximum. Finally, when the initial location of the core is at 22~au it reaches the cross-over mass in less than $10^5$~yr before it achieves the outer zero torque location. This is due to the fact of the high pebble accretion rates and the inward migration of all the solid material of the outer part of the disk.

\section{Summary \& Conclusions}
\label{sec:sec4}

In the present study we have investigated the efficiency of giant planet formation near the density/pressure maxima that may exist in protoplanetary disks. We have assumed two locations in a disk, where density/pressure maxima can be developed, these are the water snowline and the outer edge of the dead zone. The attractive features of these locations are twofold: at density maxima the torque responsible for the type 1 migration of planetary cores can vanish, thus they can act as migration traps. Moreover, near to a density maximum, a pressure maximum can also be developed. Approaching a pressure maximum, the gas orbital velocity tends to the circular Keplerian one implying that the gas drag on the solid particles becomes zero. Thus a pressure maximum is a preferential place for dust coagulation, and concentration of planetesimals. Although the pressure maximum and migration trap do not exactly coincide, their proximity makes these places favorable for the oligarchic growth of embryos, which can lead to a rapid formation of a giant planets in a considerably shorter time than the disk's lifetime. 

Our physical model incorporates (i) the time evolution of a viscosity driven axissymmetric gaseous disk, (ii) the radial drift of planetesimals due to the aerodynamic drag force arising from the ambient sub/super-Keplerian gas disk, (iii) the type 1 migration of a growing planetary embryo, and finally (iv) its hydrostatic growth by planetesimal and gas accretion until the gaseous runaway phase, which happens when the mass of the gaseous envelope becomes larger than the mass of the solid core. The differential equations describing the above processes have been numerically solved simultaneously and self-consistently. The two pressure maxima have been developed by applying a strong reduction of the $\alpha$ viscosity between the radial distances of the water snowline and the outer edge of the dead zone, mimicking the effect of the accretionally nearly inactive dead zone. In some of our simulations a random oscillation of the growing embryo's semi-major axis around the density maximum has also been incorporated. It is noteworthy to mention that such oscillations appear in HD/MHD simulations due to the gravitational interaction between the planetary core and the large scale vortex, being the latter the 2D/3D manifestation of a density/pressure maximum.

First we have studied the development of the density/pressure maxima in flared and flat disk models. We have found that at the onset of simulations, the density and pressure maxima have been developed in both cases. In the case of a flared disk, both the zero torque location and the pressure maximum have existed during the whole length of the numerical simulations (being $5\times 10^6$ years) at the water snowline. It is also clearly seen, however, that these locations do not coincide, but the distance between them remains almost constant. Regarding the outer edge of the dead zone, the zero torque location vanishes in a million years timescale, while the pressure maximum lasts for longer time, but their positions deviate relatively quickly. In the flat disk models the inner pressure maximum and the migration trap have also survived during the whole simulation time. The pressure maximum and the zero torque location have also disappeared, but in longer timescale than in the case of a flared disk. Their positions have also deviated much slower than in the case of a flared disk. Thus as a conclusion, we can state that both in flared and flat disks, at the inner density/pressure maximum the formation of a giant planet is supported during the disk's lifetime. Another conclusion is that in flat disks giant planet formation might be more likely at the outer edge of the dead zone than in flared disks, since in flat disks the migration trap and the pressure maximum exists for a longer time. 

Simulations with planetesimals of different sizes have revealed that their accumulation at the pressure maxima is size dependent. We have performed simulations with planetesimal sizes of 0.1, 1, 10, and 100~km. Since the 0.1~km sized planetesimals are the most sensitive to the drag force, their accumulation is the most effective, being followed by the other planetesimal sizes in increasing order. Since the inner pressure maximum moves slightly outwards, all of the inward drifting planetesimals are trapped there, indifferently of their sizes. The density/pressure maximum at the outer edge of the dead zone moves slightly inwards, thus only those planetesimals can be trapped there which are drifted faster than the inward shift of the outer pressure maximum. We have found that practically only the planetesimals with sizes of 0.1~km can be trapped efficiently in the outer pressure maximum. 

Having investigated planetesimal accumulation at the pressure maxima, we have studied the in situ formation of giant planets meaning that we have not considered planet-disk interaction and placed embryos to fixed locations, exactly where planetesimal accumulation occurs. We have found in all of our simulations (except the outer dead zone edge and 100~km sized planetesimals) that the formation time of a giant planet is much shorter at pressure maxima than in disk models without dead zone. (We recall that by formation time we mean the time needed until the solid core's mass equals to the gaseous envelope's mass.) In situ giant planet formation is however not physical, since due to the planet-disk interaction, the growing core either migrates or is trapped at a zero torque location, which for the outer edge of the dead zone deviates from the accumulation of planetesimals. Therefore as a next step, we have taken into account type 1 migration of the growing core. According to our results, except of the case of 100~km size planetesimals, the formation of giant planets at the water snowline happens well before the disk's lifetime. Giant planet formation at the inner density/pressure maximum is even more effective, if a random oscillation of the semi-major axis of the planetary core is assumed. At the outer edge of the dead zone the giant planet formation is only effective for 0.1~km size planetesimals. If the core's semi-major axis randomly oscillates around the outer edge of the dead zone, the formation time is also shorter than in the case without oscillation. 

Finally, we studied the formation of giant planets at the pressure maxima of the disk by the accretion of pebbles of 1 cm sizes, too. As for the case of planetesimals, pressure maxima act as locations of accumulation of solid material significantly increasing the pebble surface density. We found that due to the high pebble accretion rates, and the accumulated pebbles at the pressure maxima locations, massive cores are formed in a timescale of $10^5$~yr in the inner and outer edges of the dead zone, even for low-mass disks, meaning that giant planet formation via pebble accretion at a pressure maximum is the fastest and most efficient formation scenario in the core accretion model.    

In our study we have found that the inner pressure maximum is always a favorable place for giant planet formation for a wide range of disk's physical parameters, meaning that the formation time is much shorter that the disk's lifetime. The outer edge of the dead zone can also promote giant planet formation but only for pebbles or smaller, sub-kilometer sized planetesimals, and in disk models when the lifetime of the migration trap long enough enabling the trapped core to accrete enough material for the onset of the runaway gas accretion. 

It is important to note that during our investigations the positions of the inner and outer pressure maxima are kept fixed in time. We are aware the fact that this might be a simplification of a more complex problem not addressed in this work. As we mentioned in Sec.\ref{sec:sec1}, we locate the inner pressure maximum to the position of the water snowline that develops due to the condensation of water resulting in a sudden increase of the solid-to-gas ratio. The condensation of water reduces the number of free electrons thus increases the resistivity of the gas suppressing the MRI driven turbulence. Water condensation happens when the disk's midplane temperature drops below 170 K. During the disk's lifetime its temperature profile evolves as a function of the distance from the star, therefore the position where water condensation takes place evolves, too \citep{Garaud-Lin2007, Oka.et.al.2011}. On the other hand, the time-evolution of snowline's position might be a more complex issue than simply monitoring the place where $T(R_\text{drop}) \sim 170$K. For instance \citet{Ciesla-Cuzzi2006} also took into account the sublimation/condensation of the appropriate amount of ice or vapour to maintain the equilibrium of the vapour pressure considering the radial drift of ice rich dust and planetesimals. More recently, \citet{Morbidelli.et.al.2016} found that at certain epoch, $t_\text{crit}$ of disk evolution the radial inward velocity of gas is larger than the speed at which the condensation front moves inward. Thus at larger times $t>t_\text{crit}$, the radius of the temperature drop $R_\text{drop}$ moves in water poor gas, and no water condensation can take place. In that case the pressure maximum might be developed at the interface of the water poor and water rich gas. 

We note, however, that neither of the above scenarios are directly applicable to our case, since the pressure maximum, we assume to develop, traps icy dust grains, and also reduces the gas radial inward velocity. Therefore the above complex issue we intend to investigate in a separate work. Regarding to the outer edge of the dead zone, its location depends on the X-rays and cosmic rays penetrating depth. Therefore, when the gas surface density becomes less than some critical value, the disk can be ionized in its full vertical extent. Thus, it is expected that the outer edge of the dead zone moves in time. \citet{Matsumura.et.al.2009} showed that the outer edge of the dead zone can reach a few au in a few Myr. 

Nevertheless the motion of the snowline and the outer edge of the dead zone may not influence the growth of the embryo itself, since the planetesimals and pebbles will be trapped in the pressure maxima developed at that locations.  Being trapped, the growing cores are moving presumably together with the snowline/dead zone's outer edge as they are changing their place. Therefore an interesting issue, which deserves further investigations, is that at which distance are the snowline/dead zone's outer edge from the star when the planet traps are unable to keep locked the growing cores any longer. So the moving snowline and the outer dead zone edge will certainly affect the final formation place of the giant planet.  

The efficiency of giant planet formation at the pressure maxima in more realistic disk models, will be the subject of a forthcoming research.

\begin{acknowledgements}
We thank W. Lyra, the referee of this work, for his suggestions and comments helping us to improve the work. This work has been initiated during the visit of Zs. S\'andor at the Faculty of Astronomical and Geophysical Sciences of the National University of La Plata, therefore Zs. S\'andor thanks for Prof. P. Cincotta and Dr. N. Maffione for their kind invitation, arrangements, and hospitality. O. M. Guilera is supported by grants from the National Scientific and Technical Research Council and National University of La Plata, Argentina. Zs. S\'andor also thanks the support of the J\'anos Bolyai Research Scholarship of the Hungarian Academy of Sciences.
\end{acknowledgements}

\bibliographystyle{aa} 
\bibliography{biblio} 

\end{document}